\documentclass[prx,aps,twocolumn,superscriptaddress,10pt]{revtex4-2}

\usepackage[utf8]{inputenc}
\usepackage[T1]{fontenc}
\usepackage{color}
\usepackage{amsmath}
\usepackage{amsfonts}
\usepackage{xr-hyper}
\usepackage{hyperref}
\usepackage{amsthm}
\usepackage{bm}
\usepackage{float}
\usepackage{textcomp}
\usepackage{upgreek}
\usepackage{textgreek}
\usepackage{setspace}
\usepackage[percent]{overpic}
\usepackage[table]{xcolor}
\usepackage{placeins}
\usepackage{mathtools}
\usepackage{nameref,zref-xr}
\zxrsetup{toltxlabel}

\hypersetup{
  colorlinks = true,
  allcolors= blue,
}
\usepackage{tikz,pgfplots}
\usepackage{blkarray}
\usetikzlibrary{patterns}
\usetikzlibrary{fadings}
\usetikzlibrary{decorations.pathmorphing,patterns}
\tikzset{snake it/.style={decorate, decoration=snake}}
\usetikzlibrary{arrows, decorations.markings}
\pgfplotsset{compat=1.14}
\tikzset{
vecArrow/.style={
  thick,
  decoration={markings,mark=at position
   1 with {\arrow[scale=2,thin]{open triangle 60}}},
  double distance=1.4pt, shorten >= 10.5pt,
  preaction = {decorate},
  postaction = {draw,line width=1.4pt, white,shorten >= 4.5pt}
  },
innerWhite/.style={
  semithick,
  white,
  line width=1.4pt,
  shorten >= 4.5pt
  }
}

\definecolor{orange}{rgb}{1,0.5,0}
\definecolor{darkgreen}{rgb}{0,0.4,0.1}
\definecolor{cola}{HTML}{5FBDFF}
\definecolor{colb}{HTML}{118DFE}
\definecolor{cold}{HTML}{EEFF0C}
\definecolor{cole}{HTML}{FC9E0A}

\usepackage{tikz}
\newcommand\wye[1][]{%
    \tikz\draw[thin, line cap=round,x=1ex,y=1ex,#1]
    (0,0) -- ++(90:1)
    (0,0) -- ++(-30:1)
    (0,0) -- ++(-150:1);
}

\newcommand{\WidthFigure}{\columnwidth}

\makeatletter
\newcommand{\doublehat}[1]{%
\begingroup%
  \let\macc@kerna\z@%
  \let\macc@kernb\z@%
  \let\macc@nucleus\@empty%
  \hat{\raisebox{.3ex}{\vphantom{\ensuremath{#1}}}\smash{\hat{#1}}}%
\endgroup%
}
\makeatother

\makeatletter
\newcommand{\doublehatSub}[1]{%
\begingroup%
  \let\macc@kerna\z@%
  \let\macc@kernb\z@%
  \hat{\raisebox{-.07ex}{\vphantom{\ensuremath{#1}}}\smash{\hat{#1}}}%
\endgroup%
}
\makeatother

\makeatletter
\DeclareFontFamily{OMX}{MnSymbolE}{}
\DeclareSymbolFont{MnLargeSymbols}{OMX}{MnSymbolE}{m}{n}
\SetSymbolFont{MnLargeSymbols}{bold}{OMX}{MnSymbolE}{b}{n}
\DeclareFontShape{OMX}{MnSymbolE}{m}{n}{
    <-6>  MnSymbolE5
   <6-7>  MnSymbolE6
   <7-8>  MnSymbolE7
   <8-9>  MnSymbolE8
   <9-10> MnSymbolE9
  <10-12> MnSymbolE10
  <12->   MnSymbolE12
}{}
\DeclareFontShape{OMX}{MnSymbolE}{b}{n}{
    <-6>  MnSymbolE-Bold5
   <6-7>  MnSymbolE-Bold6
   <7-8>  MnSymbolE-Bold7
   <8-9>  MnSymbolE-Bold8
   <9-10> MnSymbolE-Bold9
  <10-12> MnSymbolE-Bold10
  <12->   MnSymbolE-Bold12
}{}

\let\llangle\@undefined
\let\rrangle\@undefined
\DeclareMathDelimiter{\llangle}{\mathopen}%
                     {MnLargeSymbols}{'164}{MnLargeSymbols}{'164}
\DeclareMathDelimiter{\rrangle}{\mathclose}%
                     {MnLargeSymbols}{'171}{MnLargeSymbols}{'171}
\makeatother
\DeclareUnicodeCharacter{300}{à}

\DeclareMathAlphabet{\mathsfit}{\encodingdefault}{\sfdefault}{m}{sl}
\SetMathAlphabet{\mathsfit}{bold}{\encodingdefault}{\sfdefault}{bx}{sl}
\newcommand{\tens}[1]{\bm{\mathsfit{#1}}}

\makeatletter
\usepackage{comment}
\let\wfs@comment@comment\comment
\let\comment\@undefined

\usepackage[final]{changes}
\let\wfs@changes@comment\comment
\let\comment\@undefined

\newcommand\comment{%
    \ifthenelse{\equal{\@currenvir}{comment}}
    {\wfs@comment@comment}
    {\wfs@changes@comment}%
}
\makeatother

\definecolor{dgreen}{rgb}{0,0.45,0}

\makeatletter
\@namedef{Changes@AuthorColor}{red}
\colorlet{Changes@Color}{red}
\setaddedmarkup{\textcolor{orange}{#1}}
\makeatother

\begin{document}

\title{
Bond-Network Entropy Governs Heat Transport in Coordination-Disordered Solids
}

\author{Kamil Iwanowski}
\affiliation{Department of Applied Physics and Applied Mathematics, Columbia University, New York (USA)}
\affiliation{Theory of Condensed Matter Group, Cavendish Laboratory, University of Cambridge (UK)}

\author{Gábor Csányi}
\affiliation{Applied Mechanics Group, Mechanics, Materials and Design, Department of Engineering, University of Cambridge (UK)}

\author{Michele Simoncelli}
\email{michele.simoncelli@columbia.edu}
\affiliation{Department of Applied Physics and Applied Mathematics, Columbia University, New York (USA)}
\affiliation{Theory of Condensed Matter Group, Cavendish Laboratory, University of Cambridge (UK)}

\begin{abstract}
Understanding how the vibrational and thermal properties of solids are influenced by atomistic structural disorder is of fundamental scientific interest, and paramount to designing materials for next-generation energy technologies. 
While several studies indicate that structural disorder strongly influences the thermal conductivity, the fundamental physics governing the disorder-conductivity relation remains elusive.
Here we show that order-of-magnitude, disorder-induced variations of conductivity in network solids can be predicted from a ‘bond-network’ entropy, an atomistic structural descriptor that quantifies heterogeneity in the topology of the atomic-bond network.
We employ the Wigner formulation of thermal transport to demonstrate the existence of a relation between the bond-network entropy, and observables such as smoothness of the vibrational density of states (VDOS) and macroscopic conductivity.
We also show that the smoothing of the VDOS encodes information about the thermal resistance induced by disorder, and can be directly related to phenomenological models for phonon-disorder scattering based on the semiclassical Peierls-Boltzmann equation.
Our findings rationalize the conductivity variations of disordered carbon polymorphs ranging from nanoporous electrodes to defective graphite used as a moderator in nuclear reactors.
\end{abstract}
\maketitle
\section{Introduction}
Carbon forms a variety of disordered allotropes useful in several industrial applications: amorphous carbon is a promising material for next-generation electronic and mechanical technologies due to its disorder-tunable electrical conductivity  \cite{tian_disorder-tuned_2023}, high elastic modulus and hardness \cite{shang_ultrahard_2021}; nanoporous carbon is employed as structural nanomaterial \cite{li_nanoporous_2024}, as well as in devices for energy harvesting \cite{statz_charge_2020,liu_structural_2024} or water desalination \cite{simoncelli_blue_2018,jeanmairet_microscopic_2022};  nuclear-grade defective graphite is used as a moderator in nuclear reactors  \cite{liu_damage_2017,farbos_time-dependent_2017,chengCeramicCompositeModerators2022}. This wide range of applications is made possible by the high variability in the macroscopic properties of carbon, which derive from its versatility at the atomic scale \cite{umemoto_body-centered_2010}, particularly its ability to form chain-like bonds between two atoms, planar (graphite-like) bonds involving three atoms, and tetrahedral (diamond-like) bonds between four atoms. The thermal conductivity is a striking example of such high variability  \cite{prasher_turning_2009, mehew_enhanced_2023, ye_anomalous_2024}, as it can change by more than four orders of magnitude --- from the ultralow (< 1 W/mK) conductivity of nanoporous carbon  \cite{kern_thermal_2016} to the ultrahigh (>2000 W/mK) conductivity of graphene \cite{pop_thermal_2012, app_Lee2015, pereira_divergence_2013, barbarino_intrinsic_2015, majee_dynamical_2018, braun_spatially_2022, han_thermal_2023}, graphite \cite{fugallo_thermal_2014, zhang_temperature-dependent_2016, huberman_observation_2019, machida_phonon_2020, Jeong2021, ding_observation_2022, huang_observation_2022, huang_graphite_2024, dragasevic_viscous_2024}, and diamond \cite{balandin_thermal_2011,goblot_imaging_2024}. 
Past works studied structural disorder and thermal conductivity in specific classes of carbon polymorphs, which include: (i) experiments, e.g., in amorphous carbon \cite{morath_picosecond_1994,hurler_determination_1995,bullen_thermal_2000, shamsa_thermal_2006, scott_thermal_2021, arlein_optical_2008, chen_thermal_2000}, graphite with variable degree of irradiation-induced defects  \cite{maruyama_neutron_1992, wu_neutron_1994, snead_reduction_1995, snead_thermal_1995, bonal_neutron_1996, ishiyama_effect_1996, barabash_effect_2002, snead_accumulation_2008, campbell_property_2016, heijna_comparison_2017, maruyama_dimensional_2019}, and nanoporous carbon with different pore-size distributions \cite{kern_thermal_2016};  (ii) molecular dynamics (MD) simulations \cite{galli_structural_1989,thomas_predicting_2010,zhou_million-atom_2025}, e.g.,  in amorphous carbon \cite{giri_atomic_2022,minamitani_relationship_2022,moon_crystal-like_2025, lv_phonon_2016, suarez-martinez_effect_2011, zhang_thermal_2017, wang_density_2025}, electron-irradiated graphite  \cite{farbos_time-dependent_2017}, and nanoporous carbon \cite{suarez-martinez_effect_2011,jung_unusually_2017,wang_density_2025}.
However, each of those past studies focused on one (or few) of the aforementioned special classes of carbon polymorphs, without investigating similarities and differences between their atomistic structure and macroscopic conductivity. As a result, the structure-conductivity relation is far from being fully understood.

Here, we address this long-standing question relying on the Wigner formulation of thermal transport  \cite{simoncelli_unified_2019,simoncelli_wigner_2022,simoncelli_thermal_2023} in conjunction with 
the machine-learned Gaussian Approximation Potential (GAP)  \cite{rowe_accurate_2020}.  We employ the latter to describe with quantitative accuracy structural, vibrational, and thermal properties of 23 disordered carbon polymorphs belonging to five different classes: amorphous, irradiated, phase separated, variable porosity, and nanoporous Carbide-Derived Carbon.
We find that disorder in the atomic bond network induces order-of-magnitude variations in the thermal conductivity.
To quantify such disorder and its relation with conductivity, we decompose the solid into a collection of local atomic environments (LAEs), and we
show that in carbon these can be comprehensively characterized using the $H_1$ barcode  ---  a ring-based structural descriptor which is formally defined in terms of the first homology group  \cite{schweinhart_statistical_2020}.
After discussing how the size (i.e., number of atoms $n$) in the LAE determines the resolution with which structural heterogeneity can be resolved, we define a bond-network entropy (BNE) that, in the presence of disorder, grows with LAE's size $n$. 
We demonstrate that the BNE's growth rate with $n$ (${\rm BNE}(n)/ n$) quantifies disorder in the atomic bond network, distinguishing the disordered phases of carbon.
Most importantly, we rely on these insights to elucidate how disorder affects 
observables such as the smoothness of the vibrational density of states (VDOS) \cite{chumakov_role_2014} and the macroscopic thermal conductivity \cite{bullen_thermal_2000,shamsa_thermal_2006, morath_picosecond_1994, scott_thermal_2021}.
Finally, we rationalize the physics underlying the correlation between ${\rm BNE}(n)/ n$ and conductivity by formally connecting the disorder-induced repulsion between vibrational eigenstates to the thermal resistance caused by disorder.
Specifically, from the smoothness of the VDOS and diffusivity, we determine the transport lengthscales for all vibrational excitations in irradiated graphite, 
showing that the characterization of transport lengthscales is not limited to low-frequency modes that approximately display a band structure blurred by disorder --- i.e., a dynamical structure factor (DSF), reminiscent of three acoustic bands with disorder-dependent linewidth \cite{fiorentino_hydrodynamic_2023, fiorentino_effects_2025, moon_crystal-like_2025,zhang_how_2022}.
Ultimately, we elucidate the disorder-conductivity relation across a comprehensive range of disordered carbon polymorphs, paving the way for the theory-driven optimization of technologies ranging from nanoporous-carbon-based supercapacitors to nuclear reactors.

\section{Wigner formulation of thermal transport}
\begin{figure*}
\vspace*{-3mm}
\includegraphics[width=\textwidth]{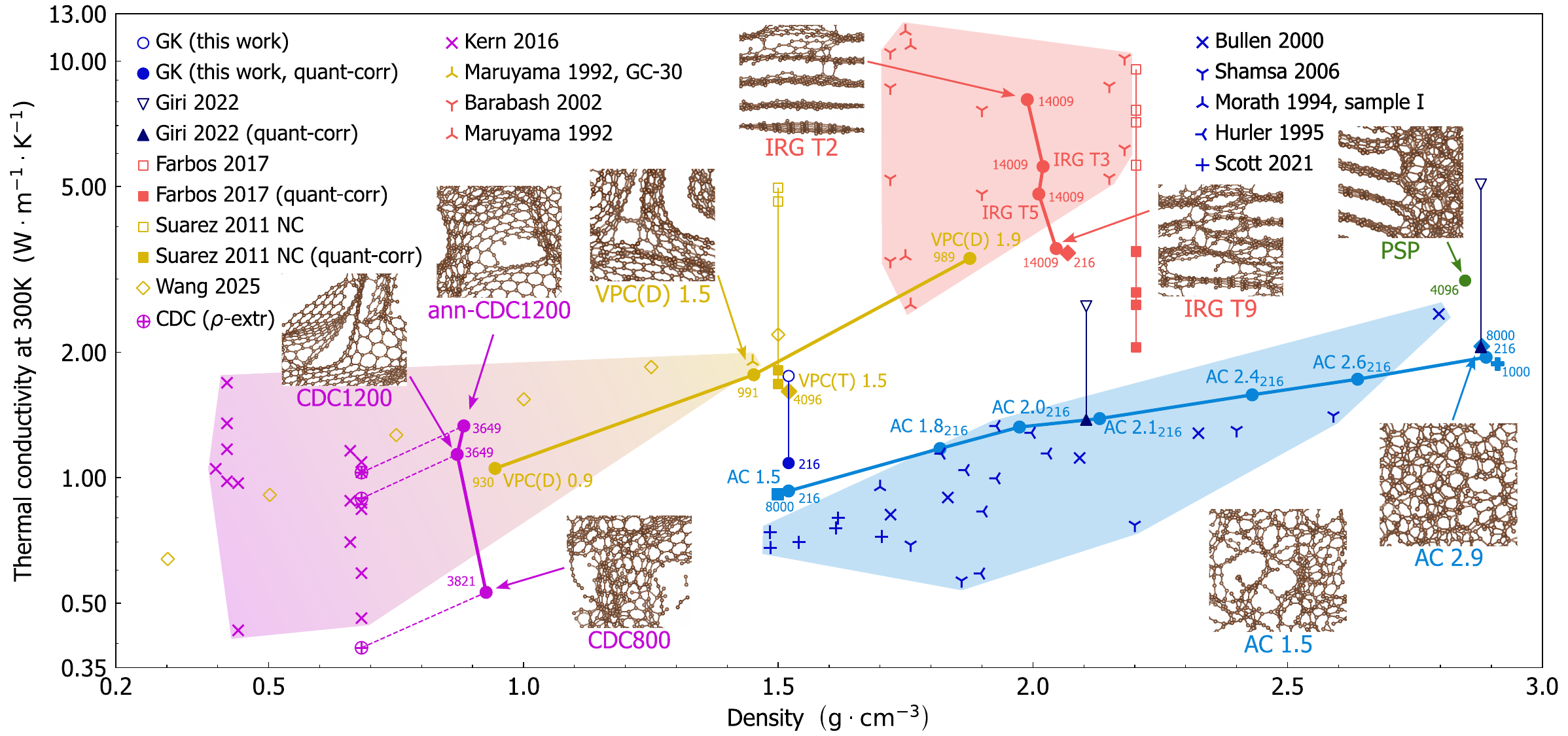}\\[-3mm]
  \caption{
  \textbf{Room-temperature conductivity as a function of density in various classes of carbon polymorphs.} 
  Carbide-derived carbon (CDC) is purple, variable-porosity carbon (VPC) is yellow, amorphous carbon (AC) is blue, irradiated graphite (IRG) is red, and phase-separated phase (PSP) is green.
  Filled markers are results from this work, empty markers are direct results from MD simulations (irradiated graphite from \citet{farbos_time-dependent_2017}, nanoporous carbon from \citet{wang_density_2025} and \citet{suarez-martinez_effect_2011}, and amorphous carbon from \citet{giri_atomic_2022} and from this work).
  The filled markers connected by solid lines to empty markers show results postprocessed from classical MD simulations \cite{farbos_time-dependent_2017, suarez-martinez_effect_2011, giri_atomic_2022}; each marker represents the conductivity after applying the correction to account for the quantum Bose–Einstein statistics of vibrations discussed in Appendix~\ref{appendix_Q_corr}.
  Purple \scalebox{0.8}{$\bigoplus$} markers, connected by dashed lines to solid circles, represent extrapolations of CDC conductivities from the simulated density ($\approx 0.9$ g/cm$^3$) to the experimental density ($\approx 0.7$ g/cm$^3$) (see text for details).
  The unfilled markers \textsf{x}, \wye[rotate=180], \wye, \wye[rotate=90], \textsf{+} are  experiments in carbide-derived carbon \cite{kern_thermal_2016}, irradiated glassy carbon \cite{maruyama_neutron_1992}, irradiated graphite \cite{barabash_effect_2002, maruyama_neutron_1992} and amorphous carbon \cite{bullen_thermal_2000, shamsa_thermal_2006, morath_picosecond_1994, hurler_determination_1995, scott_thermal_2021}. Shaded areas show convex hulls of experimental conductivities for different polymorph classes.
  Each conductivity prediction from this work is labeled with the structure name and the number of atoms in the simulation.
  }
  \label{fig:conductivity_density}
\end{figure*}

To investigate the atomistic physics underlying the conductivity of disordered carbon, we employ the Wigner Transport Equation  \cite{simoncelli_wigner_2022} (WTE).
Such an equation accounts for the quantum Bose-Einstein statistics of vibrations, anharmonicity, and disorder;
its solution yields a thermal conductivity expression that comprehensively describes solids ranging from crystals to glasses \cite{simoncelli_thermal_2023,simoncelli_temperature-invariant_2025}. 
In the following we employ the regularized WTE conductivity expression \cite{simoncelli_thermal_2023} to ensure a proper bulk-limit extrapolation in the presence of disorder:
\begin{equation}
\begin{split}
    \kappa{=}\frac{1}{\mathcal{V}{N_{\rm c}} }{\sum_{\bm{q},s,s'}}&\!
    \frac{\omega_{\bm{q}s}{+}\omega_{\bm{q}s'}}{4}\!\left(\frac{C_{\bm{q}s}}{\omega_{\bm{q}s}}{+}\frac{C_{\bm{q}s'}}{\omega_{\bm{q}s'}}\right)\!
    \frac{\rVert\tens{v}(\bm{q})_{s,s'}\lVert^2}{3}\\
    \times&\pi\mathcal{F}_{[\eta;\Gamma_{\bm{q}s}{+}\Gamma_{\bm{q}s'}]}(\omega_{\bm{q}s}-\omega_{\bm{q}s'})\;,
\label{eq:thermal_conductivity_combined}
\end{split}
\raisetag{5mm}
\end{equation}
where $\mathcal{V}$ is the volume of the reference cell of the solid,  $s$ is a mode index and $\bm{q}$ is a wavevector that labels vibrations  ---  the necessity to consider $\bm{q}$ depends on the size and degree of disorder in the solid; for strongly disordered systems described with very large ($\gtrsim 5000$ atoms) reference cell, it is sufficient to consider only $\bm{q}{=}\bm{0}$ (see Fig.~4 in Ref.~\cite{simoncelli_thermal_2023}, Sec. III.A and IV.D in Ref.~\cite{harper_vibrational_2024}, and we will discuss size effects also in Fig.~\ref{fig:conductivity_density}).
$\omega_{\bm{q}s}$ and $C_{\bm{q}s}$ are the frequency and the quantum specific heat of the vibration $\bm{q}s$:
\begin{equation} \label{eq:quantum_specific_heat}
    C_{\bm{q}s} = C[\omega_{\bm{q}s}, T] = \frac{\hbar^2 \omega_{\bm{q}s}^2}{k_{\rm B} T^2} \mathsfit{N}_{\bm{q}s} (\mathsfit{N}_{\bm{q}s} + 1),
\end{equation}
where $\mathsfit{N}_{\bm{q}s}$ is the Bose-Einstein distribution at temperature $T$: $\mathsfit{N}_{\bm{q}s} = [\exp{(\hbar \omega_{\bm{q}s} / k_{\rm B} T)} - 1]^{-1}$.
$\tens{v}(\bm{q})_{s,s'}$ is the velocity operator  --- 
its diagonal elements are the usual group velocities, while its off-diagonal non-degenerate elements describe couplings between different vibrations.
$\mathcal{F}_{[\eta;\Gamma_{\bm{q}s}{+}\Gamma_{\bm{q}s'}]}$ is a two-parameter Voigt distribution, employed to numerically preserve in finite-size models of disordered solids the physical property that neighbouring (quasi-degenerate) vibrational modes can interact and conduct heat even in the limit of vanishing intrinsic linewidths $\Gamma_{\bm{q}s}$ (due to third-order anharmonic interactions and isotope-mass impurities).
Specifically, $\eta$ is a computational parameter that has to be numerically converged \cite{simoncelli_thermal_2023,harper_vibrational_2024} and is of the order of the average energy-level spacings, see Supplementary Materials (SM) for details. When 
$\Gamma_{\bm{q}s}{+}\Gamma_{\bm{q}s'}{\gg} \eta$,
the Voigt distribution reduces to a Lorentzian {$L[\omega, \Gamma] = \frac{1}{\pi} \frac{\Gamma/2}{\omega^2 + (\Gamma/2)^2}$} with full width at half maximum (FWHM) $\Gamma$
determined by the sum of the intrinsic linewidths of mode $\bm{q}s$ and $\bm{q}s'$, while in the opposite limit $\Gamma_{\bm{q}s}{+}\Gamma_{\bm{q}s'}{\ll} \eta$ it reduces to the Gaussian representation of the Dirac delta. In formulas,
\begin{equation}
    \mathcal{F}_{[\eta;\Gamma_{\bm{q}s}{+}\Gamma_{\bm{q}s'}]} {\rightarrow}
    \begin{cases}
        L[\omega_{\bm{q}s} {-} \omega_{\bm{q}s'},\!  \Gamma_{\bm{q}s}{+}\Gamma_{\bm{q}s'}], \rm{for}\;  \Gamma_{\bm{q}s}{+}\Gamma_{\bm{q}s'}\gg \eta \\
        \frac{1}{\pi \eta} \exp{(-\frac{1}{\pi} \frac{(\omega_{\bm{q}s} - \omega_{\bm{q}s'})^2}{\eta^2})}, \rm{for}\; \Gamma_{\bm{q}s}{+}\Gamma_{\bm{q}s'} \ll \eta
    \end{cases}
\label{eq:voigt}
\end{equation}
Eq.~(\ref{eq:voigt}) implies that: for
$\Gamma_{\bm{q}s}{+}\Gamma_{\bm{q}s'}{\gg} \eta$
 Eq.~(\ref{eq:thermal_conductivity_combined}) yields the standard anharmonic WTE conductivity \cite{simoncelli_wigner_2022}, while in the opposite limit
  $\Gamma_{\bm{q}s}{+}\Gamma_{\bm{q}s'}{\ll} \eta$ Eq.~(\ref{eq:thermal_conductivity_combined}) numerically reduces to the harmonic Allen-Feldman (AF) \cite{allen_thermal_1989} conductivity.\\

\section{Conductivity of disordered carbon polymorphs} \label{sec:tc_dc}
Fig.~\ref{fig:conductivity_density} shows the room-temperature conductivity of five structurally different classes of carbon polymorphs, obtained evaluating Eq.~(\ref{eq:thermal_conductivity_combined}) with the quantum-accurate GAP potential \cite{rowe_accurate_2020}. In particular,
carbide-derived carbon (CDC) is a low-density ($\leq 1.0$ g/cm$^3$) solid consisting of nanoporous, curved graphene sheets with tunable amount of coordination defects \cite{petkov_local_1999, acharya_simulation_1999}. CDC draws its name from the carbide precursors (TiC or SiC) used in the chlorination-based synthesis procedure  \cite{palmer_modeling_2010, kern_thermal_2016}; its properties depend on the chlorination temperature, which is therefore reported in the structure name \cite{palmer_modeling_2010} (e.g., CDC800 and CDC1200 refer to materials synthesized with a chlorination temperature of 800 and 1200 $^\circ$C, respectively, and ann-CDC1200 denotes a CDC1200 structure further annealed after synthesis, see SM).
Variable-porosity carbon (VPC) is constituted of curved graphene sheets with variable porosity and rare coordination defects (see SM for details on coordination number distributions); their density ranges from 0.9 to 1.9 g/cm$^3$ \space \cite{deringer_towards_2018, de_tomas_transferability_2019, suarez-martinez_effect_2011, de_tomas_graphitization_2016}. Amorphous carbon (AC) is a packed structure with variable density from 1.5 to 3.5 g/cm$^3$, where atoms have coordination number that ranges from two to four and its average increases with density  \cite{bullen_thermal_2000,shamsa_thermal_2006, giri_atomic_2022, moon_crystal-like_2025, wang_density_2025}. Irradiated graphite (IRG) features inter-layer bonding defects (interstitial atoms, dislocations, and point defects) caused by electron or neutron irradiation, in concentration proportional to the irradiation time \cite{farbos_time-dependent_2017}.
Finally, the phase-separated phase (PSP) structure blends a mostly 3-fold coordinated graphitic phase with a mostly 4-fold coordinated amorphous-carbon phase  \cite{de_tomas_transferability_2019}.

The room-temperature conductivity of these polymorphs varies by more than one order of magnitude, from 0.5 W/mK for CDC800 to 8.1 W/mK for IRG with the lowest irradiation time (`IRG T2' corresponds to an exposure to irradiation of 2 min). Notably, we find that the thermal conductivity-density relation highly depends on the type of disorder. For AC and VPC, increasing density causes an increase in conductivity, which is faster in VPC compared to AC (the former shows 69$\%$ decrease in $\kappa$ upon decreasing density from 1.9 to 0.9 g/cm$^3$, while the latter shows 56$\%$  decrease in $\kappa$ upon decreasing density from 2.9 to 1.5 g/cm$^3$).
In CDC, the structural details --- determined by the chlorination temperature used in the synthesis \cite{palmer_modeling_2010} --- have a weak (negligible) influence on the density but strongly impact the thermal conductivity, which decreases by 53\% upon decreasing the chlorination temperature from 1200 to 800 $^\circ$C.
Similar conductivity variations at nearly constant density are observed in IRG structures, where increasing the irradiation exposure from 2 to 9 min causes a conductivity reduction of $56\%$.
Lastly, we see that annealing the maximum-density AC phase (2.9 g/cm$^3$) yields partial graphitization (PSP phase) and a conductivity increase of 44$\%$. 

Fig.~\ref{fig:conductivity_density} also shows that our nanometric  atomistic models are sufficiently large to describe the bulk limit of the conductivity \cite{allen_diffusons_1999, simoncelli_thermal_2023, harper_vibrational_2024,fiorentino_effects_2025, fiorentino_unearthing_2024, fiorentino_hydrodynamic_2023}; in fact, for all the polymorphs, we studied multiple atomistic models of sizes differing by more than one order of magnitude, and always found compatible results for their conductivity.
Importantly, we highlight that our conductivity predictions are broadly compatible in both magnitude and density dependence with several independently performed experiments and simulations.
Specifically, for AC our predictions match experiments by \citet{hurler_determination_1995} at densities of 1.8 and $\approx 2.0$ g/cm$^3$, and MD simulations by  \citet{giri_atomic_2022}, after accounting for quantum corrections \cite{puligheddu_computational_2019} (see Appendix~\ref{appendix_Q_corr} for details on the accuracy of the quantum correction applied to the MD conductivity predictions). 
Our predictions for AC also overlap with the broad density-conductivity relation experimentally observed by \citet{bullen_thermal_2000}, \citet{shamsa_thermal_2006}, \citet{morath_picosecond_1994}, \citet{hurler_determination_1995} and \citet{scott_thermal_2021}; we note that 
the dispersion of the experimental values may originate from several factors, including presence of hydrogen impurities (especially at low density \cite{bullen_thermal_2000, scott_thermal_2021}). These impurities can increase the effective mass of carbon atoms \cite{morath_picosecond_1994}, cause material softening \cite{arlein_optical_2008}, and promote disorder in the bond network \cite{balandin_thermal_2008} --- all these factors contribute to decrease in thermal conductivity \cite{hanus_thermal_2021}, and explain the tendency of our theoretical predictions for structures made purely of carbon to be an upper bound for the experimental values.
For CDC, our predictions are compatible with the experiments by \citet{kern_thermal_2016} at density $\approx 0.68$ g/cm$^3$ (samples synthesized from TiC2 precursor) matching the increase of conductivity with chlorination temperature --- specifically  
CDC synthesized at 1200 $^\circ$C (CDC1200) has a higher conductivity than CDC synthesized at 800 $^\circ$C (CDC800) in both theory and experiments. We also note that the density of the experimental samples by \citet{kern_thermal_2016} is lower than the density of our CDC models, and to extrapolate the predicted conductivity at the experimental density (dashed lines in Fig.~\ref{fig:conductivity_density}) we employ a simple density-based extrapolation originating from the inverse volume dependence in Eq.~\ref{eq:thermal_conductivity_combined}, obtaining: $\kappa = \rho_{\rm exp} \frac{\kappa_{\rm ref}}{\rho_{\rm ref}}$, where $\rho_{\rm exp}$ denotes the experimental density, $\rho_{\rm ref}$ the theoretical density, and $\kappa_{\rm ref}$ the predicted conductivity at the theoretical density. 
We have validated the accuracy of this density-based conductivity extrapolation procedure for nanoporous carbon (NPC): applying it to estimate the conductivity of low-density NPC from predictions at high density (both ours and those by \citet{wang_density_2025}) yields extrapolations that are consistent with explicit calculations in low-density NPC.
We will justify this further in the next sections, showing that for the carbon polymorphs analyzed, the value of $\kappa/\rho$ is related to the degree of structural disorder.
Focusing on our conductivity predictions for the VPC class, for ann-CDC1200 our prediction is in agreement with simulations from \citet{wang_density_2025} for densities between 0.75 and 1.0 g/cm$^3$; for VPC(D) 1.9 we are compatible with measurements in irradiated graphite by \citet{barabash_effect_2002} and \citet{maruyama_neutron_1992}; finally, for VPC(D)/(T) 1.5 we match quantum-corrected predictions from \citet{suarez-martinez_effect_2011}, and experiments in heavily irradiated glassy carbon (GC-30 \cite{maruyama_neutron_1992}, which has structural characteristics dominated by graphene-like sheets and therefore similar to NPC \cite{jurkiewicz_modelling_2017}).
For IRG, we find that our density-conductivity predictions overlap with the range of experimental results from Refs. \cite{barabash_effect_2002, maruyama_neutron_1992}, and we confirm the inverse trend between irradiation dose and conductivity previously discussed by experiments and MD simulations by \citet{farbos_time-dependent_2017}. 
In the SM we provide details on: (i) the negligible electronic contribution to the conductivity for the classes of carbon polymorphs analyzed here; (ii) the impact of irradiation characteristics (e.g., dose, temperature, type) on the conductivity of irradiated graphite; (iii) how the experimental data shown in Fig.~\ref{fig:conductivity_density} were extracted from the corresponding references.
Finally, we note that the analysis above has been limited to room temperature because this is sufficient for our goal of understanding the disorder-conductivity relation; the temperature dependence of the conductivity is discussed in detail in SM.\\

\begin{figure*}
  \centering
\includegraphics[width=\textwidth]{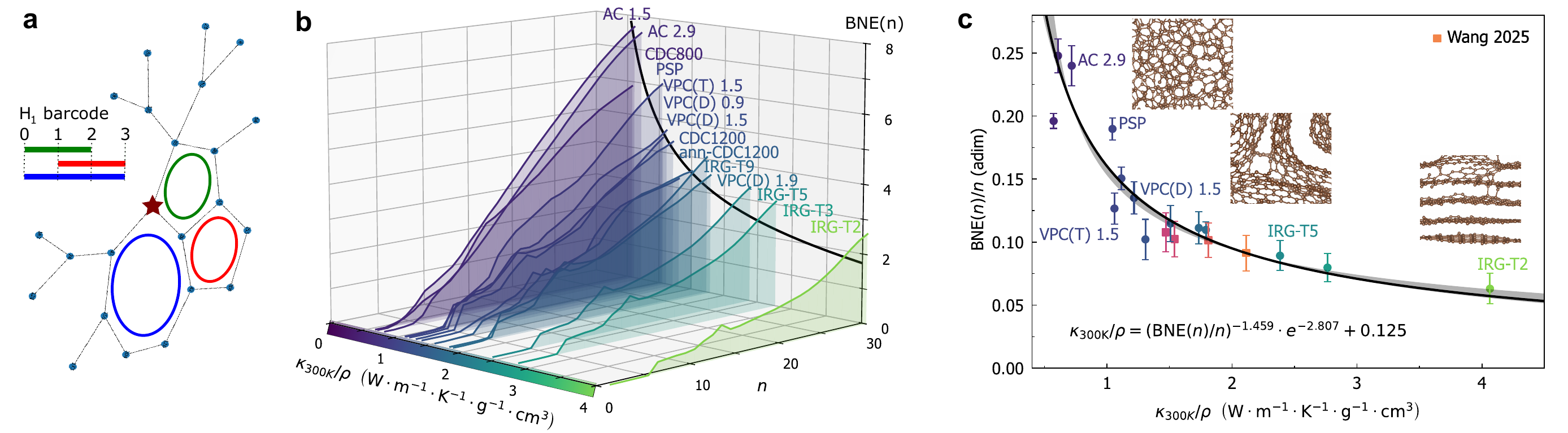}
  \caption{\textbf{Relation between bond-network entropy and thermal conductivity in disordered carbon.} Panel \textbf{a}, $H_1$ barcode of an exemplary local atomic environment (LAE) around an atom (red star) in amorphous carbon with size $n = 23$. Panel \textbf{b}, Growth of the bond-network entropy with the size of LAEs ordered by their value of room-temperature conductivity divided by density. Panel \textbf{c}, Relation between average value of bond-network entropy divided by size of LAEs, ${\rm BNE}(n)/n$, and room-temperature conductivity divided by density, $\kappa_{\rm 300K} / \rho$. The vertical error bars show the standard deviation of BNE$(n)/n$ computed for $n \in [14, 30]$. The black line is the best shifted-power-law fit ($\kappa/\rho = ({\rm BNE}(n)/n)^a \times e^b + c$, where $a,b,c$ are constants) of the relation between BNE$(n)/n$ and $\kappa_{\rm 300K} / \rho$ (best-fit parameters are shown in panel \textbf{c}). The shaded area shows changes in the best-fit power laws when the constant $c$ is changed from 0 to 0.4 $\rm W \cdot m^{-1} \cdot K^{-1} \cdot g^{-1} \cdot cm^3$. The square markers (Wang 2025) are calculated using $\kappa_{\rm 300K}$ from Ref.~\cite{wang_density_2025}, and ${\rm BNE}(n)/n$ computed from the structures used in that reference.
  }
  \label{fig:conductivity_entropy}
\end{figure*}

\section{Thermal conductivity \& bond-network entropy}
To characterize the disorder-conductivity relation, we introduce a descriptor of disorder that quantifies heterogeneity in the atomic bond network. The salient idea is to represent a solid as a collection of local atomic environments (LAEs) sampled from a certain probability distribution  ---  for a perfect crystal, this distribution will be peaked only around environments contained in crystal's primitive cell, while for a disordered solid the number of different LAEs will be higher and therefore their distribution broader.
We start by defining a LAE around a reference atom by constructing a connection graph that starts from such reference atom and reaches its $n-1$ closest neighbours, with $n$ denoting the size of the LAE. 
We consider atoms as graph vertices, which are connected if they are within the carbon bonding distance (1.8 \AA). 
We distinguish between different LAEs by looking at
their ring structure; specifically, we employ the $H_1$ barcode descriptor \cite{schweinhart_statistical_2020} to characterize the algebraically independent rings in the LAE. 
For example, the LAE in Fig. \ref{fig:conductivity_entropy}\textbf{a} contains three algebraically independent rings (green, red and blue), which are classified by $H_1$ barcode in terms of their minimum and maximum edge distance from the reference-atom vertex (0-2 for green, 1-3 for red, and 0-3 for blue); see Appendix~\ref{barcode_details} for technical details.
Changing the LAE's size $n$ allows us to change the resolution with which we resolve structural disorder in a solid.  For small values of $n$ (e.g., $n \lesssim 4$) the LAE contains information only about connectivity between nearest neighbours, and therefore describes short-range order (SRO). For $14{\lesssim} n{\lesssim}30$, the LAEs are large enough to describe structural features in the lengthscale range 5-20\AA, known as medium-range order (MRO) \cite{elliott_medium-range_1991}.

To describe disorder through the statistical distribution of LAEs, for each atom in our system we consider an $n$-sized LAE centered around it, we calculate its $H_1$ barcode (one for each atom in the system) and construct a distribution of $H_1$ barcodes:
\begin{equation}
    P(H_1, n) = \frac{1}{N_{\rm at}} \sum_b \delta_{H_1[{\rm LAE}(b, n)], H_1} ,
    \label{eq:prob_dist}
\end{equation}
where $N_{\rm at}$ is the number of atoms in the atomistic model, $H_1[\rm{LAE}(b, n)]$ is the $H_1$ barcode of the LAE having size $n$ and centered at atom $b$, and $\delta_{H_1[{\rm LAE}(b, n)], H_1}$ is an indicator function equal to one if $H_1[{\rm LAE}(b, n)]$ is equal to the given barcode $H_1$, and zero otherwise. 
The $H_1$ barcode distribution~(\ref{eq:prob_dist}) becomes broader as disorder in the topology of the atomic bond network increases. Therefore, it is natural to quantify disorder through the information entropy of such a distribution, which we will hereafter intuitively refer to as bond-network entropy (BNE):
\begin{equation}
    {\rm BNE}(n) = -\sum_{H_1} P(H_1, n) \ln{P(H_1, n)}.
    \label{eq:BNE}
\end{equation}
To understand the information provided by BNE, it is useful to evaluate it in two limiting cases: (i) a perfectly ordered, idealized crystal with one atom per primitive cell; 
(ii) a strongly disordered bulk glass.
In the idealized crystalline case, BNE$(n)=0\;\forall \,n$, since the crystal order implies that each atom in the system has the same LAE for all $n$, and therefore $P(H_1,n)$ is a Kronecker delta.
In contrast, in the second case, the presence of disorder implies that different LAEs are present, so the LAEs' distribution becomes broader and the corresponding information entropy larger.

We note that the choice of an appropriate descriptor to distinguish between LAEs is crucial for quantifying disorder in an accurate and numerically amenable way \cite{schweinhart_statistical_2020,wei_assessing_2019,schwalbe-koda_model-free_2025}. The choice of employing the $H_1$ barcode descriptor here is motivated by the recent work of \citet{schweinhart_statistical_2020}, which showed that a classification based on the $H_1$ barcode resolves the problem of the proliferation of equivalence classes \cite{schweinhart_statistical_2020}. This problem arises when established methods such as graph isomorphism \cite{vink_configurational_2002,masonStatisticalTopologyCellular2012} are used to distinguish LAEs, since with these methods nearly every environment is considered unique, making it necessary to sample a prohibitively large number of environments to estimate the probability distribution of LAEs and its information entropy. See also the SM for a discussion of the advantages of describing LAEs with the $H_1$ barcode rather than with other descriptors.

Here we focus on the scaling of BNE($n$) with respect to LAE size $n$, in particular on the following key findings:
(i) BNE($n$) is a size-dependent quantity that grows linearly with $n$ across all carbon polymorphs considered;
(ii) the growth rate, ${\rm BNE}(n)/n$, is a size-independent, intensive quantity determined by the degree of disorder in the topology of the atomic bond network;
(iii) ${\rm BNE}(n)/n$ correlates with the macroscopic thermal conductivity and determines the smoothness of the vibrational density of states.

In Fig.~\ref{fig:conductivity_entropy}\textbf{b} we show how BNE's growth rate 
distinguishes different classes of carbon polymorphs.
In particular, BNE($n$) displays the fastest growth with $n$ in the phases having the lowest conductivity (e.g., AC and CDC800). As disorder decreases, the growth rate becomes lower: a medium growth rate is found for CDC1200, VPC(D) 1.5 or IRG T9, and the lowest growth rate emerges in weakly irradiated graphite (IRG T3 and T2). 
We note, in passing, that the atomistic models discussed here have a size much larger than the range of LAE's size $n$ reported in Fig.~\ref{fig:conductivity_entropy}; therefore this analysis 
is not affected by periodic boundary conditions (see Fig.~\ref{fig:top_entropy_convergence} in Appendix~\ref{app_BNE_finitesize} for details).
In Fig. \ref{fig:conductivity_entropy}\textbf{c} we show that 
BNE's average growth rate (computed averaging ${\rm BNE}(n)/ n$ for $n\in [14,30]$) and room-temperature conductivity divided by density are inversely correlated (Spearman's rank-correlation coefficient equal to -0.947). 
For the materials analyzed here ($0.9\lesssim  \rho \lesssim 2.9$ g/cm$^3$) the correlation is approximately a shifted power law ($\kappa/\rho = ({\rm BNE}(n)/n)^a \times e^b + c$). As ${\rm BNE}(n)/ n$ decreases, the derivative of our fitting function decreases, i.e., fixed changes in ${\rm BNE}(n)/ n$ correlate with increasingly larger changes in the conductivity. 
Finally, to evaluate the robustness of our findings, we computed ${\rm BNE}(n)/n$ for the atomistic models of nanoporous carbon released by \citet{wang_density_2025}, using the density and conductivity values reported in that reference to add additional points — independent from our study — to Fig.~\ref{fig:conductivity_entropy}\textbf{c}. The perfect overlap between these external data and the relation between $\text{BNE}(n)/n$ and $\kappa/\rho$ established in this work further validates our results.

The correlation in Fig. \ref{fig:conductivity_entropy}\textbf{c} provides insights on the influence of atomic disorder on the conductivities in Fig.~\ref{fig:conductivity_density}.
Starting from a comparison between AC and VPC at density 1.5 g/cm$^3$, we rationalize the lower conductivity of the former as originating from a higher degree of disorder in the bond network (i.e., higher BNE($n$)/$n$ for $n {\in} [14, 30]$). Analogous considerations hold for CDC and IRG, where samples with very similar density show conductivities that are very different and inversely related to BNE($n$)/$n$.
Focusing on the conductivity variations observed within a certain class upon changing density, we see that in AC, increasing density from 1.5 to 2.9 g/cm$^3$ has negligible effect on BNE($n$)/$n$; therefore, the conductivity differences between the various phases of AC can be explained mainly in terms of density --- we will see later that the higher the density, the higher the number of vibrational modes per unit volume, and in disordered systems this contributes to increasing the conductivity.
In contrast, in VPC increasing density implies also
a decrease in disorder (BNE($n$)/$n$); therefore, the conductivity increase observed in VPC upon increasing density is stronger than in AC.

Overall, these findings suggest that conductivity and density measurements can be used to quantify the structural heterogeneity of disordered solids, and motivate us to investigate further the relationship between the structural-disorder descriptor ${\rm BNE}(n)/ n$ and both thermal and vibrational properties.

In the SM, we analyze the relations between the descriptor BNE($n$)/$n$ we introduced and other properties, including density, heat capacity, and thermal conductivities at low and high temperature (100 and 700 K, respectively).
We also investigate the relation between BNE($n$)/$n$ and conductivity for polycrystalline graphene with different grain sizes \cite{zhou_million-atom_2025}.

\begin{figure*}
\centering
\includegraphics[width=\textwidth]{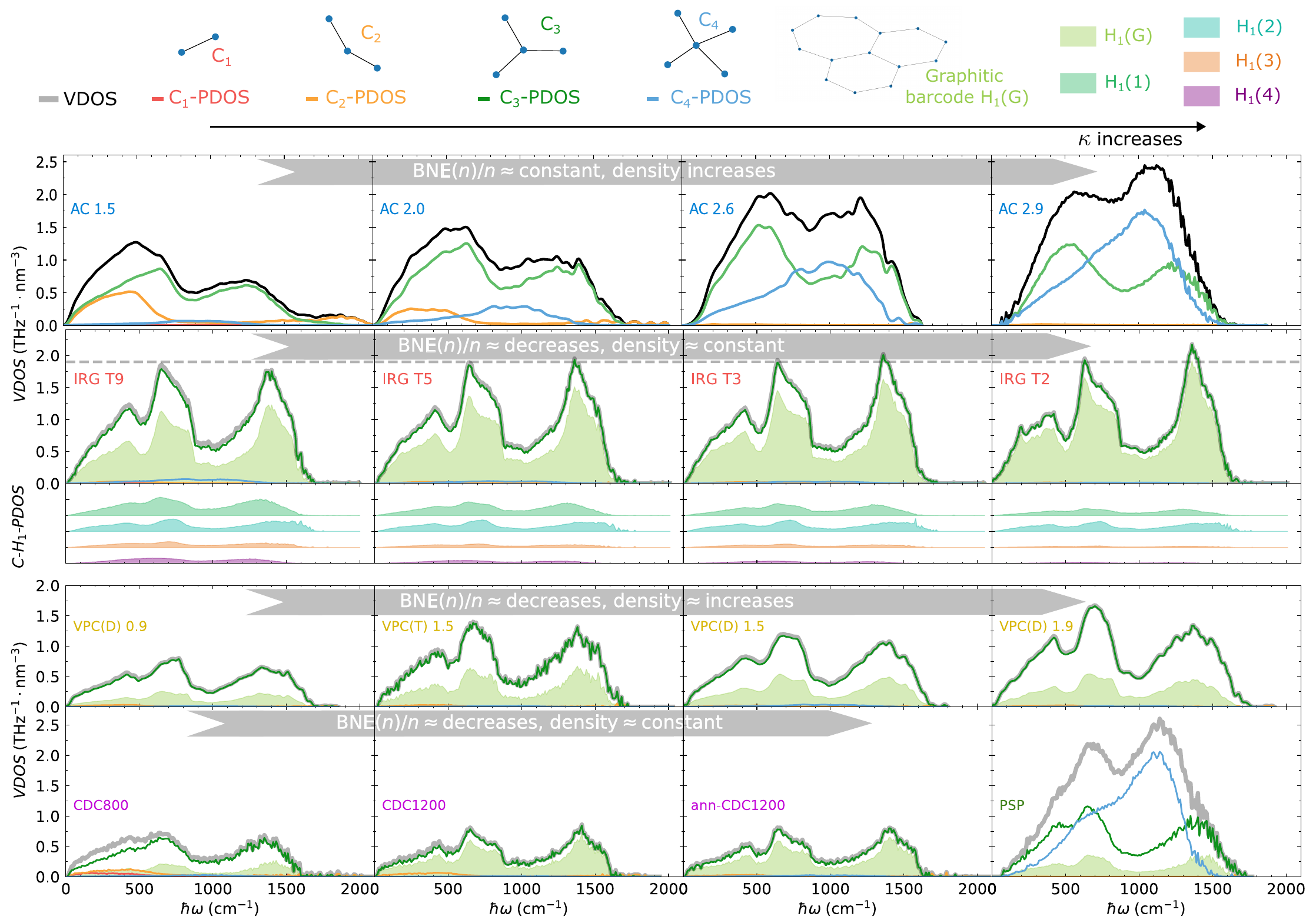}
  \caption{\textbf{Dependence of VDOS on short- and medium-range order.} The panels are split by different groups of structures: first row is AC, second row is IRG, third row is VPC, and fourth row shows CDC \& PSP. The total VDOS is thick gray, its decomposition into $C$-PDOS contributions from atoms having different coordination (SRO) is given by thin lines of different colors: red for dangling bonds (present appreciably only in CDC800, albeit practically invisible), while orange, green, and blue show the PDOS for two-, three-, and four-fold coordinated atoms, respectively. 
  The shaded areas show the $C_3$-$H_1$-PDOS: light green corresponds to the graphitic barcode, and the other four colors show other common barcodes in IRG. The dashed gray horizontal line in the IRG panel shows that the VDOS becomes smoother as irradiation-induced structural disorder increases.
  }
  \label{fig:VDOS_barcode_coordination}
\end{figure*}

\section{Atomic vibrations \& bond-network entropy}
\label{sec:vibration_BNE}
In this section we show that a relation between ${\rm BNE}(n)/ n$ and microscopic vibrational properties exists. We start by characterizing atomic vibrations with the VDOS:
\begin{equation}
    g(\omega)=\frac{1}{\mathcal{V} N_{\rm c}} \sum_{{\bm{q},s}} \delta(\omega - \omega_{\bm{q}s}).
\label{eq:bare_VDOS}
\end{equation}
We also resolve how atoms with a certain coordination contribute to the VDOS by decomposing Eq.~(\ref{eq:bare_VDOS}) into coordination-resolved partial VDOS \cite{milkus_interpretation_2018, harper_vibrational_2024} ($C$-PDOS), 
\begin{equation}
       g_{C_x}(\omega) = \frac{1}{\mathcal{V} N_{\rm c}} \sum_{{\bm{q},s}} \delta(\omega - \omega_{\bm{q}s}) \sum_{b, \alpha} |\mathcal{E}^{b\alpha}_{{\bm{q}s}}|^2 \delta_{b , C_x},
       \label{eq:PDOS}
\end{equation}
where $\delta_{b , C_x}$ is an indicator function equal to one if atom $b$ has coordination number equal to $x$, and zero otherwise;  $\mathcal{E}^{b\alpha}_{{\bm{q}s}}$ is the eigendisplacement that describes how atom $b$ oscillates in direction $\alpha$ when the mode $\bm{q}s$ is excited  \cite{wallace_thermodynamics_1972,simoncelli_wigner_2022}.
Eq.~(\ref{eq:PDOS}) is directly related to the VDOS via marginalization over the coordination variable: $\sum_{C_x}g_{C_x}(\omega){=}g(\omega)$.
Importantly, we can go beyond the SRO nearest-neighbour (coordination) analysis and look at how MRO properties influence the VDOS. To this aim, we further decompose the $C$-PDOS into contributions from different $H_1$ barcodes ($C$-$H_1$-PDOS):
\begin{equation}\label{eq:PDOS_H1}
\begin{split}
    & g_{C_x, H_1}(\omega) =  \\ 
    & \frac{1}{\mathcal{V} N_{\rm c}}  \sum_{{\bm{q},s}}  \delta(\omega - \omega_{\bm{q}s}) \sum_{b, \alpha} |\mathcal{E}^{b\alpha}_{{\bm{q}s}}|^2 \delta_{b , C_x} \delta_{H_1[{\rm LAE}(b, n)], H_1},
\end{split}
\end{equation}
where we choose the size of the LAE $n$ to be equal to 14 for the following reasons: (i) $n = 14$ is sufficiently large to capture changes in the MRO features related to 3 graphitic 6-fold rings, even in the presence of perturbations in the graphite interlayer distance, or changes in the number of atoms within one of the rings; (ii) the range $n \in [14, 30]$ is sufficient to capture signatures of MRO, and since Fig.~\ref{fig:conductivity_entropy} shows that the BNE's growth with $n$ is practically constant for $n \in [14, 30]$, we can choose the lowest value of $n$.
Finally, it can be verified that marginalizing the $C$-$H_1$-PDOS~(\ref{eq:PDOS_H1}) with respect to the barcode variable yields the $C$-PDOS~(\ref{eq:PDOS}):
$\sum_{H_1}g_{C_x, H_1}(\omega)=g_{C_x}(\omega)$. 

In Fig.~\ref{fig:VDOS_barcode_coordination} we plot the VDOS and its decomposition into $C$-PDOS and $C$-$H_1$-PDOS for all the classes of carbon polymorphs studied.
Starting from AC, we see that increasing density yields an increase in magnitude of the VDOS, which trivially follows from the appearance of the system's cell volume $\mathcal{V}$ in the denominator of Eq.~(\ref{eq:bare_VDOS}). Moreover, the changes in the shape of the VDOS with density are determined by having different proportions of $C_2$, $C_3$ and $C_4$ coordination environments, and we highlight how the shape of the $C$-PDOS contributions are almost unchanged across all amorphous carbon structures (see Fig.~SF 6 in SM for details). In particular, at low density (1.5 g/cm$^3$) the  VDOS is bimodal with the low-frequency peak stronger than the high-frequency peak --- the $C$-PDOS decomposition shows that these are determined by a superimposition of a monomodal low-frequency $C_2$ PDOS and a bimodal $C_3$ PDOS. Upon increasing density, the magnitude of the low-frequency monomodal $C_2$ contribution is progressively replaced by a high-frequency monomodal $C_4$ PDOS, resulting in a bimodal distribution with the high-frequency peak stronger than the low-frequency peak.
In contrast, in less disordered polymorphs such as CDC, VPC and IRG, 
SRO is dominated by the $C_3$ coordination environment, hence VDOS $\approx C_3$-PDOS --- this indicates that these solids are ordered over the SRO lengthscale, and therefore the visible VDOS changes must originate from disorder over a larger lengthscale.
Therefore, in Fig.~\ref{fig:VDOS_barcode_coordination} we further decompose the $C_3$-PDOS into contributions from different MRO using the $C_3$-$H_1$-PDOS. We show that the `graphitic $H_1(G)$' barcode --- which describes MRO due to three hexagonal 6-fold rings --- allows us to resolve whether the $C_3$ SRO is associated with high density of graphitic rings (e.g., CDC1200 and IRG-T2) or not (e.g., CDC800). 
We note, in passing, that the $C_3$-$H_1 (G)$-PDOS has a triple-modal shape in all IRG structures, as well as in CDC1200 and in medium-high density VPC --- specifically, we show in the SM Fig.~SF 7 that for a finite concentration of graphitic $H_1(G)$ environments, the shape of $C_3$-$H_1 (G)$-PDOS is practically independent from their concentration. 
More generally, the decomposition of IRG's $C_3$-PDOS into the five most frequent barcodes (shaded areas in Fig.~\ref{fig:VDOS_barcode_coordination}) shows that increasing disorder yields an increase in the number of different barcodes at a fixed LAE size that contribute to the VDOS, consistently with ${\rm BNE}(n)/ n$'s increase shown in Fig.~\ref{fig:conductivity_entropy}.
Most importantly, we note that different barcode environments have $C_3$-$H_1$-PDOS with different shapes; their superimposition leads to an overall smoothing of the VDOS, and hence a decrease in the magnitude of its peaks.
These findings generally apply to all carbon polymorphs, as we observe VDOS smoothing when we compare: (i) CDC1200 and CDC800; (ii) VPC(D) 1.9, 1.5, and 0.9; (iii) PSP and AC 2.9. In summary, increasing disorder in the atomic bond network causes an
increase in BNE's growth rate and in the VDOS' smoothness; in the next section we show that the latter can be related to an effective model of a solid, where heat transport is limited by phonon scattering due to structural disorder, whose special limit is Kittel's empirical interpretation of conductivity in glasses \cite{kittel_interpretation_1949}.

\section{Explicitly Disordered Glass vs Perturbatively Disordered Crystal}
In this section we elucidate a relation between our WTE-based conductivity predictions and those obtained employing a model that generalizes Kittel's phenomenological treatment of the thermal resistance induced by disorder  \cite{kittel_interpretation_1949, casimir_note_1938}. We will show how this relation allows us to connect ${\rm BNE}(n)/ n$, smoothness of the VDOS, and conductivity to the lengthscales of disorder and of the heat-transport mechanisms, thus fundamentally rationalize the correlation between ${\rm BNE}(n)/ n$ and thermal conductivity shown in Fig.~\ref{fig:conductivity_entropy}.

We start by resolving the thermal conductivity with the usual frequency-dependent decomposition \cite{simoncelli_thermal_2023}
\begin{equation} 
  \kappa = \int_0^{\omega_{\rm max}} d\omega \, g(\omega)C(\omega) D(\omega),
\label{eq:kappa_omega}
\end{equation}
where $\omega_{\rm max}$ is the maximum vibrational frequency of the solid, $g(\omega)$ is the VDOS~(\ref{eq:bare_VDOS}), $C(\omega)$ is the specific heat of a vibration with frequency $\omega$ (Eq.~(\ref{eq:quantum_specific_heat})), and $D(\omega)$ is its diffusivity, describing the rate at which the heat carried by a vibration spreads \cite{allen_thermal_1989}.
The first description of heat transport in the presence of structural disorder was done relying on the semiclassical Peierls-Boltzmann transport equation (BTE)  \cite{peierls1955quantum} and phenomenologically considering disorder-induced thermal resistance in a `Perturbatively Disordered Crystal' (PDC). The PDC picture accounts for the frequency dependence of the diffusivity, interpreting it in terms of particle-like excitations having energy $\hbar\omega$ and propagating isotropically with velocities $v_{\rm eff}(\omega)$ over disorder-limited transport lengthscales (mean free paths) $\lambda_{\rm eff}(\omega)$:
\begin{equation}     
D_{\rm PDC}(\omega) = v_{\rm eff}(\omega) \lambda_{\rm eff}(\omega) = \frac{v^2_{\rm eff}(\omega)}{\Gamma_{\rm dis}(\omega)},
\label{eq:diffusivity_crystal}
\end{equation}
where the last equality follows from the relation $v_{\rm eff}(\omega)/\lambda_{\rm eff}(\omega){=}\Gamma_{\rm dis}(\omega)$, with $\Gamma_{\rm dis}(\omega)$ being the `disorder linewidth' or 
inverse scattering time $\Gamma_{\rm dis}(\omega) {=} [\tau_{\rm dis}(\omega)]^{-1} $ that describes the thermal resistance encountered by a vibration $\omega$ due to the presence of structural disorder. 

A well known special case of Eq.~(\ref{eq:diffusivity_crystal}) is Kittel's `phonon liquid' picture \cite{kittel_interpretation_1949}, which attempted to phenomenologically explain the conductivity of glasses by combining the BTE with Casimir's model \cite{casimir_note_1938} for phonon-interface scattering around its physical lower bound. Specifically, this picture considers: (i) the mean free path (MFP) $\lambda_{\rm eff}(\omega)$ to be a temperature-dependent, but frequency-independent lengthscale $\lambda_{\rm eff}(\omega){\to} \Lambda_0$ determined by the type of disorder (e.g., for silica glass \cite{kittel_interpretation_1949} $\Lambda_0$ is the ring size ${\approx}7$\AA); 
(ii) the propagation velocity $v_{\rm eff}(\omega)$ as frequency-independent average velocity of sound ($v_{\rm eff}(\omega){\to} v_{\rm sound}$).
Kittel's model is top-down interpretative but not bottom-up predictive; in fact, it 
allows to estimate the value of the microscopic $\Lambda_0$ from the knowledge of measured values of $\kappa$ (as well as of $v_{\rm sound}$ and $C$), but does not provide a rigorous prescription for determining the value of $\Lambda_0$ from first principles that can be used for the bottom-up prediction of $\kappa$.
To address the limitations of Kittel's phenomenological explanation, Allen and Feldman  \cite{allen_thermal_1989} (AF) introduced an alternative definition of diffusivity for harmonic glasses based on a Zener-like tunnelling transport mechanism between quasi-degenerate vibrational modes. 
In contrast to considering disorder as originating from perturbations of long-range crystalline order, AF describes heat transport in an Explicitly Disordered Glass (EDG) that does not need a crystalline precursor, or a relation to it. 
As already mentioned, the harmonic AF formalism emerges as a special case of the more general anharmonic WTE framework \cite{simoncelli_wigner_2022}, and in the following we show that the latter allows to shed light on the connection between atomic disorder, VDOS smoothing, macroscopic $\kappa$ and Kittel's `phonon liquid' interpretation.

We start by showing that the WTE exposes a proportionality relation between the diffusivity of a generic (EDG or PDC) disordered system, its VDOS and quasi-degenerate velocity operator elements ($\tens{v}_{s,s'}$ with $\omega_s {\simeq} \omega_{s'}$).
This is apparent when we consider Eq.~(\ref{eq:thermal_conductivity_combined}) in the EDG limit, i.e.: 
(i) we take the bulk-disordered limit (in practice using $\mathcal{V}$ containing thousands of atoms and thus considering $\bm{q} {=}\bm{0}$ only); (ii) we consider linewidths slowly varying with frequency, larger than the average energy-level spacing, and far from the overdamped regime. This implies that the Voigt distribution reduces to a Lorentzian with FWHM determined by the sum of intrinsic linewidths $\Gamma_s {+}\Gamma_{s'}$ of vibrations with practically equal frequencies $\omega_{s}{\sim}\omega_{s'}$ and specific heats $C_{s}{\sim} C_{s'}$. Under these conditions, the WTE diffusivity~\cite{simoncelli_thermal_2023} reduces to:
\begin{equation} 
\begin{split}
\label{eq:diff_intermediate}
D_{\rm EDG} (\omega)\! &= \!\frac{\pi}{\mathcal{V}g(\omega)}\!\sum_{s,s'} \!\frac{\rVert \tens{v}_{ss'} \lVert^2}{3}\! L[\omega_s {-} \omega_{s'}\!, \Gamma_s {+}\Gamma_{s'}] \delta(\omega-\omega_s)\\
&\approx  \pi {\Upsilon^2(\omega)}  \frac{1}{\rho_{n}} \left[\int  d\omega' L[\omega {-} \omega' \!, \Gamma(\omega)]  g_d(\omega')\right],\\
\end{split}
\raisetag{10.5mm}
\end{equation}
where the simplification from the first to the second line is justified in strongly disordered systems \cite{simoncelli_thermal_2023,harper_vibrational_2024} that feature velocity-operator elements negligibly dependent on the frequency difference between the modes $s$ and $s'$ {(these are denoted by ${\Upsilon^2(\omega_s)}{\approx} N_{\rm at}{\rVert \tens{v}_{ss'} \lVert^2}/{3}$, see Appendix~\ref{app_velop})}. Moreover, in the second line we have denoted the atom number density with $\rho_{n} {=} \frac{N_{\rm at}}{\mathcal{V}}$, and  rewritten the mode linewidth $\Gamma_s$ as a function of frequency $\Gamma(\omega_s)$ using the bijective mapping between frequency and mode arising from lack of symmetries (and hence lack of perfectly degenerate frequencies) in disordered systems  \cite{maradudin_symmetry_1968}.
Eq.~(\ref{eq:diff_intermediate}) shows that in an EDG the diffusivity is practically determined by a convolution between a Lorentzian $L[\omega - \omega', \Gamma(\omega)]$ and the `dressed' VDOS
\begin{equation}
\begin{split}
\label{eq:relationDOS}
    g_d(\omega) \!=\! \frac{1}{\mathcal{V}} \!\sum_s \!L[\omega {-} \omega_s,\Gamma(\omega_s)] {=} 
    {\int} {d\omega'} L[\omega {-} \omega', \Gamma(\omega')] g(\omega'),
    \raisetag{12mm}
\end{split}
\end{equation}
whose name derives from its differences relative to the `bare' VDOS $g(\omega)$ defined in Eq.~(\ref{eq:bare_VDOS}).
Specifically, within the many-body Green's function formalism, the Dirac deltas $\delta(\omega{-}\omega_s)$ appearing in the bare VDOS~(\ref{eq:bare_VDOS}) can be seen as resulting from the integration of
bare non-interacting phonon spectral functions, while in the dressed VDOS~(\ref{eq:relationDOS}) we have the integration of Lorentzian spectral functions $b(\omega)_s{=}\frac{1}{\pi} \!\frac{\Gamma_s / 2}{(\omega {-} \omega_s)^2 {+} (\Gamma_s / 2)^2}$, i.e., $g_d(\omega){=}\frac{1}{\mathcal{V}} \sum_s b(\omega)_s$  \cite{prat_semiclassical_2016, chandrasekaran_effect_2022} where the broadenings $\Gamma_s$ are determined by interactions  (due to, e.g., disorder or anharmonicity).
We highlight that the second equivalence in Eq.~(\ref{eq:relationDOS}) shows that the dressed VDOS is related to the `bare' VDOS $g(\omega)$ via a `dressing' integral, which practically is a convolution with a Lorentzian having frequency-dependent broadening (linewidth) and implies that the dressed VDOS becomes smoother as linewidths (proportional to the interaction strength) become larger \footnote{This statement is valid in the Lorentzian spectral function regime where the WTE can be applied \cite{caldarelli_many-body_2022,simoncelli_wigner_2022}.}.

In strongly disordered systems (EDG), the vibrational energy levels significantly repel each other \cite{simkin_minimum_2000}, implying that the bare VDOS is already very smooth and practically indistinguishable from the dressed VDOS, $g_{\rm EDG}(\omega)\approx g_d(\omega)$.
This also implies that we can neglect the convolution between dressed VDOS and Lorentzian in the second line of Eq.~(\ref{eq:diff_intermediate}), obtaining that in this (non-interacting,  harmonic) disordered limit the diffusivity of an EDG is directly proportional to the bare VDOS:
\begin{equation}
    D_{\rm EDG}(\omega) \approx D_{\rm AF} (\omega) = \pi {\Upsilon^2(\omega)}  \frac{1}{\rho_n} g_{\rm EDG}(\omega).
\label{eq:diff_vdos}
\end{equation}
Eq.~(\ref{eq:diff_vdos}) shows that the EDG diffusivity arises from a vibration mixing with a dense set of quasi-degenerate vibrations, and the strength of this mixing is described by the square of the quasi-degenerate velocity operator $\Upsilon^2(\omega)$. 
Finally, we note that Eq.~(\ref{eq:diff_vdos}) can be equivalently obtained from Eq.~(\ref{eq:diff_intermediate}) by taking the ordered limit $\mathcal{V} {\to} \infty$ and $\Gamma_s{\to}\eta{\to} 0$, and coincides with the AF diffusivity \cite{allen_thermal_1989}.\\

\section{Disorder-induced VDOS smoothing \& thermal resistance.} 
\label{sec:disorder}
In the BTE picture for conduction in a PDC, $\kappa$ (Eqs.(\ref{eq:kappa_omega}, \ref{eq:diffusivity_crystal})) is influenced by structural disorder through a disorder-limited MFP determined phenomenologically.
In contrast, in the rigorous WTE treatment of an EDG (Eqs.(\ref{eq:kappa_omega}, \ref{eq:diff_intermediate})), disorder impacts both VDOS and diffusivity.
It is therefore natural to ask whether it is possible to obtain two compatible physical descriptions of heat transport using the rigorous WTE for an EDG and BTE treatment for a PDC.
Here we demonstrate that this is indeed possible, showing that PDC's phenomenological disorder linewidth can be formally derived from the WTE EDG through algebraic manipulations; then, we use these insights to formally determine disorder-limited MFP.
 
We start from the PDC picture, which prescribes specific heat and VDOS to be those of an Unperturbed Ordered Crystal (UOC), and the influence of disorder to be accounted by the diffusivity term, which is formally unknown and usually determined by phenomenological arguments.
Since the WTE in the EDG limit shows that in the presence of structural disorder the diffusivity is related to the VDOS (Eq.~(\ref{eq:diff_intermediate})), we hypothesize that the unknown PDC diffusivity also assumes such functional form, which we highlight contains a convolution between the dressed VDOS and a Lorentzian distribution having FWHM that is unknown (and in principle frequency-dependent). Importantly, through algebraic manipulations one can recast such convolution to apply on the UOC VDOS appearing at the very beginning of PDC conductivity expression.
These manipulations imply that the disorder-mediated interaction that limits thermal transport within PDC model can be formally related to the smoothing of the dressed VDOS.
In particular, the unknown disorder linewidth $\Gamma_{\rm dis}(\omega)$ in Eq.~(\ref{eq:diffusivity_crystal}) can be determined as the broadening that within the convolution (Eq.~(\ref{eq:relationDOS})) transforms the bare UOC VDOS into a dressed PDC VDOS equal to the bare EDG VDOS. In formulas:
\begin{equation}\label{eq:kappa_PDC_rewritten}
\begin{split}
    \kappa_{\rm PDC} & = \int d\omega \, g_{\rm UOC}(\omega) C(\omega) D_{\rm unknown}(\omega)\\
    & = \int  d\omega \, g_{\rm UOC}(\omega) C(\omega) \pi {\Upsilon^2(\omega)}  \frac{1}{\rho_{n}} \\ & \qquad \qquad \; \times \left[\int  d\omega' L[\omega {-} \omega' \!, \Gamma_{\rm dis}(\omega)]  g_{\rm PDC}(\omega')\right] \\
    & \approx \int d\omega \, \underbrace{g_{\rm PDC}(\omega)}_{\approx g_{\rm EDG}(\omega)} C(\omega) \underbrace{\left[ \pi \Upsilon^2(\omega) \frac{1}{\rho_{n}} g_{\rm PDC}(\omega) \right]}_{\approx D_{\rm EDG}(\omega) \approx D_{\rm AF}(\omega)},
\end{split}
\raisetag{40mm}
\end{equation}
where the last approximated equivalence 
holds under the assumptions used to obtain Eq.~(\ref{eq:diff_intermediate}) and Eq.~(\ref{eq:diff_vdos}), see Appendix~\ref{app_derivation_diffusivity} for details. 
From a mathematical viewpoint, Eq.~(\ref{eq:kappa_PDC_rewritten}) shows that the algebraic manipulations used to rewrite the PDC conductivity in WTE form are self-consistent, i.e. by rearranging the integration of the Lorentzian appearing in the WTE conductivity~(\ref{eq:thermal_conductivity_combined}) one obtains an expression in which the dressed PDC VDOS appears twice, consistently with the form of the WTE EDG conductivity~(\ref{eq:diff_vdos}) (see Appendix \ref{app_derivation_diffusivity} for details).
\begin{figure}[b]
  \centering
\includegraphics[width=\WidthFigure]{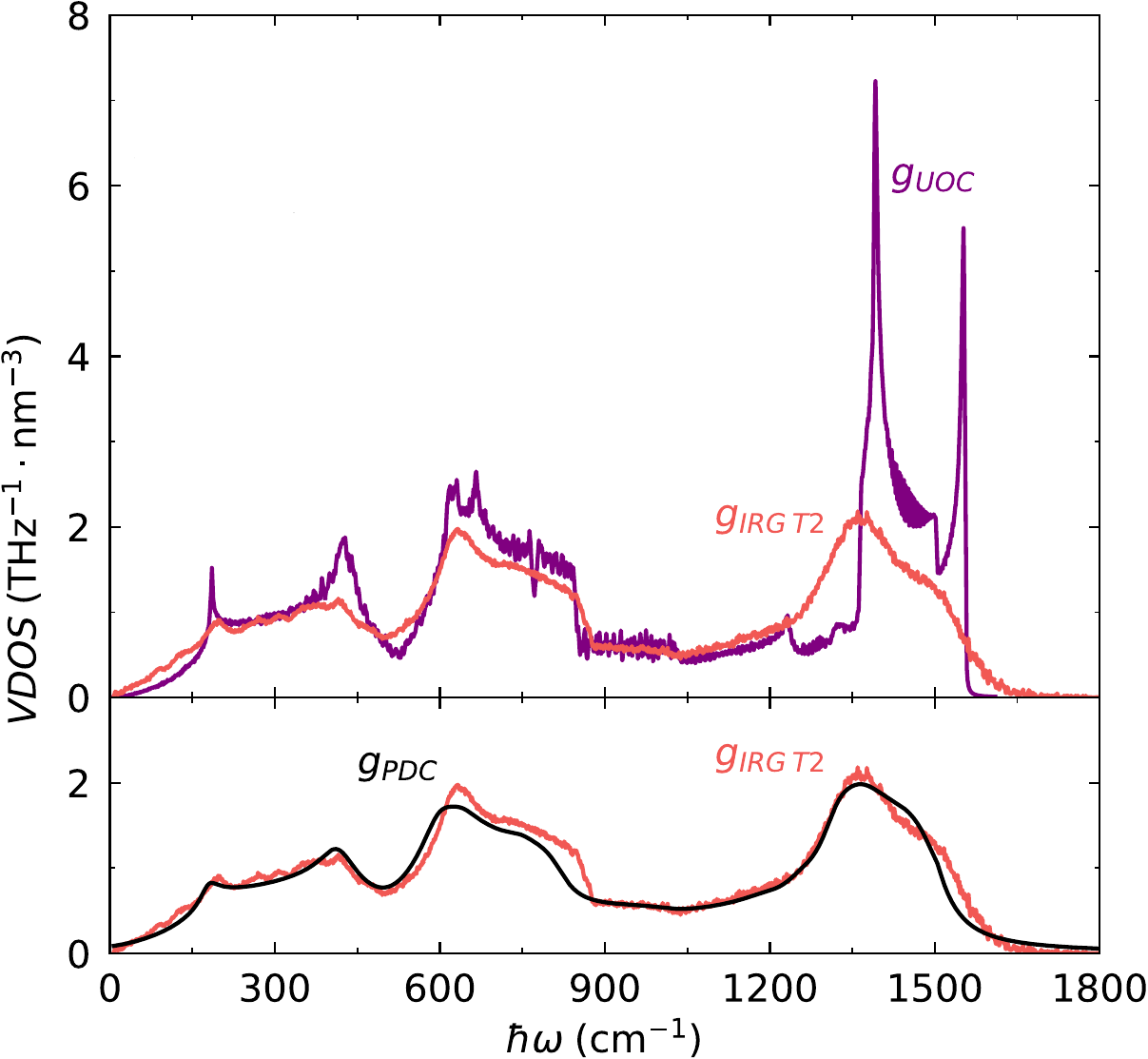}
  \caption{\textbf{Equivalence map between PDC and EDG.} Top, VDOS of unperturbed, ordered crystalline (UOC) graphite (purple) and of EDG irradiated graphite (IRG T2, red).  Bottom, PDC VDOS $g_{\rm PDC}(\omega)$ obtained perturbing the VDOS of UOC graphite (black), and VDOS of EDG IRG T2 (red).
  }
  \label{fig:PDC_VDOS_theory}
\end{figure}
\begin{figure*}
  \centering
\includegraphics[width=\textwidth]{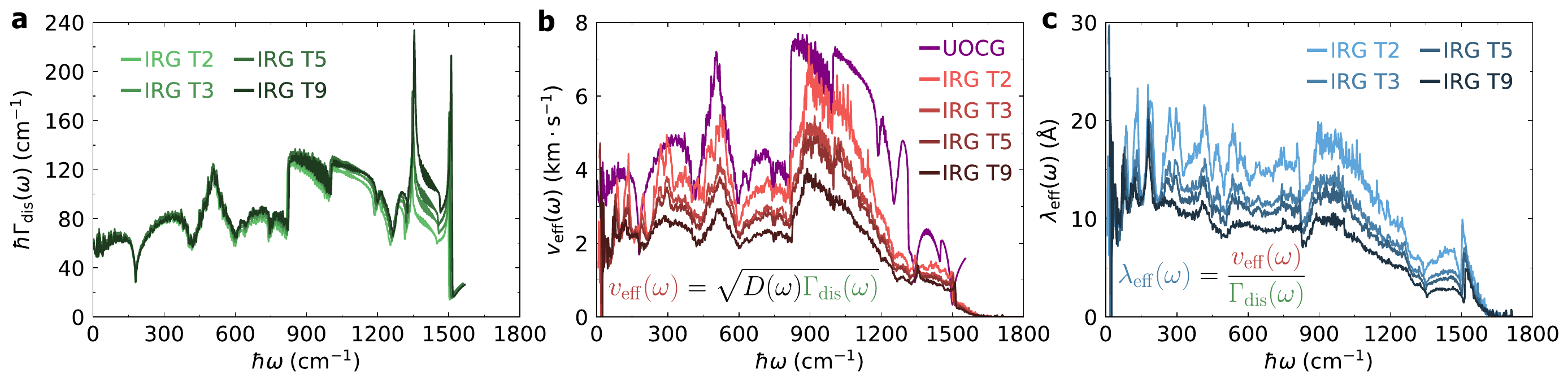}\\[-3mm]
  \caption{\textbf{Influence of structural disorder on (a) linewidths, (b) propagation velocity, and (c) vibrations' mean free paths.} Starting from Unperturbed Ordered Crystalline Graphite (UOCG), structural disorder is progressively induced by irradiation in samples IRG T2, T3, T5, and T9 (see text). Increasing disorder yields larger linewidths and lower velocities; consequently, the vibrational mean free paths decrease upon increasing disorder. 
  }
  \label{fig:IRG_diffusivity_decomposition}
\end{figure*}
In practice, $\Gamma_{\rm dis}(\omega)$ is determined starting from the established \cite{ziman_electrons_1960,hanus_thermal_2021} frequency-linewidth relations for phonon-disorder scattering,
\begin{equation}
    \Gamma_{\rm dis}(\omega) = \frac{v(\omega)}{L} + R\; \omega^2 g^{\rm DR}_{\rm UOC}(\omega).
\label{eq:linewidth_expression}
\end{equation}
The first term weakly depends on frequency (constant for $\omega\to 0$); it describes wavepackets of atomic vibrations having velocity $v(\omega)$ and scattering with structural nonhomogeneities of lengthscale $L$ --- e.g., grain boundaries  \cite{casimir_note_1938}, dislocations or interstitial atoms \cite{osipov_disclination_1998} that have been observed in irradiated graphite  \cite{karthik_situ_2011}.
The second term, instead, strongly depends on frequency and in IRG is in practice most relevant for  $\hbar\omega{\gtrsim}$600 cm$^{-1}$; it accounts for point-like defects  \cite{hanus_thermal_2021} or mass impurities  \cite{tamura_isotope_1983} having density proportional to the parameter $R$. Importantly, in such a term we also account for how the density reduction (DR) caused by irradiation disorder (porosity increase) changes the VDOS: 
$g^{\rm DR}_{\rm UOC}(\omega){=}\tfrac{\rho_{\rm EDG}}{\rho_{\rm UOC}} g_{\rm UOC}\big(\omega / a \big) 1/a$, where $a {=} \sqrt[3]{\frac{\rho_{\rm EDG}}{\rho_{\rm UOC}}}$ accounts for the frequency shift due to density changes through an elementary Debye-like model \footnote{Specifically, the form of the factor $a$ derives from: (i) considering an elementary Debye model having speed of sound $v_{\rm sound}$ and Debye frequency $\omega_D^3 = 6 \pi^2 \frac{N_{at}}{\mathcal{V}} v_{\rm sound}^3$; (ii) assuming the main change in the frequencies when lowering the density of graphite is mainly due to changes in the volume and has negligible effect on $v_{\rm sound}$. This implies that the frequencies scale with the cube root of the density: $\omega \propto \rho^{1/3}$.} 
(see Appendix~\ref{app_VDOS_smoothing} for computational details); the factor $1/a$ ensures that the integral of the VDOS is preserved by the shift transformation;
finally, the density ratio ${\rho_{\rm EDG}}/{\rho_{\rm UOC}}$ 
ensures the proportionality between density and VDOS magnitude (Eq.~(\ref{eq:bare_VDOS})).

Then, the two parameters ($L$ and $R$) in the frequency-linewidth relation (Eq.~(\ref{eq:linewidth_expression})) are fitted to map, through the dressing (convolution) (\ref{eq:relationDOS}) and DR transformations, the bare UOC VDOS into a dressed PDC VDOS that overlaps with the bare EDG VDOS.
Fig.~\ref{fig:PDC_VDOS_theory} demonstrates compatibility between the EDG VDOS (IRG T2) and the corresponding PDC VDOS (see Fig.~\ref{fig:VDOS_all_IRG_vs_model} in the Appendix for analogous results in other structures). It is now important to summarize and highlight two insights.
First, the equivalence between PDC and EDG VDOS shows that the disorder-induced repulsion between vibrational eigenstates, which yields a smooth bare VDOS in an EDG, can be interpreted in terms of PDC's disorder linewidth $\Gamma_{\rm dis}(\omega)$ (appearing in Eq.~\ref{eq:diffusivity_crystal}) that yields a smooth dressed VDOS in a PDC.
Second, the relation~(\ref{eq:kappa_PDC_rewritten}) also implies that when PDC and EDG conductivities are compatible, PDC and EDG diffusivities have not only the same functional form but also assume the same numerical values.

These two insights allow us to connect and interpret the rigorous WTE predictions with Kittel's intuitive `phonon liquid' picture. 
In particular, Eq.~(\ref{eq:diffusivity_crystal}) shows that the knowledge of the diffusivity and of the disorder linewidth $\Gamma_{\rm dis}(\omega)$ fully determines the disorder-limited MFP 
\begin{equation}
    \lambda_{\rm eff}(\omega){=}\sqrt{\frac{D_{\rm EDG}(\omega)}{\Gamma_{\rm dis}(\omega)}},
\end{equation}
as well as the propagation velocity $v_{\rm eff}(\omega){=}\sqrt{D_{\rm EDG}(\omega)\Gamma_{\rm dis}(\omega)}$.
In Fig. \ref{fig:IRG_diffusivity_decomposition} we show all these quantities, evaluated in IRG using the quantum-accurate GAP machine-learning potential \cite{rowe_accurate_2020}, and also discussing how they are impacted by the amount of structural defects. 
We highlight that the disorder linewidth $\Gamma_{\rm dis}(\omega)$ generally increases upon increasing structural disorder, with visible variations at high vibrational energy (the parameter $L\approx 20$\AA\space that practically controls the low-frequency behavior of Eq.~(\ref{eq:linewidth_expression}) is negligibly affected by the increase in disorder from IRG T2 to IRG T9, while the parameter $R$ that controls the high-frequency behavior increases from $R = 9.5 \times 10^{-6}$ in IRG T2 to $R = 17.4 \times 10^{-6}$ $\text{THz}{\cdot} \text{cm}{\cdot} \text{nm}^3$ in IRG T9).
This increase of disorder linewidth upon irradiation is consistent with the increase in BNE's growth rate highlighted in Fig.~\ref{fig:conductivity_entropy}, confirming that ${\rm BNE}(n)/ n$ is an informative descriptor for the defect density in IRG.
We also note that the propagation velocity of vibrations in IRG structures follows similar trends with frequency as the average group velocity of UOC graphite, and smoothly decreases upon increasing disorder --- this quantitatively confirms the intuitive expectation that disorder induces repulsion between phonon bands \cite{simkin_minimum_2000}, and hence reduces the average phonon group velocity. In Appendix~\ref{app_Kittel_limit} we clarify the meaning of propagation velocity in disordered systems and also show how to obtain Kittel's picture \cite{kittel_interpretation_1949} as a limiting case of the PDC framework.
Finally, the MFP is frequency dependent and generally decreases upon increasing structural disorder. 

Overall these insights improve our understanding of transport beyond Kittel's picture for two reasons. 
First, they explain how increasing structural disorder reduces the phonon propagation velocity, through strongly decreasing diffusivity and weakly increasing linewidth.
This is an important difference from Kittel's model, in which the sound velocity was phenomenologically set relying on experimental values.
Second, they extend Kittel's single temperature-dependent MFP picture by showing explicitly how MFPs depend on both frequency and degree of disorder; specifically, heat-transport lengthscales at room temperature are not necessarily limited to the upper-bound SRO lengthscale, but can be continuously engineered in the MRO range through control of structural disorder. 

\section{Conclusions}
We have rationalized the fundamental mechanisms governing the thermal-conductivity variations induced by atomistic disorder in a broad density range of structurally diverse disordered carbon polymorphs, solving the Wigner Transport Equation (WTE) \cite{simoncelli_thermal_2023} with quantum accuracy using a machine-learned interatomic potential \cite{rowe_accurate_2020}.
Specifically, we have introduced the bond-network entropy growth rate ${\rm BNE}(n)/ n$ as a structural descriptor that quantifies heterogeneity in the topology of the atomic-bond network, and demonstrated that it predicts variations in the conductivity of coordination-disordered solids.
We have shown that ${\rm BNE}(n)/ n$ determines the smoothness of the vibrational density of states (VDOS), and relied on all these findings to show how ${\rm BNE}(n)/ n$, VDOS smoothness, and conductivity are related to the lengthscales of structural disorder, and to the thermal resistance they induce.
These fundamental insights on the relationship between atomistic structure and macroscopic conductivity address the long-standing problem of formally determining the lengthscales underlying heat transport in disordered materials. 
In particular, we demonstrated the existence of a mapping between the WTE's bottom-up predictive theory for the conductivity of an explicitly disordered glass (EDG) and a generalized version of Kittel's top-down `phonon liquid' interpretative model for the thermal conductivity of a perturbatively disordered crystal (PDC) \cite{kittel_interpretation_1949}. 

We note that the established approaches used to estimate heat-transport lengthscales in disordered solids --- based on the Dynamical Structure Factor (DSF) \cite{fiorentino_hydrodynamic_2023} and its vibrational extension \cite{fiorentino_effects_2025}, or velocity-current correlations \cite{moon_crystal-like_2025} --- all rely on an approximate (lossy) mapping of the vibrational properties of a glass into the Brillouin zone of a reference crystal, i.e., identification of a blurred band structure. 
This identification is often (albeit not always \cite{conyuh_random_2021}) possible at low frequency, but in general very challenging at high frequency in structurally disordered materials where disorder cannot be directly obtained from a reference crystalline structure \cite{ciarella_finding_2023}, limiting the application of these approaches. 
The `VDOS smoothing' approach we introduced overcomes these limitations, as  it does not require to identify a reference periodic crystal structure and a Brillouin zone to approximately define bands --- it only requires to know the variations in the smoothness between two (observable) VDOS. Most importantly, the VDOS smoothing approach relies on a formal PDC-EDG mapping to determine from (physically observable) diffusivity and VDOS smoothness the lengthscales underlying heat transport in  disordered solids.

Finally, we note that the established interpretation of transport in glasses --- which classifies vibrations into propagons, diffusons, and locons \cite{allen_diffusons_1999} --- is based on band blurring, and phenomenologically relies on the property that the transition between propagon (intraband-like or propagation-like) transport and diffuson (interband-like or tunneling-like) transport is often centered around one single Ioffe-Regel frequency in the THz regime \cite{allen_diffusons_1999,fiorentino_effects_2025}, which can be estimated from the DSF.
Within this picture, transport lengthscales can be attributed exclusively to low-frequency propagon excitations. In contrast, within the VDOS smoothing picture transport lengthscales can be determined for all excitations in the vibrational spectrum, and we have found that some low-frequency vibrations can have transport lengthscales similar to those of some high-frequency vibrations.

Overall, we proposed an explanation of transport in disordered solids based on VDOS smoothing that is alternative to the established band-blurring picture and related propagon/diffuson classification. As such, this work calls for future studies to characterize conduction phenomena in solids in terms of 
heterogeneity of local atomic environments, VDOS smoothing and transport lengthscales, and the relation between them.
From a technological viewpoint, this study shows that it is possible to extract information on atomistic structural properties and heat-transport lengthscales from VDOS smoothness \cite{chumakov_role_2014} or conductivity measurements \cite{billinge_problem_2007}, and establishes the bond-network entropy as a fundamental degree of freedom to control and engineer the thermal properties of materials for energy-management applications \cite{luo_vibrational_2020, legenstein_heat_2025, hurley_thermal_2022, dennett_integrated_2021, islamov_high-throughput_2023, haque_halide_2020, bosoni_atomistic_2020, PhysRevB.102.201201, verdi2021thermal, yang_inter-channel_2022, xia_unified_2023, knoop_anharmonicity_2023, dangic_lattice_2025, jasrasaria_strong_2025}.
In particular, the conductivity-BNE relation discussed here may be especially relevant for:
(i) quantifying atomic disorder in supercapacitors, which strongly influences capacitance \cite{liu_structural_2024, jin_structural_2023};
(ii) assessing structural degradation and reliability of ceramic plasma-facing materials exposed to neutron irradiation in nuclear-fusion reactors \cite{fedrigucciComprehensiveScreeningPlasmaFacing2024, lin_perspectives_2025, nygren_new_2016, abrams_evaluation_2021, linke_challenges_2019}.


\section{Acknowledgments} 
We thank Prof. Jean-Marc Leyssale for providing us the structures of irradiated graphite. We gratefully acknowledge Prof. Mike C. Payne, Dr Nikita S. Shcheblanov, and Dr Mikhail E. Povarnitsyn for the useful discussions.
K.I. acknowledges support from Winton \& Cavendish Scholarship at the Department of Physics, University of Cambridge. M. S. acknowledges support from: (i) Gonville and Caius College; (ii) the Swiss National Science Foundation (SNSF) project P500PT\_203178. 
The computational resources were provided by: (i) the Sulis Tier 2 HPC platform (funded by EPSRC Grant EP/T022108/1 and the HPC Midlands+consortium); (ii) the UK National Supercomputing Service ARCHER2, for which access was obtained via the UKCP consortium and funded by EPSRC [EP/X035891/1];
(iii) the Kelvin2 HPC platform at the NI-HPC Centre (funded by EPSRC and jointly managed by Queen’s University Belfast and Ulster University). G.C. has equity interest of Symmetric Group LLP that licenses force fields commercially and also in Ångstrom AI. The other authors declare that they have no competing interest.\\

\appendix
\section{Quantum corrections on conductivity from previous molecular dynamics simulations}
\label{appendix_Q_corr}
Several previous studies have computed the thermal conductivity of coordination-disordered carbon polymorphs using molecular dynamics (MD) simulations \cite{giri_atomic_2022,suarez-martinez_effect_2011,farbos_time-dependent_2017}, in which vibrational energy is distributed among vibrational modes according to classical equipartition \cite{puligheddu_computational_2019}.
To compare our quantum-accurate thermal conductivity predictions with these earlier classical results, we must account for differences in the energy distribution of the microscopic vibrational degrees of freedom. 
In particular, in quantum-accurate approaches such as the WTE, the atomic vibrational energy is distributed among microscopic vibrational degrees of freedom according to the Bose-Einstein statistics; therefore, the specific heat $C(\omega, T)$ depends on frequency and temperature (Eq.~(\ref{eq:quantum_specific_heat})).
In contrast, in MD simulations each vibrational mode has, regardless of its frequency, a constant specific heat equal to the Boltzmann constant $k_B$ (i.e., the infinite-temperature limit of Eq.~(\ref{eq:quantum_specific_heat})). 
Consequently, classical MD simulations tend to overestimate the thermal conductivity at low temperature; to correct this and compare with our quantum calculations, we have adopted a phenomenological correction inspired by past work \cite{lv_non-negligible_2016}.

Specifically, the thermal conductivity analysis reported in \citet{giri_atomic_2022} allowed us to implement the following frequency-dependent corrections:
\begin{enumerate}
  \item Starting from the frequency-dependent VDOS and AF diffusivity for densities 2.1 and 2.88 g/cm$^3$ (see Fig. 3(a) and 4(a) of Ref.~\cite{giri_atomic_2022}), we transformed VDOS to have density-dependent integral equal to $\int_0^{\omega_{max}} g(\omega) d\omega = \frac{3 N_{at}}{\mathcal{V}}$. 
  \item We computed the conductivity at room temperature using the equation $\kappa(300{\rm K}) {=} \int_0^{\omega_{max}} d\omega g(\omega) C(\omega, 300{\rm K}) D(\omega)$ and its classical limit according to $\kappa_{\rm{classical}} {=} \int_0^{\omega_{max}} d\omega g(\omega) k_B D(\omega)$. 
  \item From those quantum (room-temperature) and classical values, we obtained a correction factor (CF) for each density, ${\rm CF} = \frac{\kappa(300{\rm K})}{\kappa_{\rm{classical}}}$.
  \item We multiplied the classical MD conductivity predictions (which used a modified Tersoff potential, see Fig. 4 (b) in ref. \cite{giri_atomic_2022}) by the corresponding CF, and reported them with filled markers in Fig.~\ref{fig:conductivity_density}.
\end{enumerate}

The works by \citet{farbos_time-dependent_2017}, and \citet{suarez-martinez_effect_2011} report the conductivities of IRG and VPC, respectively. Both these studies focus on the macroscopic conductivity and do not provide frequency-dependent mode diffusivity or vibrational density of states (VDOS).
To implement a frequency-dependent correction as described above, we approximately treated the structures in those references as having a VDOS equivalent to that of our corresponding (or most similar) structures.
For  \citet{farbos_time-dependent_2017} the VDOS of IRG T2, and for  \citet{suarez-martinez_effect_2011} we used VPC(D) 1.5 g/cm$^3$. 
Then, we implemented a correction based only on specific heat:
\begin{enumerate}
   \item We computed the specific heats according to $C(300{\rm K}) = \int_0^{\omega_{max}} d\omega g(\omega) C(\omega, 300{\rm K})$ and $C_{\rm classical} = \int_0^{\omega_{max}} d\omega g(\omega) k_B$, obtaining a correction factor ${\rm CF} = \frac{C(300{\rm K})}{C_{\rm{classical}}}$;
   \item We multiplied the conductivities reported in Refs  \cite{suarez-martinez_effect_2011, farbos_time-dependent_2017} by CF, and reported them with filled markers in Fig.~\ref{fig:conductivity_density}.
 \end{enumerate}

\begin{figure*}
  \centering
\includegraphics[width=\textwidth]{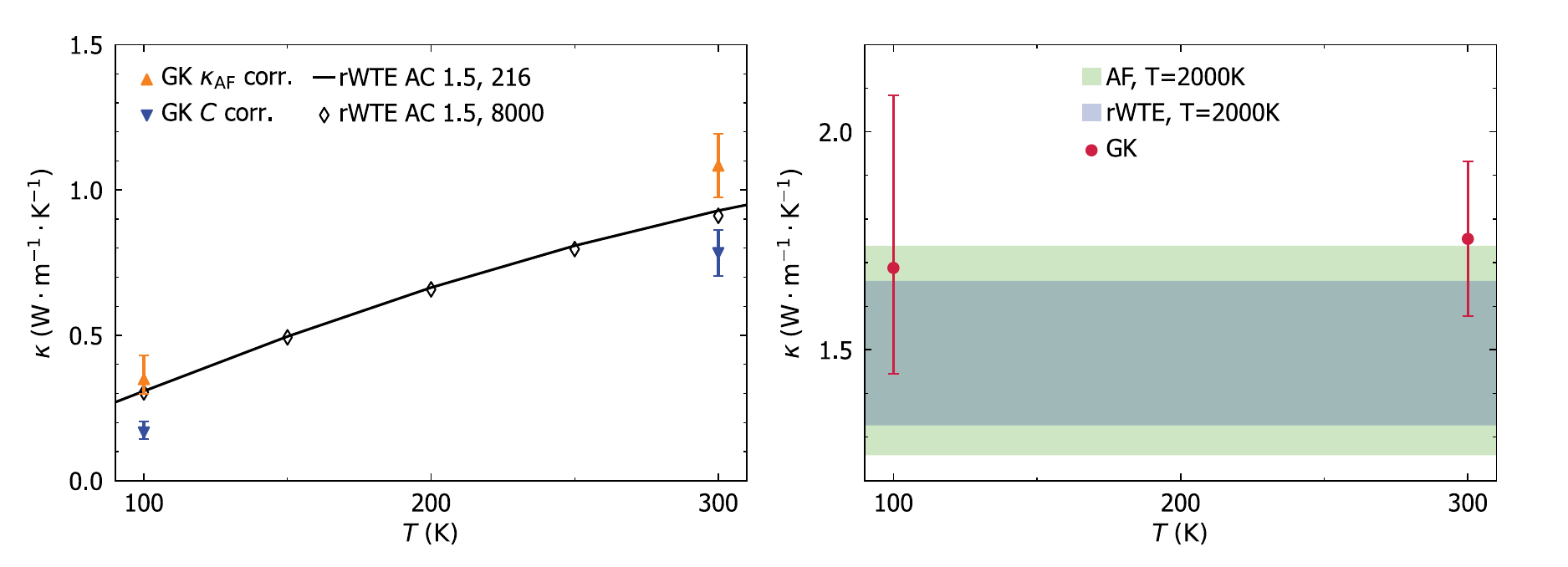}\\[-3mm]
  \caption{\textbf{Comparison between GK and WTE predictions for amorphous carbon at density 1.5 g/cm$^3$.} Left, comparison between rWTE conductivity predictions for AC 1.5 models containing 216 and 8000 atoms (solid line and diamond scatter points, respectively) and GK conductivity for AC 1.5 with 216 atoms with the quantum correction schemes discussed in the text, derived either from Allen-Feldman conductivity ($\kappa_{AF}$, orange triangles) or from specific heat ($C$, blue triangles).  Right, comparison in AC 1.5 with 216 atoms between the trace of uncorrected GK conductivity, and AF and rWTE predictions at 2000 K. The spread of AF and rWTE predictions denote the variations in the eigenvalues of the conductivity tensors. The GK error bars in the right panel are derived from both directional uncertainty and from standard deviation of GK conductivity in the last 250 ps of the simulation, and they have been scaled down by the corresponding correction factors in the left panel.
  }
  \label{fig:GK_vs_ALD}
\end{figure*}

In Fig.~\ref{fig:GK_vs_ALD}, we assess the accuracy of the two MD-conductivity correction protocols discussed above. Specifically in structure of AC 1.5 with 216 atoms, we compare Green-Kubo (GK) equilibrium molecular dynamics (EMD) conductivity calculations corrected either by the AF conductivity ($\kappa_{\rm AF}$) or specific heat ($C$) correction factors with direct rWTE calculations. For details of the simulation, see SM. We find that direct GK predictions at 100 and 300 K are in satisfactory agreement with rWTE and AF conductivity predictions in the classical limit (T$=$2000 K). After corrections, in AC 1.5 the $\kappa_{\rm AF}$-corrected GK conductivity slightly overestimates the rWTE predictions and $C$-corrected GK conductivity slightly underestimates it. At room temperature, the GK-corrected predictions are within 20\% of the rWTE value; at 100 K, the $\kappa_{\rm AF}$-corrected GK conductivity is within the error bars of rWTE result, and the upper error bar of $C$-corrected value is within 35\% of the rWTE result. 
A discussion on the agreement between quantum-corrected MD methods and rWTE predictions is also reported in Ref.~\cite{liang_mechanisms_2023}, and other correction schemes are discussed in Refs.~\cite{moon_crystal-like_2025, wang_density_2025}.

\section{Determination of the H$_1$ barcode}
\label{barcode_details}

\begin{figure}[b]
  \centering
\includegraphics[width=\WidthFigure]{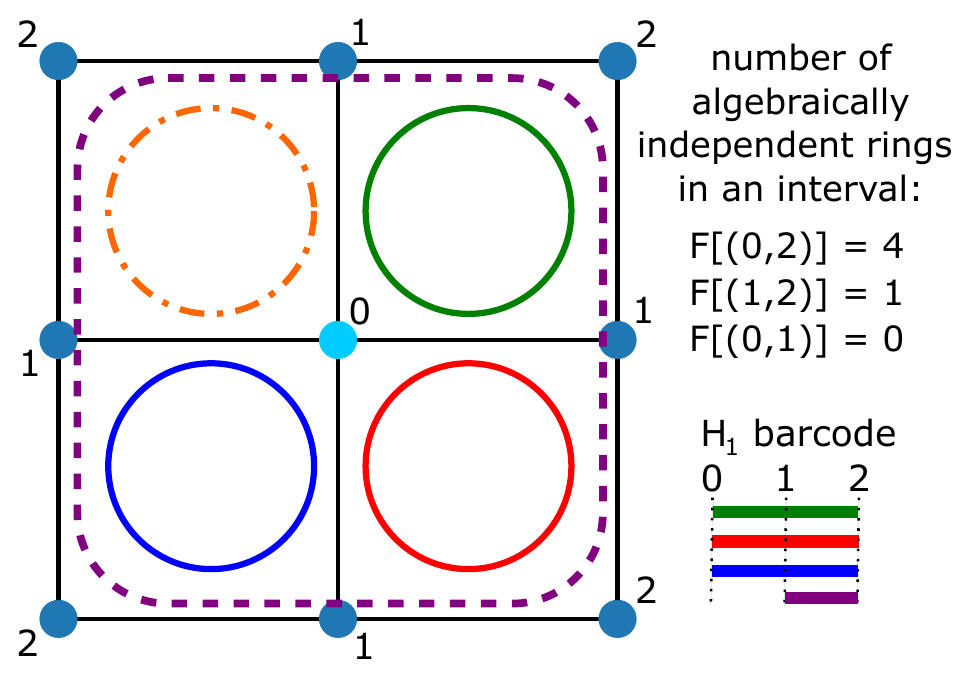}
  \caption{\textbf{Algebraically independent vs primitive rings.} Example of a local atomic environment (LAE) formed by blue atoms, black bonds, and having root atom cyan. 
The primitive rings \cite{le_roux_ring_2010}, are solid green, red, blue and dashed-dotted orange.
  In contrast, the algebraically independent rings in the LAE are solid green, red, blue, and dashed purple; each of these corresponds to a bar in the $H_1$ barcode. 
The matrix $F[(c, d)]$ (see text) contains information about the 
 the number of algebraically independent rings in the $(c, d)$-shell annulus, $S(c, d)$, of the LAE, where $c$ and $d$ are edge distances of atoms from the root atom, denoted with numbers in the figure. 
 Specifically, the number of algebraically independent rings in the annulus that contains the entire LAE is equal to four ($F[(0, 2)] = 4$, the four primitive rings). The number of rings in the annulus that excludes the root atom is equal to one (only purple dashed ring, $F[(1, 2)] = 1$). There are no rings in the annuli containing atoms with edge distance equal to only 0, 1 or 2, or either 0 or 1 ($F[(0, 0)] = 0$, $F[(1, 1)] = 0$, $F[(2, 2)] = 0$, $F[(0, 1)] = 0$). The $H_1$ barcode reproduces the number of algebraically independent rings according to Eq.~(\ref{eq:form_of_F}).
  The dashed-dotted orange ring is not contained in the $H_1$ barcode, since it is not algebraically independent from the solid green, red, blue, and dashed purple rings.
  } 
  \label{fig:combination_barcode}
\end{figure}

Here we summarize the algorithm discussed in Ref. \cite{schweinhart_statistical_2020} to compute the $H_1$ barcode of the atoms-bonds graph obtained from a certain LAE.
We denote with $r$ the root atom (center) of the LAE. Given nonnegative integers $c \le d$, we define the ($c$, $d$)-shell annulus, $S(c, d)$, which is a subgraph of the LAE composed of all atoms (and bonds connecting them) with edge distance (number of bonds on shortest path) between $c$ and $d$ from $r$. To characterize the bond network of the subgraph, it is useful to determine the number of algebraically independent rings in $S(c, d)$, given by:
\begin{equation}
\begin{split}
    F[(c, d)] =&\; \text{number of components}[S(c, d)]\\ &-\; \text{number of atoms}[S(c, d)]\\ &+\; \text{number of bonds}[S(c, d)].
\end{split}
\end{equation}
The number of algebraically independent rings is related to the $H_1$ barcode as \cite{schweinhart_statistical_2020}
\begin{equation}
  F [(c, d)] = {\rm rank}\big\{H_1[S(c, d)]\big\},
\end{equation}
i.e., it is the size of the set of rings $H_1[S(c, d)]$, that allow to express all other rings by linear combinations. $F[(c, d)]$ can also be resolved in terms of the number of intervals of the form (a, b) (\textit{e.g.}, shown as bars in Fig. \ref{fig:conductivity_entropy}\textbf{a}) in the $H_1$ barcode, denoted with $G[(a, b)]$:
\begin{equation}
    F[(c, d)] = \sum_{(a, b) \leq (c, d)} G[(a, b)],
    \label{eq:form_of_F}
\end{equation}
where the sum is taken over annuli ($a$, $b$) contained within or equal to the ($c$, $d$) annulus.
As discussed in Ref. \cite{schweinhart_statistical_2020}, the form of Eq.~(\ref{eq:form_of_F}) implies that $G[(c, d)]$ can be obtained from $F[(a, b)]$ using Möbius inversion:
\begin{equation}
    G[(c, d)] = \sum_{(a, b) \leq (c, d)} F[(a, b)] \mu[(a, b), (c, d)],
\end{equation}
where $\mu[(a, b), (c, d)]$ is a Möbius function defined as:
\begin{equation}
    \mu[(a, b), (a, b)] = 1,
\end{equation}
\begin{equation}
    \mu[(a, b), (c, d)] = - \sum_{(a, b) \leq (e, f) < (c, d)} \mu[(a, b), (e, f)],
    \label{eq:mu}
\end{equation}
which can be solved recursively. We note that the sum in Eq.~(\ref{eq:mu}) is computed over $(a, b) \leq (e, f) < (c, d)$ (excluding the case where $(e, f) = (c, d)$, as implemented in the \texttt{Swatches} software \footnote{\url{https://github.com/bschweinhart/Swatches}}), correcting a typo present in the original paper \cite{schweinhart_statistical_2020}.

We also note that, in order to characterize and  compare disorder across structures with different density, in this work we define LAEs in terms of number of atoms; this differs from the convention adopted by Schweinhart et al.  \cite{schweinhart_statistical_2020}, who defined LAEs as a function of number of layers (atoms with the same edge distance to the root).
Finally, we note that algebraically independent rings obtained from the intervals $G[(c, d)]$ are not necessarily primitive rings (i.e., rings that cannot be decomposed into smaller rings, see Ref. \cite{le_roux_ring_2010} for details) present in the LAE, see Fig. \ref{fig:combination_barcode} for an example.\\


\section{BNE and finite-size effects}
\label{app_BNE_finitesize}

\begin{figure}
  \centering
\includegraphics[width=\WidthFigure]{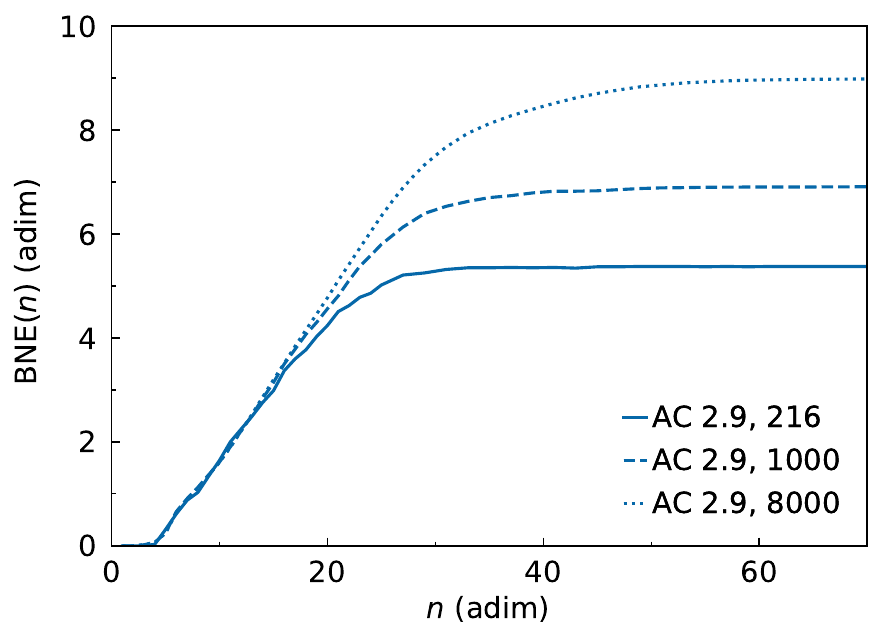}\\[-3mm]
  \caption{\textbf{BNE and finite-size effects.} Solid, dashed, and dotted lines correspond to 216, 1000, and 8000-atom models of AC with density $\sim$2.9 g/cm$^3$, respectively. $n$ is the number of atoms in the local environments.
 We note that for $5\lesssim n\lesssim 15$, BNE's growth rate is approximately constant and indistinguishable between the three physically equivalent representations of AC considered. 
 We highlight how the finite-size saturation of BNE$(n)$ occurs for larger value of $n$ in larger models.
 This shows that BNE's growth rate can be used to characterize structural disorder in AC over lengthscales smaller than the model's size. 
  } 
  \label{fig:top_entropy_convergence}
\end{figure}

In Fig. \ref{fig:top_entropy_convergence} we show the behavior of BNE as a function of LAE's size $n$ for three models of amorphous carbon that are expected to represent the same physical system, as they have all density $\sim$2.9 g/cm$^3$ and they differ only by simulation-cell size. 
We see that BNE grows with the LAE's size $n$, as expected from the extensivity property of the entropy, and then saturates to a constant value ${\rm BNE}_{\rm sat} = \ln{(N_{\rm at})}$ determined by the number of atoms in the simulation cell ($N_{\rm at}$). 
From a mathematical viewpoint, the saturation occurs when the $H_1$ barcode distinguishes all atomic environments, and therefore the distribution of barcodes approaches a uniform distribution with value $1/N_{\rm at}$ for each atom.
Most importantly, before the finite-size effect saturation occurs, BNE's growth rate is practically indistinguishable between the three models of AC, suggesting that it can be used to characterize structural disorder in AC over lengthscales smaller than the model's size. This motivates using BNE's growth rate as a descriptor for disorder in the bond network of solids.

We additionally note that the saturation is the reason we limit range of $n$ in Fig.~\ref{fig:conductivity_entropy} to $n \le 30$, as higher $n$ require exponentially bigger models. For example, in amorphous carbon to obtain converged results for $n = 30$ compared with $n = 15$, one requires a model which is $\approx 40$ times bigger (8000 vs 216 atoms). Hence, the estimated model size to obtain converged results for $n = 45$ in amorphous carbon is around 320,000 atoms. As we mentioned in section IV, the range of $n \in [14, 30]$ is large enough to capture medium-range order features and all structures in Fig.~\ref{fig:conductivity_entropy} have disorder in the short to medium range. The region of $n > 30$ might be a subject of future research analyzing very weakly-disordered solids, where conductivity is both significantly affected by structural order as well as intrinsic anharmonicity, going beyond the regime analyzed in this paper.

\section{Quasi-degenerate velocity operator}
\label{app_velop}
The quasi-degenerate velocity-operator in the frequency representation $\Upsilon(\omega)$ appearing in Eq.~(\ref{eq:diff_vdos}) is defined as:
\begin{equation}
\label{eq:quasi_deg_vel}
\begin{split}
    &\Upsilon^2(\omega) {=} [\mathcal{G}(\omega)]^{-1} \frac{1}{\mathcal{V}} \sum_{s, s'} \frac{\rVert \tens{v}_{ss'} \lVert^2}{3}
    \delta(\omega {-} \frac{\omega_s {+} \omega_{s'}}{2}) \delta(\omega_s {-} \omega_{s'}),
\end{split}
\end{equation}
where $\mathcal{G}(\omega)$ is a density of states defined as
\begin{equation}
    \mathcal{G}(\omega) = \frac{1}{N_{at}} \frac{1}{\mathcal{V}} \sum_{s,s'} \delta(\omega - \frac{\omega_s + \omega_{s'}}{2}) \delta(\omega_s - \omega_{s'}).
\end{equation}

\section{Derivation of the WTE diffusivity in the Explicitly Disordered Glass limit}
\label{app_derivation_diffusivity}
We start from the first line of Eq.~(\ref{eq:diff_intermediate}) and we use Eq.~(\ref{eq:quasi_deg_vel}) to approximate the velocity operator elements $\frac{\rVert \tens{v}_{ss'} \lVert^2}{3}$ by a single-frequency dependent function $\frac{\Upsilon^2(\omega)}{N_{at}}$,  
\begin{equation} 
\begin{split}
\label{eq:WDL_derivation1}
&D_{\rm EDG} (\omega)\!= \!\frac{\pi}{\mathcal{V}g(\omega)}\!\sum_{s,s'} \!\frac{\rVert \tens{v}_{ss'} \lVert^2}{3}\! L[\omega_s {-} \omega_{s'}\!, \Gamma_s {+}\Gamma_{s'}] \delta(\omega-\omega_s)\\
&\approx  \frac{\pi}{g(\omega)} \frac{\Upsilon^2(\omega)}{N_{at}} \left[\frac{1}{\mathcal{V}}\sum_{s,s'}\! L[\omega_s {-} \omega_{s'}\!, \Gamma_s {+}\Gamma_{s'}] \delta(\omega-\omega_s)\right].
\end{split}
\end{equation}
Then, we rewrite the quantity in the square brackets in terms of the convolution:
\begin{equation} 
\begin{split}
\label{eq:WDL_derivation2}
    & \left[\frac{1}{\mathcal{V}}\sum_{s,s'}\! L[\omega_s {-} \omega_{s'}\!, \Gamma_s {+}\Gamma_{s'}] \delta(\omega-\omega_s)\right] = \\
    = & \frac{1}{\mathcal{V}}\sum_{s,s'} \! \int \!d(\omega' {-} \omega_{s'})  L[(\omega_s {-} \omega_{s'}) - (\omega' {-} \omega_{s'}) \!, \Gamma_s] \\ & \qquad \qquad \times \! L[\omega' {-} \omega_{s'} \!, \Gamma_{s'}] \delta(\omega-\omega_s) \\
    = & \frac{1}{\mathcal{V}}\sum_{s,s'} \! \int \!d\omega'  L[\omega_s {-} \omega' \!, \Gamma_s]  L[\omega' {-} \omega_{s'} \!, \Gamma_{s'}] \delta(\omega-\omega_s) \\
    = & \sum_{s} \delta(\omega-\omega_s) \int \!d\omega'  L[\omega {-} \omega' \!, \Gamma(\omega)]  \frac{1}{\mathcal{V}} \sum_{s'} \! L[\omega' {-} \omega_{s'} \!, \Gamma_{s'}]  \\
    = & \mathcal{V} g(\omega) \int \! d\omega' L[\omega {-} \omega' \!, \Gamma(\omega)] g_d(\omega') = \mathcal{V} g(\omega) g_{dd}(\omega)
\end{split}
\end{equation}
where $g_{dd}(\omega)$ denotes the convolution of the dressed VDOS with the Lorentzian with FWHM $\Gamma(\omega)$. 
In Eq.~(\ref{eq:WDL_derivation2}), to go from the first to the second line, we used the property that a Lorentzian with FWHM  $\Gamma_s {+}\Gamma_{s'}$ can be obtained by convolving two Lorentzians with FWHM $\Gamma_s$ and $\Gamma_{s'}$, respectively. 
To transform the second line into the third, we changed the variable of integration from $\omega' - \omega_{s'}$ to $\omega'$. From third to fourth line, we exploited the delta function $\delta(\omega- \omega_s)$ to replace $\omega_s$ and $\Gamma_s$ in the first Lorentzian to $\omega$ and $\Gamma(\omega)$. $\Gamma(\omega)$ denotes the linewidth evaluated at frequency $\omega$, this notation is possible here because the presence of structural disorder forbids degeneracies and thus allows a bijective mapping between frequency and mode, i.e., $\Gamma_s = \Gamma(\omega_s) = \Gamma(\omega = \omega_s)$. In the fourth line we also regrouped the terms, and from the fourth to the fifth line we used the definitions of VDOS $\mathcal{V} g(\omega) = \sum_s \delta(\omega - \omega_s)$ and dressed VDOS $g_d(\omega')$ (Eq.~(\ref{eq:relationDOS})). 
By combining Eq.~(\ref{eq:WDL_derivation1}) and Eq.~(\ref{eq:WDL_derivation2}), we obtain the second line of Eq.~(\ref{eq:diff_intermediate}).

\begin{figure}[t]
  \centering
\includegraphics[width=\WidthFigure]{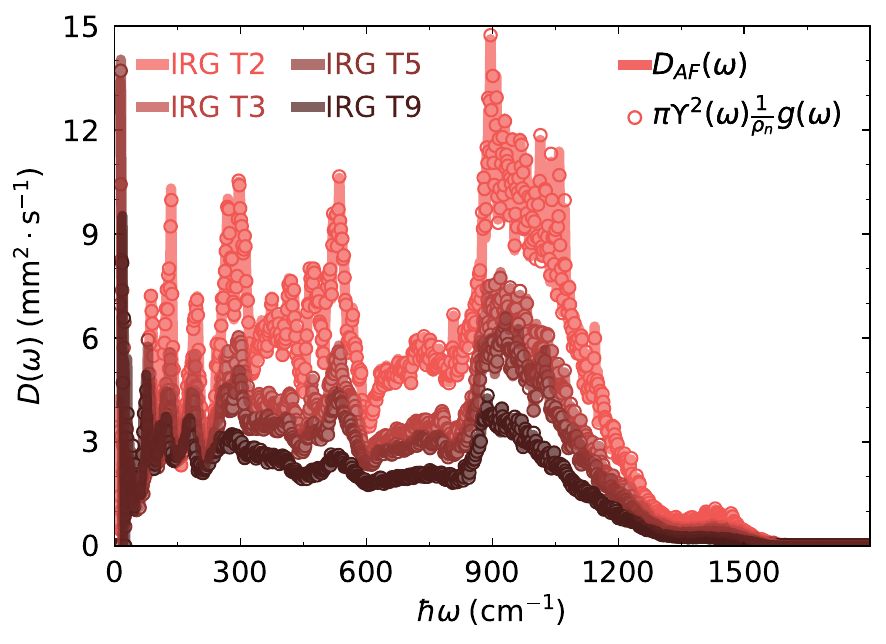}\\[-3mm]
  \caption{\textbf{Equivalence between EDG and AF diffusivity in IRG.} Solid lines are AF diffusivities in various IRG samples, and the overlapping empty circles are the corresponding WTE diffusivities in the EDG limit (Eq.~(\ref{eq:diff_vdos})).
  }
  \label{fig:D_vs_upsilon2g}
\end{figure}

If we neglect the influence of the intrinsic linewidth on the broadening of $g_{dd}(\omega)$, then we can approximate the disordered solid to be harmonic, leading to the alternative expression for AF diffusivity shown by Eq.~(\ref{eq:diff_vdos}).
In Fig. \ref{fig:D_vs_upsilon2g}, we show the numerical equivalence between Eq.~(\ref{eq:diff_vdos}) and the AF diffusivity in various IRG models.

Finally, we discuss how to obtain the self-consistent relation discussed in Eq.~(\ref{eq:kappa_PDC_rewritten}).
We start considering the frequency decomposition equation (Eq.~(\ref{eq:kappa_omega})), plug in it the expression~(\ref{eq:diff_intermediate}) for the EDG diffusivity $D(\omega)$, and perform the following algebraic manipulations:
\begin{equation} 
\begin{split}
    \kappa & = \int d\omega \, g(\omega) C(\omega) D(\omega) \\
    & = \!\int\! d\omega g(\omega) C(\omega) \pi \Upsilon^2(\omega) \frac{1}{\rho_n} \!\left[\int\!\!  d\omega' L[\omega {-} \omega' \!, \Gamma(\omega)]  g_d(\omega')\right] \\
    & = \pi \frac{1}{\rho_n} \iint d\omega d\omega' C(\omega) \Upsilon^2(\omega) g(\omega) L[\omega {-} \omega' \!, \Gamma(\omega)]  g_d(\omega') \\
    & \approx \pi \frac{1}{\rho_n}\! \int \!d\omega' g_d(\omega') C(\omega') \Upsilon^2(\omega')\! \int \! d\omega g(\omega) L[\omega {-} \omega' \!, \Gamma(\omega)]  \\
    & = \pi \frac{1}{\rho_n} \int d\omega' g_d(\omega') C(\omega') \Upsilon^2(\omega') g_d(\omega') \\
    & = \!\!\! \int \!\!\!d\omega g_d(\omega) C(\omega) \pi \Upsilon^2(\omega) \frac{1}{\rho_n} g_d(\omega) \!= \!\!\!\int \!\!\!d\omega g_d(\omega) C(\omega) D_d(\omega),
\end{split}
\label{eq:kappa_omega_dressed}
\end{equation}
{where we denote the dressed diffusivity $D_d(\omega) = \pi \Upsilon^2(\omega) \frac{1}{\rho_n} g_d(\omega)$ (i.e., having the same form of Eq.~(\ref{eq:diff_vdos}), but with bare VDOS replaced by the dressed VDOS).}
In the third line we reordered the integrals and in the fourth line we used the fact that in derivation of Eq.~(\ref{eq:diff_intermediate}) we used an approximation that the interacting modes have similar frequency due to value of the linewidth $\Gamma(\omega)$ being much smaller than the value of the frequency $\omega$. We used this approximation again here to set $C(\omega) \sim C(\omega')$ and $\Upsilon(\omega) \sim \Upsilon(\omega')$ for interacting modes. In the fifth line we identify $\int \! d\omega g(\omega) L[\omega {-} \omega' \!, \Gamma(\omega)]$ with the dressed VDOS $g_d(\omega')$. In the sixth line we renamed $\omega'$ to $\omega$.
The expression in the final line of Eq.~(\ref{eq:kappa_omega_dressed}) is particularly insightful, since the VDOS in the first term and VDOS in the diffusivity are dressed, preserving the property that the diffusivity in disordered systems is proportional to the VDOS.
We conclude by noting that the relation between the PDC and EDG conductivities is summarized in Tab.~\ref{tab:compare}.

\begin{table*}[!ht]
\caption{Comparison between the WTE treatments of PDC and EDG.\\}
\label{tab:compare}
\hspace*{-4mm}
\begin{tabular}{c|c|c}
\hline
    \hline
    & \begin{minipage}{0.3\textwidth}  
    \vspace*{2mm}  
    \textbf{Perturbatively Disordered Crystal (PDC)} 
    \vspace*{2mm} \end{minipage} 
    & \textbf{Explicitly Disordered Glass (EDG)} \\
    \hline
\begin{minipage}{0.12\textwidth}    
    \vspace*{2mm} 
    Transport mechanism 
    \vspace*{2mm} \end{minipage}
    & Particle-like propagation
    & Wave-like tunnelling \\
    \hline
\begin{minipage}{0.12\textwidth}    
    \vspace*{2mm} 
    VDOS 
    \vspace*{2mm} \end{minipage}
    & \begin{minipage}{0.42\textwidth}
    \vspace*{2mm}
    Not broadened \\
    $g(\omega) = g^{\rm DR}_{\rm UOC}(\omega)$
    \vspace*{2mm} \end{minipage}
    & \begin{minipage}{0.42\textwidth}    
    \vspace*{2mm}
    Broadened by structural disorder \\
    $g(\omega) = g_{\rm EDG}(\omega)$
    \vspace*{2mm} \end{minipage} \\
    \hline
\begin{minipage}{0.12\textwidth}    
    \vspace*{2mm} 
    Dressed VDOS
    \vspace*{2mm} \end{minipage}
    & \begin{minipage}{0.42\textwidth}
    \vspace*{2mm}
    Broadened by structural disorder \\[1mm]
    $g_{\rm PDC}(\omega) = \int  d\omega' L[\omega {-}\omega', \Gamma_{\rm dis}(\omega')] g^{\rm DR}_{\rm UOC}(\omega')$ \\[1mm]
    $g_d(\omega) \equiv g_{\rm PDC}(\omega) \approx g_{\rm EDG}(\omega)$
    \vspace*{2mm} \end{minipage}
    & \begin{minipage}{0.42\textwidth}    
    \vspace*{2mm}
    Broadened by disorder and anharmonicity \\[1mm]
    $g_{d}(\omega) = \int  d\omega' L[\omega {-}\omega', \Gamma(\omega')] g(\omega')$ \\[1mm]
    $g_d(\omega) \approx g(\omega) = g_{\rm EDG}(\omega)$
    \vspace*{2mm} \end{minipage} \\
    \hline
\begin{minipage}{0.12\textwidth}
\vspace*{2mm} 
 Origin of linewidth
 \vspace*{2mm} 
    \end{minipage}
    & \begin{minipage}{0.42\textwidth}  
    \vspace*{2mm} 
Structural disorder
    \vspace*{2mm} \end{minipage}
    &  \begin{minipage}{0.42\textwidth} 
    \vspace*{2mm} 
    Intrinsic, e.g., due to isotopes or anharmonicity
    \vspace*{2mm} 
    \end{minipage} \\
    \hline
\begin{minipage}{0.12\textwidth}    
    \vspace*{2mm} 
    Diffusivity
    \vspace*{2mm} \end{minipage}
    & \begin{minipage}{0.42\textwidth}    
    \vspace*{2mm} 
    Eq.~(\ref{eq:diffusivity_crystal}), \\[1mm]
    $D_{\rm PDC}(\omega) = v_{\rm eff}(\omega) \lambda_{\rm eff}(\omega) = \frac{v^2_{\rm eff}(\omega)}{\Gamma_{\rm dis}(\omega)}$ \\
    and Eq.~(\ref{eq:kappa_PDC_rewritten}), \\ [1mm]
    $D_{\rm PDC}(\omega) = \pi \Upsilon^2(\omega) \frac{1}{\rho_n} g_{\rm PDC}(\omega)$
    \vspace*{2mm} \end{minipage}
    & \begin{minipage}{0.42\textwidth}    
    \vspace*{2mm} 
    Eqs.~(\ref{eq:diff_intermediate}) and (\ref{eq:diff_vdos}), \\[2mm]
    $D_{\rm EDG}(\omega) {\approx} \pi {\Upsilon^2(\omega)}  \frac{1}{\rho_n}\!  
    \left[\!\int  d\omega' L[\omega {-} \omega' \!, \Gamma(\omega)]  g_d(\omega')\right]$ \\[1mm]
    $\approx \pi {\Upsilon^2(\omega)}  \frac{1}{\rho_n} g_{\rm EDG}(\omega)$
    \vspace*{2mm} \end{minipage} \\
    \hline
\begin{minipage}{0.12\textwidth}
    \vspace*{2mm} 
    Compatible
    observables 
    \vspace*{2mm} \end{minipage}
    & \multicolumn{2}{c}{\begin{minipage}{0.88\textwidth} 
    \vspace*{2mm} 
    Thermal conductivity $\kappa$, dressed VDOS $g_d(\omega)$, and diffusivity $D(\omega)$
    \vspace*{2mm} \end{minipage}} \\
    \hline
\hline
\end{tabular}
\end{table*}

\section{Relation between PDC and Kittel's interpretation of conductivity in glasses}
\label{app_Kittel_limit}

We find that the PDC-EDG mapping is a generalization of Kittel's interpretation, since the conductivity of a PDC reduces to the Kittel's expression $\kappa = \frac{1}{4} C_{\mathcal{V}} \, v_{\rm sound} \, \Lambda_0$ \cite{kittel_interpretation_1949} by averaging propagation velocity and transport lengthscale over frequency:
\begin{equation}\label{eq:avg}
\begin{split}
    \kappa_{\rm PDC} & = \int d\omega \; g_d(\omega) C(\omega) v_{\rm eff}(\omega) \lambda_{\rm eff}(\omega)
    \\ & = \left[ \int d\omega g_d(\omega) C(\omega) \right] \frac{\int d\omega g_d(\omega) C(\omega) v_{\rm eff}(\omega) \lambda_{\rm eff}(\omega)}{\int d\omega  g_d(\omega) C(\omega)} 
    \\ & = C_{\mathcal{V}} \, v_{\rm eff} \lambda_{\rm eff} 
    \\ & \rightarrow \frac{1}{4} C_{\mathcal{V}} \, v_{\rm sound} \, \Lambda \rightarrow \frac{1}{4} C_{\mathcal{V}} \, v_{\rm sound} \, \Lambda_0,
\end{split}
\end{equation}
where in the second and third lines $v_{\rm eff}(\omega)$ and $\lambda_{\rm eff}(\omega)$ are averaged and become frequency-independent; in the fourth line the propagation velocity is changed to velocity of sound and a factor of 4 is absorbed into the product $v_{\rm sound} \,\Lambda$ to account for the phenomenological factor $\frac{1}{4}$ in the original Kittel's paper \cite{kittel_interpretation_1949}. The final arrow represents the limit of heavily disordered glasses at non-cryogenic temperatures \cite{kittel_interpretation_1949}, where $\Lambda$ is assumed to be independent of temperature and hence is redefined as $\Lambda_0$. 

We note that increase of $\Lambda$ with decrease in temperature for a quartz glass observed by Ref.~\cite{kittel_interpretation_1949}, naturally follows from Eq.~\ref{eq:avg}, since diffusivity can depend on temperature and is averaged over frequency with temperature-dependent weights:
\begin{equation}
    \Lambda(T) = \frac{4}{v_{\rm sound}} \frac{\int d\omega g_d(\omega) C(\omega, T) D(\omega, T)}{\int d\omega  g_d(\omega) C(\omega, T)}
\end{equation}

Here we also note that the name `propagation velocity' for $v_{\rm eff}(\omega)$ was used to point out the relationship to the Boltzmann transport equation (BTE), as we use a BTE-like expression for the diffusivity $D(\omega) = v_{\rm eff}(\omega) \lambda_{\rm eff}(\omega)$, and propagation is the transport mechanism described by the BTE \cite{simoncelli_wigner_2022}. Due to the frequency dependence, this quantity generalizes the frequency-independent Debye-like sound velocity $v_{\rm sound}$ used in Kittel's interpretation of conductivity in glasses.
We note that defining a frequency-dependent velocity does not require resolving phonon wavevectors or bandstructures, and is effectively determined here from the off-diagonal elements of the velocity operator that assume non-zero values in glasses; interestingly, we note that resolving the velocity in frequency allows one to connect it a vibrational mode in strongly disordered systems where disorder forbids perfect degeneracies and therefore one has bijective mapping between frequency and mode.
In summary, the meaning of frequency-dependent effective propagation velocity for disordered systems is different from either the sound velocity or from group velocity.
The propagation velocity $v_{\rm eff}(\omega)$ and transport lengthscale $\lambda_{\rm eff}(\omega)$ are defined based on physical observables of frequency-dependent VDOS and diffusivity. Their product is equal to the diffusivity $D(\omega) = v_{\rm eff}(\omega) \lambda_{\rm eff}(\omega)$ and their quotient is the disorder linewidth $\Gamma_{\rm dis}(\omega) = v_{\rm eff}(\omega) / \lambda_{\rm eff}(\omega)$ that determines the smoothness of the dressed PDC VDOS $g_d(\omega) \!=\! {\int} {d\omega'} L[\omega {-} \omega', \Gamma_{\rm dis}(\omega')] g(\omega')$ equal to the VDOS of a disordered solid.

Through PDC–EDG mapping, we can represent the diffusivity of a glass (EDG) in terms of propagation velocities and transport lengthscales of an effective crystal (PDC), in which vibrations interact in a way determined by the disorder within the EDG. The existence of a formal mapping between PDC and EDG does not imply that the vibrational modes in the actual disordered system (EDG) propagate as in a PDC --- this is evident from the fact that the effective velocities and transport lengthscales emerging from this mapping result from combining several quasi-degenerate off-diagonal elements of the EDG’s velocity operator.

The usefulness of the EDG-PDC mapping lies in its ability to determine the microscopic physical lengthscale underlying heat transport mediated by vibrations of frequency $\omega$ (albeit without resolving whether they contribute through propagation or tunneling mechanisms). Because the effective frequency-dependent velocities and transport lengthscales are derived from observables such as the VDOS and diffusivity, they have a physical meaning. Their determination is made possible by the PDC-EDG mapping, and we have shown that one can obtain fundamental insights into structure-conductivity relations by comparing the transport lengthscales with material-structure lengthscales (e.g., short- and medium-range order).

\section{Determination of the disorder linewidth from PDC VDOS}
\label{app_VDOS_smoothing}

\begin{figure}[h]
  \centering
\includegraphics[width=\WidthFigure]{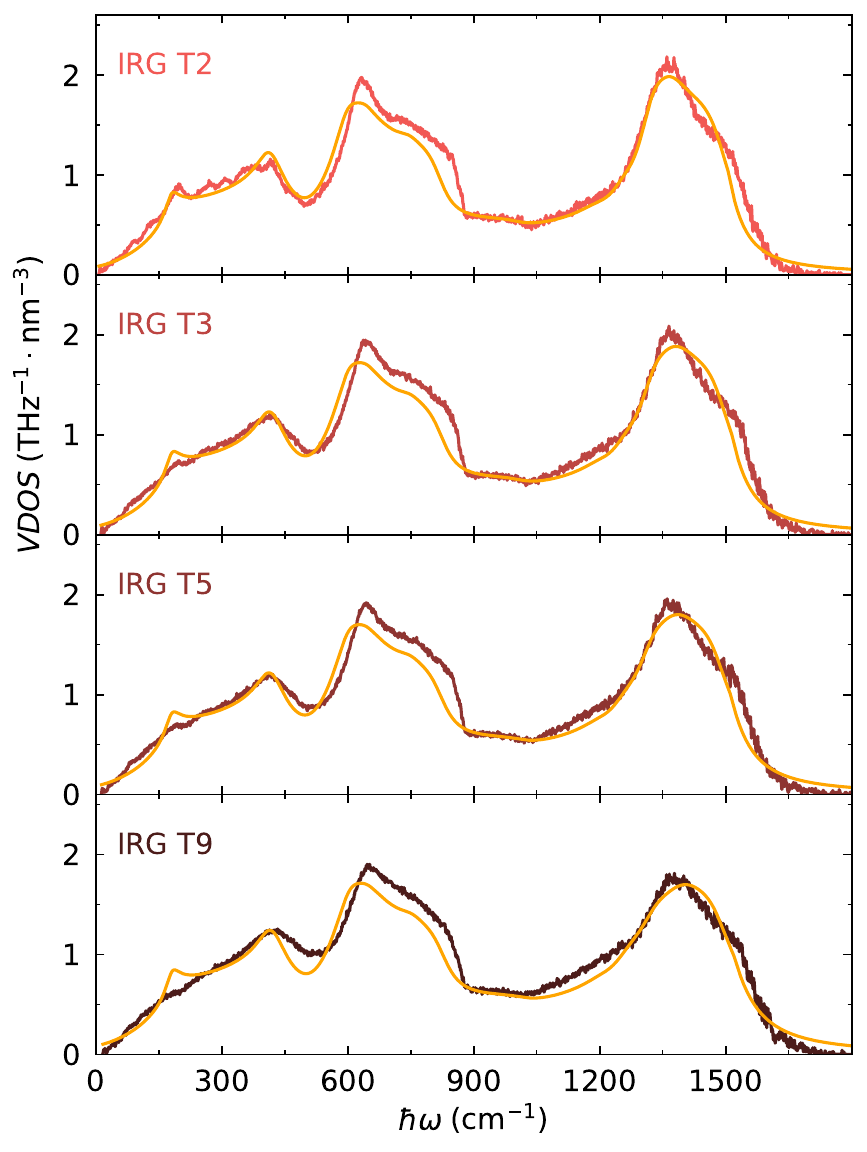}
  \caption{\textbf{Bare VDOS of IRG vs PDC VDOS.} Yellow solid lines are the PDC VDOS obtained dressing the bare VDOS of pristine UOC graphite, and red-to-black lines are bare EDG VDOS of IRG graphite.
  }
  \label{fig:VDOS_all_IRG_vs_model}
\end{figure}

In this appendix we discuss the numerical details on how to determine the PDC disorder linewidths $\Gamma(\omega')$ that have to be used in Eq.~(\ref{eq:relationDOS}) to transform the VDOS of pristine UOC graphite into the VDOS of IRG. 

To obtain the density-renormalized VDOS discussed in Sec.~\ref{sec:disorder}, $g^{\rm DR}_{\rm UOC}(\omega)$, we proceeded as follows. First, we used the GAP potential \cite{rowe_accurate_2020} to compute second-order interatomic forces in a 8x8x2 supercell, then used these and Fourier-interpolation \cite{togo_first-principles_2023} to compute vibrational frequencies and group velocities on a dense 128x128x32 $\bm{q}$-mesh.
By multiplying the frequencies by the factor $a$, as discussed in Sec.~\ref{sec:disorder}, 
we take into account the effect that the density reduction induced by irradiation has on the vibrational frequencies. The bare VDOS was computed using a Lorentzian broadening for the delta functions with FWHM equal to $2 \times 0.6$ cm$^{-1}$, where 0.6 cm$^{-1}$ is the value of the convergence plateau parameter $\eta$ used for thermal conductivity calculations in IRG (see SM for details).

After determining the bare $g^{\rm DR}_{\rm UOC}(\omega)$, we apply to it the dressing transformation~\ref{eq:relationDOS} to obtain the dressed VDOS of the PDC.
When performing such transformation, we assume that the disorder linewidths appearing in the dressing integral assume the established functional form~(\ref{eq:linewidth_expression}); 
specifically, we calculated the modulus of the group velocity in the frequency representation, $v(\omega)$, which appears in such an expression as:
\begin{equation}
    v(\omega) = \frac{\sum_{\bm{q}s} \sqrt{\frac{v^2_{\bm{q}s}}{3}} \delta(\omega - a\omega_{\bm{q}s})}{\sum_{\bm{q}s} \delta(\omega - a\omega_{\bm{q}s})},
    \label{eq:vel_omega}
\end{equation}
where sum $\sum_{\bm{q}s}$ runs over the aforementioned 128x128x32 $\bm{q}$-mesh, $\omega_{\bm{q}s}$ is the frequency of the mode $\bm{q}$,$s$ in UOC graphite, and $v_{\bm{q}s}$ the modulus of its group velocity.  To evaluate numerically Eq.~(\ref{eq:vel_omega}), the Dirac delta distributions are broadened with a Lorentzian having FWHM equal to twice the aforementioned value of the convergence plateau parameter $\eta$ in IRG ($2 \times 0.6$ cm$^{-1}$).
The values for the parameter L and R appearing in Eq.~(\ref{eq:linewidth_expression}) are determined as the values that  minimize the Mean Squared Deviation (MSD) between the PDC VDOS and EDG VDOS of irradiated graphite:
\begin{equation}
    {\rm MSD} = \sum_i \left(g_{\rm PDC}(\omega_i) - g_{\rm EDG}(\omega_i) \right)^2.
\end{equation}
This minimization is performed using a stochastic gradient descent optimizer implemented in \texttt{PyTorch} \cite{NEURIPS2019_9015}.

We show the results of the fits for disorder linewidth $\Gamma_{\rm dis}(\omega)$ in Fig. \ref{fig:IRG_diffusivity_decomposition}a and the corresponding predictions for the PDC VDOS in Fig. \ref{fig:VDOS_all_IRG_vs_model}. We find that PDC VDOS reasonably replicates the variation with frequency and irradiation dose of the VDOS for all structures of IRG analysed in this work. 
Finally, we note that after shifting the frequencies of UOC graphite, there are no vibrational modes above ${\sim}1550$ cm$^{-1}$. 
These modes have negligible impact on thermal transport; to decompose IRG's diffusivity above ${\sim}1550$ cm$^{-1}$, we determine the disorder linewidth of frequencies above ${\sim}1550$ cm$^{-1}$  as by fitting the linewidth value that matches the tail of disorder linewidth plot in Fig.~\ref{fig:IRG_diffusivity_decomposition}a, finding a linewidth of 24 cm$^{-1}$.\\

\section{Influence of anharmonicity on thermal conductivity in IRG}
\begin{figure}[ht]
  \centering
\includegraphics[width=\WidthFigure]{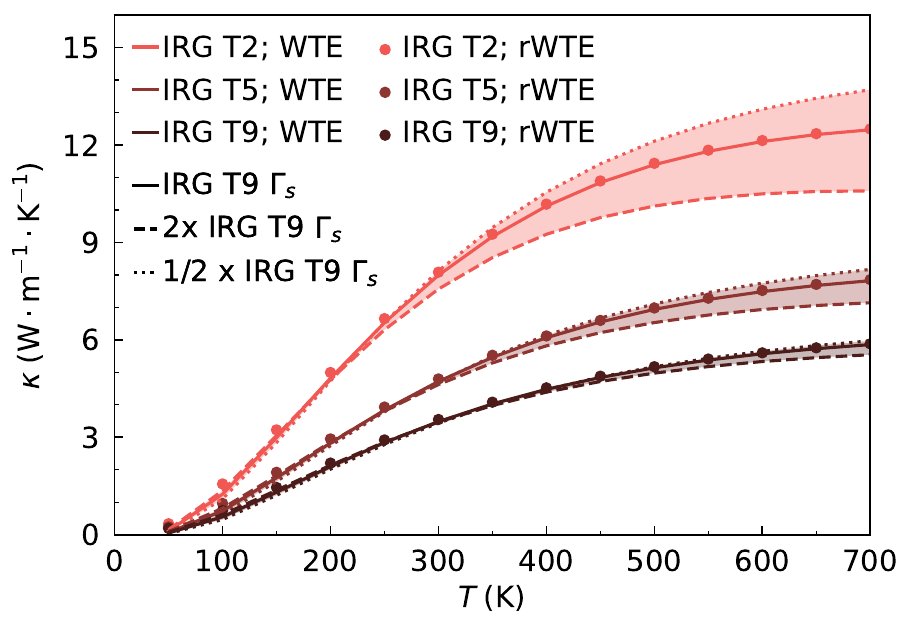}
  \caption{\textbf{Influence of anharmonicity on thermal conductivity in IRG.} Solid, dashed and dotted lines show WTE conductivity calculations for IRG with linewidth equal to linewidth derived from 216-atom model of IRG T9, linewidth of twice this value and linewidth of half this value respectively. The scatter points are rWTE conductivity calculations with linewidth derived from IRG T9 216. Red, brown and black colors denote IRG T2, T5 and T9 respectively.
  }
  \label{fig:anharmonicity}
\end{figure}
Fig.~\ref{fig:conductivity_density} shows that the rWTE yields results in agreement with experiments and previous studies based on molecular dynamics. We recall that the rWTE accounts for the influence of anharmonicity only when these are not spuriously affected by finite-size effects. It has been shown in Refs.~\cite{simoncelli_thermal_2023,harper_vibrational_2024} that the rWTE allows to converge thermal-conductivity calculations faster compared to the bare WTE in `strongly disordered' solids having: (i) disorder-induced repulsion between energy levels \cite{simkin_minimum_2000} comparable to the intrinsic linewidth \footnote{We recall that stronger disorder-induced repulsion between energy levels promotes a smoother VDOS.}; (ii) average velocity-operator elements nearly constant with respect to the energy difference between two eigenstates \cite{simoncelli_thermal_2023}. 
In this `strongly disordered' regime, the rWTE evaluated in a small atomistic model (containing hundreds of atoms) is expected to yield results in agreement with the bare WTE evaluated in a large model (containing several thousands of atoms).
In this section we provide additional information to strengthen the claim that in the disordered carbon polymorphs that we studied, structural disorder is the dominant source of thermal resistance and hence the rWTE applied to small models yields results compatible with the WTE applied to very large models.  
We focus on irradiated graphite, recalling that in Fig.~\ref{fig:conductivity_density} we showed that for IRG T9 the room-temperature rWTE conductivity of a 216-atom model is practically indistinguishable from the bare WTE conductivity of a 14009-atom model. 
To confirm that disorder is the dominant source of thermal resistance, we evaluate the bare WTE on very large 14009-atom atomistic models and artificially rescale the anharmonic linewidths, enlarging or reducing them by a factor 2. 
Fig.~\ref{fig:anharmonicity} shows that the lower is structural disorder in IRG, the higher the conductivity and the more relevant are the effects of anharmonicity. 
Importantly, for all the IRG structures considered, this artificial amplification or reduction of anharmonicity has negligible effect on the conductivity in the temperature range 50<T$\lesssim$300 K. Since IRG graphite is the class of carbon polymorphs with the largest conductivity and hence most sensitive to anharmonicity, we infer from Fig.~\ref{fig:anharmonicity} that for the polymorphs studied, the conductivity in the temperature range 50<T$\lesssim$300 K is mainly limited by structural disorder.
Our findings are in broad agreement with Refs.~\cite{thebaud_breaking_2023,fiorentino_effects_2025}, which discussed how compositional disorder can yield damping linewidth stronger than those due to anharmonic effects, resulting in a convergent conductivity in the low-temperature limit where anharmonicity phases out.

Finally, we note that upon increasing temperature anharmonicity becomes more relevant---we see that at 700 K it is important to use the actual physical values of the anharmonic linewidths. We also highlight how in this high-temperature regime the rWTE and WTE yield indistinguishable results, confirming that anharmonicity is strong enough to dominate over the computational broadening $\eta$ employed in the rWTE (i.e., anharmonic effects are negligibly affected by finite-size effects).\\


\begin{thebibliography}{138}%
\makeatletter
\providecommand \@ifxundefined [1]{%
 \@ifx{#1\undefined}
}%
\providecommand \@ifnum [1]{%
 \ifnum #1\expandafter \@firstoftwo
 \else \expandafter \@secondoftwo
 \fi
}%
\providecommand \@ifx [1]{%
 \ifx #1\expandafter \@firstoftwo
 \else \expandafter \@secondoftwo
 \fi
}%
\providecommand \natexlab [1]{#1}%
\providecommand \enquote  [1]{``#1''}%
\providecommand \bibnamefont  [1]{#1}%
\providecommand \bibfnamefont [1]{#1}%
\providecommand \citenamefont [1]{#1}%
\providecommand \href@noop [0]{\@secondoftwo}%
\providecommand \href [0]{\begingroup \@sanitize@url \@href}%
\providecommand \@href[1]{\@@startlink{#1}\@@href}%
\providecommand \@@href[1]{\endgroup#1\@@endlink}%
\providecommand \@sanitize@url [0]{\catcode `\\12\catcode `\$12\catcode `\&12\catcode `\#12\catcode `\^12\catcode `\_12\catcode `\%12\relax}%
\providecommand \@@startlink[1]{}%
\providecommand \@@endlink[0]{}%
\providecommand \url  [0]{\begingroup\@sanitize@url \@url }%
\providecommand \@url [1]{\endgroup\@href {#1}{\urlprefix }}%
\providecommand \urlprefix  [0]{URL }%
\providecommand \Eprint [0]{\href }%
\providecommand \doibase [0]{https://doi.org/}%
\providecommand \selectlanguage [0]{\@gobble}%
\providecommand \bibinfo  [0]{\@secondoftwo}%
\providecommand \bibfield  [0]{\@secondoftwo}%
\providecommand \translation [1]{[#1]}%
\providecommand \BibitemOpen [0]{}%
\providecommand \bibitemStop [0]{}%
\providecommand \bibitemNoStop [0]{.\EOS\space}%
\providecommand \EOS [0]{\spacefactor3000\relax}%
\providecommand \BibitemShut  [1]{\csname bibitem#1\endcsname}%
\let\auto@bib@innerbib\@empty
\bibitem [{\citenamefont {Tian}\ \emph {et~al.}(2023)\citenamefont {Tian}, \citenamefont {Ma}, \citenamefont {Li}, \citenamefont {Cheng}, \citenamefont {Ning}, \citenamefont {Han}, \citenamefont {Xu}, \citenamefont {Zhang}, \citenamefont {Zhao}, \citenamefont {Li}, \citenamefont {Zou}, \citenamefont {Liao}, \citenamefont {Yu}, \citenamefont {Li}, \citenamefont {Wang}, \citenamefont {Liu}, \citenamefont {Li}, \citenamefont {Huang}, \citenamefont {Yao}, \citenamefont {Ding}, \citenamefont {Guo}, \citenamefont {Huang}, \citenamefont {Lu}, \citenamefont {Han}, \citenamefont {Wang}, \citenamefont {Cheng}, \citenamefont {Liu}, \citenamefont {Xu}, \citenamefont {Liu}, \citenamefont {Gao}, \citenamefont {Jiang}, \citenamefont {Lin}, \citenamefont {Zhao}, \citenamefont {Wang}, \citenamefont {Bai}, \citenamefont {Fu}, \citenamefont {Wang}, \citenamefont {Li}, \citenamefont {Lei}, \citenamefont {Zhang}, \citenamefont {Hou}, \citenamefont {Pei}, \citenamefont {Pennycook}, \citenamefont {Wang}, \citenamefont {Chen}, \citenamefont {Zhou},\ and\ \citenamefont {Liu}}]{tian_disorder-tuned_2023}%
  \BibitemOpen
  \bibfield  {author} {\bibinfo {author} {\bibfnamefont {H.}~\bibnamefont {Tian}}, \bibinfo {author} {\bibfnamefont {Y.}~\bibnamefont {Ma}}, \bibinfo {author} {\bibfnamefont {Z.}~\bibnamefont {Li}}, \bibinfo {author} {\bibfnamefont {M.}~\bibnamefont {Cheng}}, \bibinfo {author} {\bibfnamefont {S.}~\bibnamefont {Ning}}, \bibinfo {author} {\bibfnamefont {E.}~\bibnamefont {Han}}, \bibinfo {author} {\bibfnamefont {M.}~\bibnamefont {Xu}}, \bibinfo {author} {\bibfnamefont {P.-F.}\ \bibnamefont {Zhang}}, \bibinfo {author} {\bibfnamefont {K.}~\bibnamefont {Zhao}}, \bibinfo {author} {\bibfnamefont {R.}~\bibnamefont {Li}}, \bibinfo {author} {\bibfnamefont {Y.}~\bibnamefont {Zou}}, \bibinfo {author} {\bibfnamefont {P.}~\bibnamefont {Liao}}, \bibinfo {author} {\bibfnamefont {S.}~\bibnamefont {Yu}}, \bibinfo {author} {\bibfnamefont {X.}~\bibnamefont {Li}}, \bibinfo {author} {\bibfnamefont {J.}~\bibnamefont {Wang}}, \bibinfo {author} {\bibfnamefont {S.}~\bibnamefont {Liu}}, \bibinfo {author} {\bibfnamefont {Y.}~\bibnamefont {Li}}, \bibinfo {author} {\bibfnamefont {X.}~\bibnamefont {Huang}}, \bibinfo {author} {\bibfnamefont {Z.}~\bibnamefont {Yao}}, \bibinfo {author} {\bibfnamefont {D.}~\bibnamefont {Ding}}, \bibinfo {author} {\bibfnamefont {J.}~\bibnamefont {Guo}}, \bibinfo {author} {\bibfnamefont {Y.}~\bibnamefont {Huang}}, \bibinfo {author} {\bibfnamefont {J.}~\bibnamefont {Lu}}, \bibinfo {author} {\bibfnamefont {Y.}~\bibnamefont {Han}}, \bibinfo {author} {\bibfnamefont {Z.}~\bibnamefont {Wang}}, \bibinfo {author} {\bibfnamefont {Z.~G.}\ \bibnamefont {Cheng}}, \bibinfo {author} {\bibfnamefont {J.}~\bibnamefont {Liu}}, \bibinfo {author} {\bibfnamefont {Z.}~\bibnamefont {Xu}}, \bibinfo {author} {\bibfnamefont {K.}~\bibnamefont {Liu}}, \bibinfo {author} {\bibfnamefont {P.}~\bibnamefont {Gao}}, \bibinfo {author} {\bibfnamefont {Y.}~\bibnamefont {Jiang}}, \bibinfo {author} {\bibfnamefont {L.}~\bibnamefont {Lin}}, \bibinfo {author} {\bibfnamefont {X.}~\bibnamefont {Zhao}}, \bibinfo {author} {\bibfnamefont {L.}~\bibnamefont {Wang}}, \bibinfo {author} {\bibfnamefont {X.}~\bibnamefont {Bai}}, \bibinfo {author} {\bibfnamefont {W.}~\bibnamefont {Fu}}, \bibinfo {author} {\bibfnamefont {J.-Y.}\ \bibnamefont {Wang}}, \bibinfo {author} {\bibfnamefont {M.}~\bibnamefont {Li}}, \bibinfo {author} {\bibfnamefont {T.}~\bibnamefont {Lei}}, \bibinfo {author} {\bibfnamefont {Y.}~\bibnamefont {Zhang}}, \bibinfo {author} {\bibfnamefont {Y.}~\bibnamefont {Hou}}, \bibinfo {author} {\bibfnamefont {J.}~\bibnamefont {Pei}}, \bibinfo {author} {\bibfnamefont {S.~J.}\ \bibnamefont {Pennycook}}, \bibinfo {author} {\bibfnamefont {E.}~\bibnamefont {Wang}}, \bibinfo {author} {\bibfnamefont {J.}~\bibnamefont {Chen}}, \bibinfo {author} {\bibfnamefont {W.}~\bibnamefont {Zhou}},\ and\ \bibinfo {author} {\bibfnamefont {L.}~\bibnamefont {Liu}},\ }\bibfield  {title} {\bibinfo {title} {Disorder-tuned conductivity in amorphous monolayer carbon},\ }\href {https://doi.org/10.1038/s41586-022-05617-w} {\bibfield  {journal} {\bibinfo  {journal} {Nature}\ }\textbf {\bibinfo {volume} {615}},\ \bibinfo {pages} {56} (\bibinfo {year} {2023})},\ \bibinfo {note} {number: 7950 Publisher: Nature Publishing Group}\BibitemShut {NoStop}%
\bibitem [{\citenamefont {Shang}\ \emph {et~al.}(2021)\citenamefont {Shang}, \citenamefont {Liu}, \citenamefont {Dong}, \citenamefont {Yao}, \citenamefont {Yang}, \citenamefont {Li}, \citenamefont {Zhai}, \citenamefont {Shen}, \citenamefont {Hou}, \citenamefont {Wang}, \citenamefont {Zhang}, \citenamefont {Zhang}, \citenamefont {Fu}, \citenamefont {Ji}, \citenamefont {Zhang}, \citenamefont {Lin}, \citenamefont {Fei}, \citenamefont {Sundqvist}, \citenamefont {Wang},\ and\ \citenamefont {Liu}}]{shang_ultrahard_2021}%
  \BibitemOpen
  \bibfield  {author} {\bibinfo {author} {\bibfnamefont {Y.}~\bibnamefont {Shang}}, \bibinfo {author} {\bibfnamefont {Z.}~\bibnamefont {Liu}}, \bibinfo {author} {\bibfnamefont {J.}~\bibnamefont {Dong}}, \bibinfo {author} {\bibfnamefont {M.}~\bibnamefont {Yao}}, \bibinfo {author} {\bibfnamefont {Z.}~\bibnamefont {Yang}}, \bibinfo {author} {\bibfnamefont {Q.}~\bibnamefont {Li}}, \bibinfo {author} {\bibfnamefont {C.}~\bibnamefont {Zhai}}, \bibinfo {author} {\bibfnamefont {F.}~\bibnamefont {Shen}}, \bibinfo {author} {\bibfnamefont {X.}~\bibnamefont {Hou}}, \bibinfo {author} {\bibfnamefont {L.}~\bibnamefont {Wang}}, \bibinfo {author} {\bibfnamefont {N.}~\bibnamefont {Zhang}}, \bibinfo {author} {\bibfnamefont {W.}~\bibnamefont {Zhang}}, \bibinfo {author} {\bibfnamefont {R.}~\bibnamefont {Fu}}, \bibinfo {author} {\bibfnamefont {J.}~\bibnamefont {Ji}}, \bibinfo {author} {\bibfnamefont {X.}~\bibnamefont {Zhang}}, \bibinfo {author} {\bibfnamefont {H.}~\bibnamefont {Lin}}, \bibinfo {author} {\bibfnamefont {Y.}~\bibnamefont {Fei}}, \bibinfo {author} {\bibfnamefont {B.}~\bibnamefont {Sundqvist}}, \bibinfo {author} {\bibfnamefont {W.}~\bibnamefont {Wang}},\ and\ \bibinfo {author} {\bibfnamefont {B.}~\bibnamefont {Liu}},\ }\bibfield  {title} {\bibinfo {title} {Ultrahard bulk amorphous carbon from collapsed fullerene},\ }\href {https://doi.org/10.1038/s41586-021-03882-9} {\bibfield  {journal} {\bibinfo  {journal} {Nature}\ }\textbf {\bibinfo {volume} {599}},\ \bibinfo {pages} {599} (\bibinfo {year} {2021})},\ \bibinfo {note} {publisher: Nature Publishing Group}\BibitemShut {NoStop}%
\bibitem [{\citenamefont {Li}\ \emph {et~al.}(2024)\citenamefont {Li}, \citenamefont {Bhardwaj}, \citenamefont {He}, \citenamefont {Zhang}, \citenamefont {Tran}, \citenamefont {Li}, \citenamefont {McClung}, \citenamefont {Nuguri}, \citenamefont {Watkins},\ and\ \citenamefont {Lee}}]{li_nanoporous_2024}%
  \BibitemOpen
  \bibfield  {author} {\bibinfo {author} {\bibfnamefont {Z.}~\bibnamefont {Li}}, \bibinfo {author} {\bibfnamefont {A.}~\bibnamefont {Bhardwaj}}, \bibinfo {author} {\bibfnamefont {J.}~\bibnamefont {He}}, \bibinfo {author} {\bibfnamefont {W.}~\bibnamefont {Zhang}}, \bibinfo {author} {\bibfnamefont {T.~T.}\ \bibnamefont {Tran}}, \bibinfo {author} {\bibfnamefont {Y.}~\bibnamefont {Li}}, \bibinfo {author} {\bibfnamefont {A.}~\bibnamefont {McClung}}, \bibinfo {author} {\bibfnamefont {S.}~\bibnamefont {Nuguri}}, \bibinfo {author} {\bibfnamefont {J.~J.}\ \bibnamefont {Watkins}},\ and\ \bibinfo {author} {\bibfnamefont {S.-W.}\ \bibnamefont {Lee}},\ }\bibfield  {title} {\bibinfo {title} {Nanoporous amorphous carbon nanopillars with lightweight, ultrahigh strength, large fracture strain, and high damping capability},\ }\href {https://doi.org/10.1038/s41467-024-52359-6} {\bibfield  {journal} {\bibinfo  {journal} {Nature Communications}\ }\textbf {\bibinfo {volume} {15}},\ \bibinfo {pages} {8151} (\bibinfo {year} {2024})},\ \bibinfo {note} {publisher: Nature Publishing Group}\BibitemShut {NoStop}%
\bibitem [{\citenamefont {Statz}\ \emph {et~al.}(2020)\citenamefont {Statz}, \citenamefont {Schneider}, \citenamefont {Berger}, \citenamefont {Lai}, \citenamefont {Wood}, \citenamefont {Abdi-Jalebi}, \citenamefont {Leingang}, \citenamefont {Himmel}, \citenamefont {Zaumseil},\ and\ \citenamefont {Sirringhaus}}]{statz_charge_2020}%
  \BibitemOpen
  \bibfield  {author} {\bibinfo {author} {\bibfnamefont {M.}~\bibnamefont {Statz}}, \bibinfo {author} {\bibfnamefont {S.}~\bibnamefont {Schneider}}, \bibinfo {author} {\bibfnamefont {F.~J.}\ \bibnamefont {Berger}}, \bibinfo {author} {\bibfnamefont {L.}~\bibnamefont {Lai}}, \bibinfo {author} {\bibfnamefont {W.~A.}\ \bibnamefont {Wood}}, \bibinfo {author} {\bibfnamefont {M.}~\bibnamefont {Abdi-Jalebi}}, \bibinfo {author} {\bibfnamefont {S.}~\bibnamefont {Leingang}}, \bibinfo {author} {\bibfnamefont {H.-J.}\ \bibnamefont {Himmel}}, \bibinfo {author} {\bibfnamefont {J.}~\bibnamefont {Zaumseil}},\ and\ \bibinfo {author} {\bibfnamefont {H.}~\bibnamefont {Sirringhaus}},\ }\bibfield  {title} {\bibinfo {title} {Charge and {Thermoelectric} {Transport} in {Polymer}-{Sorted} {Semiconducting} {Single}-{Walled} {Carbon} {Nanotube} {Networks}},\ }\href {https://doi.org/10.1021/acsnano.0c06181} {\bibfield  {journal} {\bibinfo  {journal} {ACS Nano}\ }\textbf {\bibinfo {volume} {14}},\ \bibinfo {pages} {15552} (\bibinfo {year} {2020})},\ \bibinfo {note} {publisher: American Chemical Society}\BibitemShut {NoStop}%
\bibitem [{\citenamefont {Liu}\ \emph {et~al.}(2024)\citenamefont {Liu}, \citenamefont {Lyu}, \citenamefont {Merlet}, \citenamefont {Leesmith}, \citenamefont {Hua}, \citenamefont {Xu}, \citenamefont {Grey},\ and\ \citenamefont {Forse}}]{liu_structural_2024}%
  \BibitemOpen
  \bibfield  {author} {\bibinfo {author} {\bibfnamefont {X.}~\bibnamefont {Liu}}, \bibinfo {author} {\bibfnamefont {D.}~\bibnamefont {Lyu}}, \bibinfo {author} {\bibfnamefont {C.}~\bibnamefont {Merlet}}, \bibinfo {author} {\bibfnamefont {M.~J.~A.}\ \bibnamefont {Leesmith}}, \bibinfo {author} {\bibfnamefont {X.}~\bibnamefont {Hua}}, \bibinfo {author} {\bibfnamefont {Z.}~\bibnamefont {Xu}}, \bibinfo {author} {\bibfnamefont {C.~P.}\ \bibnamefont {Grey}},\ and\ \bibinfo {author} {\bibfnamefont {A.~C.}\ \bibnamefont {Forse}},\ }\bibfield  {title} {\bibinfo {title} {Structural disorder determines capacitance in nanoporous carbons},\ }\href {https://doi.org/10.1126/science.adn6242} {\bibfield  {journal} {\bibinfo  {journal} {Science}\ }\textbf {\bibinfo {volume} {384}},\ \bibinfo {pages} {321} (\bibinfo {year} {2024})},\ \bibinfo {note} {publisher: American Association for the Advancement of Science}\BibitemShut {NoStop}%
\bibitem [{\citenamefont {Simoncelli}\ \emph {et~al.}(2018)\citenamefont {Simoncelli}, \citenamefont {Ganfoud}, \citenamefont {Sene}, \citenamefont {Haefele}, \citenamefont {Daffos}, \citenamefont {Taberna}, \citenamefont {Salanne}, \citenamefont {Simon},\ and\ \citenamefont {Rotenberg}}]{simoncelli_blue_2018}%
  \BibitemOpen
  \bibfield  {author} {\bibinfo {author} {\bibfnamefont {M.}~\bibnamefont {Simoncelli}}, \bibinfo {author} {\bibfnamefont {N.}~\bibnamefont {Ganfoud}}, \bibinfo {author} {\bibfnamefont {A.}~\bibnamefont {Sene}}, \bibinfo {author} {\bibfnamefont {M.}~\bibnamefont {Haefele}}, \bibinfo {author} {\bibfnamefont {B.}~\bibnamefont {Daffos}}, \bibinfo {author} {\bibfnamefont {P.-L.}\ \bibnamefont {Taberna}}, \bibinfo {author} {\bibfnamefont {M.}~\bibnamefont {Salanne}}, \bibinfo {author} {\bibfnamefont {P.}~\bibnamefont {Simon}},\ and\ \bibinfo {author} {\bibfnamefont {B.}~\bibnamefont {Rotenberg}},\ }\bibfield  {title} {\bibinfo {title} {Blue {Energy} and {Desalination} with {Nanoporous} {Carbon} {Electrodes}: {Capacitance} from {Molecular} {Simulations} to {Continuous} {Models}},\ }\href {https://doi.org/10.1103/PhysRevX.8.021024} {\bibfield  {journal} {\bibinfo  {journal} {Physical Review X}\ }\textbf {\bibinfo {volume} {8}},\ \bibinfo {pages} {021024} (\bibinfo {year} {2018})},\ \bibinfo {note} {publisher: American Physical Society}\BibitemShut {NoStop}%
\bibitem [{\citenamefont {Jeanmairet}\ \emph {et~al.}(2022)\citenamefont {Jeanmairet}, \citenamefont {Rotenberg},\ and\ \citenamefont {Salanne}}]{jeanmairet_microscopic_2022}%
  \BibitemOpen
  \bibfield  {author} {\bibinfo {author} {\bibfnamefont {G.}~\bibnamefont {Jeanmairet}}, \bibinfo {author} {\bibfnamefont {B.}~\bibnamefont {Rotenberg}},\ and\ \bibinfo {author} {\bibfnamefont {M.}~\bibnamefont {Salanne}},\ }\bibfield  {title} {\bibinfo {title} {Microscopic {Simulations} of {Electrochemical} {Double}-{Layer} {Capacitors}},\ }\href {https://doi.org/10.1021/acs.chemrev.1c00925} {\bibfield  {journal} {\bibinfo  {journal} {Chemical Reviews}\ }\textbf {\bibinfo {volume} {122}},\ \bibinfo {pages} {10860} (\bibinfo {year} {2022})},\ \bibinfo {note} {publisher: American Chemical Society}\BibitemShut {NoStop}%
\bibitem [{\citenamefont {Liu}\ \emph {et~al.}(2017)\citenamefont {Liu}, \citenamefont {Gludovatz}, \citenamefont {Barnard}, \citenamefont {Kuball},\ and\ \citenamefont {Ritchie}}]{liu_damage_2017}%
  \BibitemOpen
  \bibfield  {author} {\bibinfo {author} {\bibfnamefont {D.}~\bibnamefont {Liu}}, \bibinfo {author} {\bibfnamefont {B.}~\bibnamefont {Gludovatz}}, \bibinfo {author} {\bibfnamefont {H.~S.}\ \bibnamefont {Barnard}}, \bibinfo {author} {\bibfnamefont {M.}~\bibnamefont {Kuball}},\ and\ \bibinfo {author} {\bibfnamefont {R.~O.}\ \bibnamefont {Ritchie}},\ }\bibfield  {title} {\bibinfo {title} {Damage tolerance of nuclear graphite at elevated temperatures},\ }\href {https://doi.org/10.1038/ncomms15942} {\bibfield  {journal} {\bibinfo  {journal} {Nature Communications}\ }\textbf {\bibinfo {volume} {8}},\ \bibinfo {pages} {15942} (\bibinfo {year} {2017})},\ \bibinfo {note} {publisher: Nature Publishing Group}\BibitemShut {NoStop}%
\bibitem [{\citenamefont {Farbos}\ \emph {et~al.}(2017)\citenamefont {Farbos}, \citenamefont {Freeman}, \citenamefont {Hardcastle}, \citenamefont {Da~Costa}, \citenamefont {Brydson}, \citenamefont {Scott}, \citenamefont {Weisbecker}, \citenamefont {Germain}, \citenamefont {Vignoles},\ and\ \citenamefont {Leyssale}}]{farbos_time-dependent_2017}%
  \BibitemOpen
  \bibfield  {author} {\bibinfo {author} {\bibfnamefont {B.}~\bibnamefont {Farbos}}, \bibinfo {author} {\bibfnamefont {H.}~\bibnamefont {Freeman}}, \bibinfo {author} {\bibfnamefont {T.}~\bibnamefont {Hardcastle}}, \bibinfo {author} {\bibfnamefont {J.-P.}\ \bibnamefont {Da~Costa}}, \bibinfo {author} {\bibfnamefont {R.}~\bibnamefont {Brydson}}, \bibinfo {author} {\bibfnamefont {A.~J.}\ \bibnamefont {Scott}}, \bibinfo {author} {\bibfnamefont {P.}~\bibnamefont {Weisbecker}}, \bibinfo {author} {\bibfnamefont {C.}~\bibnamefont {Germain}}, \bibinfo {author} {\bibfnamefont {G.~L.}\ \bibnamefont {Vignoles}},\ and\ \bibinfo {author} {\bibfnamefont {J.-M.}\ \bibnamefont {Leyssale}},\ }\bibfield  {title} {\bibinfo {title} {A time-dependent atomistic reconstruction of severe irradiation damage and associated property changes in nuclear graphite},\ }\href {https://doi.org/10.1016/j.carbon.2017.05.009} {\bibfield  {journal} {\bibinfo  {journal} {Carbon}\ }\textbf {\bibinfo {volume} {120}},\ \bibinfo {pages} {111} (\bibinfo {year} {2017})}\BibitemShut {NoStop}%
\bibitem [{\citenamefont {Cheng}\ \emph {et~al.}(2022)\citenamefont {Cheng}, \citenamefont {Duchnowski}, \citenamefont {Sprouster}, \citenamefont {Snead}, \citenamefont {Brown},\ and\ \citenamefont {Trelewicz}}]{chengCeramicCompositeModerators2022}%
  \BibitemOpen
  \bibfield  {author} {\bibinfo {author} {\bibfnamefont {B.}~\bibnamefont {Cheng}}, \bibinfo {author} {\bibfnamefont {E.~M.}\ \bibnamefont {Duchnowski}}, \bibinfo {author} {\bibfnamefont {D.~J.}\ \bibnamefont {Sprouster}}, \bibinfo {author} {\bibfnamefont {L.~L.}\ \bibnamefont {Snead}}, \bibinfo {author} {\bibfnamefont {N.~R.}\ \bibnamefont {Brown}},\ and\ \bibinfo {author} {\bibfnamefont {J.~R.}\ \bibnamefont {Trelewicz}},\ }\bibfield  {title} {\bibinfo {title} {Ceramic composite moderators as replacements for graphite in high temperature microreactors},\ }\href {https://doi.org/10.1016/j.jnucmat.2022.153591} {\bibfield  {journal} {\bibinfo  {journal} {Journal of Nuclear Materials}\ }\textbf {\bibinfo {volume} {563}},\ \bibinfo {pages} {153591} (\bibinfo {year} {2022})}\BibitemShut {NoStop}%
\bibitem [{\citenamefont {Umemoto}\ \emph {et~al.}(2010)\citenamefont {Umemoto}, \citenamefont {Wentzcovitch}, \citenamefont {Saito},\ and\ \citenamefont {Miyake}}]{umemoto_body-centered_2010}%
  \BibitemOpen
  \bibfield  {author} {\bibinfo {author} {\bibfnamefont {K.}~\bibnamefont {Umemoto}}, \bibinfo {author} {\bibfnamefont {R.~M.}\ \bibnamefont {Wentzcovitch}}, \bibinfo {author} {\bibfnamefont {S.}~\bibnamefont {Saito}},\ and\ \bibinfo {author} {\bibfnamefont {T.}~\bibnamefont {Miyake}},\ }\bibfield  {title} {\bibinfo {title} {{Body-{Centered} {Tetragonal} C4: {A} {Viable} sp3 {Carbon} {Allotrope}}},\ }\href {https://doi.org/10.1103/PhysRevLett.104.125504} {\bibfield  {journal} {\bibinfo  {journal} {Physical Review Letters}\ }\textbf {\bibinfo {volume} {104}},\ \bibinfo {pages} {125504} (\bibinfo {year} {2010})}\BibitemShut {NoStop}%
\bibitem [{\citenamefont {Prasher}\ \emph {et~al.}(2009)\citenamefont {Prasher}, \citenamefont {Hu}, \citenamefont {Chalopin}, \citenamefont {Mingo}, \citenamefont {Lofgreen}, \citenamefont {Volz}, \citenamefont {Cleri},\ and\ \citenamefont {Keblinski}}]{prasher_turning_2009}%
  \BibitemOpen
  \bibfield  {author} {\bibinfo {author} {\bibfnamefont {R.~S.}\ \bibnamefont {Prasher}}, \bibinfo {author} {\bibfnamefont {X.~J.}\ \bibnamefont {Hu}}, \bibinfo {author} {\bibfnamefont {Y.}~\bibnamefont {Chalopin}}, \bibinfo {author} {\bibfnamefont {N.}~\bibnamefont {Mingo}}, \bibinfo {author} {\bibfnamefont {K.}~\bibnamefont {Lofgreen}}, \bibinfo {author} {\bibfnamefont {S.}~\bibnamefont {Volz}}, \bibinfo {author} {\bibfnamefont {F.}~\bibnamefont {Cleri}},\ and\ \bibinfo {author} {\bibfnamefont {P.}~\bibnamefont {Keblinski}},\ }\bibfield  {title} {\bibinfo {title} {Turning {Carbon} {Nanotubes} from {Exceptional} {Heat} {Conductors} into {Insulators}},\ }\href {https://doi.org/10.1103/PhysRevLett.102.105901} {\bibfield  {journal} {\bibinfo  {journal} {Physical Review Letters}\ }\textbf {\bibinfo {volume} {102}},\ \bibinfo {pages} {105901} (\bibinfo {year} {2009})},\ \bibinfo {note} {publisher: American Physical Society}\BibitemShut {NoStop}%
\bibitem [{\citenamefont {Mehew}\ \emph {et~al.}(2023)\citenamefont {Mehew}, \citenamefont {Timmermans}, \citenamefont {Saleta~Reig}, \citenamefont {Sergeant}, \citenamefont {Sledzinska}, \citenamefont {Chávez-Ángel}, \citenamefont {Gallagher}, \citenamefont {Sotomayor~Torres}, \citenamefont {Huyghebaert},\ and\ \citenamefont {Tielrooij}}]{mehew_enhanced_2023}%
  \BibitemOpen
  \bibfield  {author} {\bibinfo {author} {\bibfnamefont {J.~D.}\ \bibnamefont {Mehew}}, \bibinfo {author} {\bibfnamefont {M.~Y.}\ \bibnamefont {Timmermans}}, \bibinfo {author} {\bibfnamefont {D.}~\bibnamefont {Saleta~Reig}}, \bibinfo {author} {\bibfnamefont {S.}~\bibnamefont {Sergeant}}, \bibinfo {author} {\bibfnamefont {M.}~\bibnamefont {Sledzinska}}, \bibinfo {author} {\bibfnamefont {E.}~\bibnamefont {Chávez-Ángel}}, \bibinfo {author} {\bibfnamefont {E.}~\bibnamefont {Gallagher}}, \bibinfo {author} {\bibfnamefont {C.~M.}\ \bibnamefont {Sotomayor~Torres}}, \bibinfo {author} {\bibfnamefont {C.}~\bibnamefont {Huyghebaert}},\ and\ \bibinfo {author} {\bibfnamefont {K.-J.}\ \bibnamefont {Tielrooij}},\ }\bibfield  {title} {\bibinfo {title} {Enhanced {Thermal} {Conductivity} of {Free}-{Standing} {Double}-{Walled} {Carbon} {Nanotube} {Networks}},\ }\href {https://doi.org/10.1021/acsami.3c09210} {\bibfield  {journal} {\bibinfo  {journal} {ACS Applied Materials \& Interfaces}\ }\textbf {\bibinfo {volume} {15}},\ \bibinfo {pages} {51876} (\bibinfo {year} {2023})},\ \bibinfo {note} {publisher: American Chemical Society}\BibitemShut {NoStop}%
\bibitem [{\citenamefont {Ye}\ \emph {et~al.}(2024)\citenamefont {Ye}, \citenamefont {Park}, \citenamefont {Zhou}, \citenamefont {Montano}, \citenamefont {Pandey}, \citenamefont {Zhang}, \citenamefont {Lin},\ and\ \citenamefont {Wang}}]{ye_anomalous_2024}%
  \BibitemOpen
  \bibfield  {author} {\bibinfo {author} {\bibfnamefont {Z.}~\bibnamefont {Ye}}, \bibinfo {author} {\bibfnamefont {J.}~\bibnamefont {Park}}, \bibinfo {author} {\bibfnamefont {Y.}~\bibnamefont {Zhou}}, \bibinfo {author} {\bibfnamefont {R.~D.}\ \bibnamefont {Montano}}, \bibinfo {author} {\bibfnamefont {T.}~\bibnamefont {Pandey}}, \bibinfo {author} {\bibfnamefont {Y.}~\bibnamefont {Zhang}}, \bibinfo {author} {\bibfnamefont {J.-F.}\ \bibnamefont {Lin}},\ and\ \bibinfo {author} {\bibfnamefont {Y.}~\bibnamefont {Wang}},\ }\bibfield  {title} {\bibinfo {title} {Anomalous {Thermal} {Transport} in {Compressed} {Carbon} {Phases}},\ }\href {https://doi.org/10.1103/PhysRevLett.133.206301} {\bibfield  {journal} {\bibinfo  {journal} {Physical Review Letters}\ }\textbf {\bibinfo {volume} {133}},\ \bibinfo {pages} {206301} (\bibinfo {year} {2024})},\ \bibinfo {note} {publisher: American Physical Society}\BibitemShut {NoStop}%
\bibitem [{\citenamefont {Kern}\ \emph {et~al.}(2016)\citenamefont {Kern}, \citenamefont {Zierath}, \citenamefont {Haertlé}, \citenamefont {Fey},\ and\ \citenamefont {Etzold}}]{kern_thermal_2016}%
  \BibitemOpen
  \bibfield  {author} {\bibinfo {author} {\bibfnamefont {A.~M.}\ \bibnamefont {Kern}}, \bibinfo {author} {\bibfnamefont {B.}~\bibnamefont {Zierath}}, \bibinfo {author} {\bibfnamefont {J.}~\bibnamefont {Haertlé}}, \bibinfo {author} {\bibfnamefont {T.}~\bibnamefont {Fey}},\ and\ \bibinfo {author} {\bibfnamefont {B.~J.~M.}\ \bibnamefont {Etzold}},\ }\bibfield  {title} {\bibinfo {title} {Thermal and {Electrical} {Conductivity} of {Amorphous} and {Graphitized} {Carbide}-{Derived} {Carbon} {Monoliths}},\ }\href {https://doi.org/10.1002/ceat.201600011} {\bibfield  {journal} {\bibinfo  {journal} {Chemical Engineering \& Technology}\ }\textbf {\bibinfo {volume} {39}},\ \bibinfo {pages} {1121} (\bibinfo {year} {2016})}\BibitemShut {NoStop}%
\bibitem [{\citenamefont {Pop}\ \emph {et~al.}(2012)\citenamefont {Pop}, \citenamefont {Varshney},\ and\ \citenamefont {Roy}}]{pop_thermal_2012}%
  \BibitemOpen
  \bibfield  {author} {\bibinfo {author} {\bibfnamefont {E.}~\bibnamefont {Pop}}, \bibinfo {author} {\bibfnamefont {V.}~\bibnamefont {Varshney}},\ and\ \bibinfo {author} {\bibfnamefont {A.~K.}\ \bibnamefont {Roy}},\ }\bibfield  {title} {\bibinfo {title} {Thermal properties of graphene: {Fundamentals} and applications},\ }\href {https://doi.org/10.1557/mrs.2012.203} {\bibfield  {journal} {\bibinfo  {journal} {MRS Bulletin}\ }\textbf {\bibinfo {volume} {37}},\ \bibinfo {pages} {1273} (\bibinfo {year} {2012})}\BibitemShut {NoStop}%
\bibitem [{\citenamefont {Lee}\ \emph {et~al.}(2015)\citenamefont {Lee}, \citenamefont {Broido}, \citenamefont {Esfarjani},\ and\ \citenamefont {Chen}}]{app_Lee2015}%
  \BibitemOpen
  \bibfield  {author} {\bibinfo {author} {\bibfnamefont {S.}~\bibnamefont {Lee}}, \bibinfo {author} {\bibfnamefont {D.}~\bibnamefont {Broido}}, \bibinfo {author} {\bibfnamefont {K.}~\bibnamefont {Esfarjani}},\ and\ \bibinfo {author} {\bibfnamefont {G.}~\bibnamefont {Chen}},\ }\bibfield  {title} {\bibinfo {title} {Hydrodynamic phonon transport in suspended graphene},\ }\href {https://doi.org/10.1038/ncomms7290} {\bibfield  {journal} {\bibinfo  {journal} {Nat. Commun.}\ }\textbf {\bibinfo {volume} {6}},\ \bibinfo {pages} {6290} (\bibinfo {year} {2015})}\BibitemShut {NoStop}%
\bibitem [{\citenamefont {Pereira}\ and\ \citenamefont {Donadio}(2013)}]{pereira_divergence_2013}%
  \BibitemOpen
  \bibfield  {author} {\bibinfo {author} {\bibfnamefont {L.~F.~C.}\ \bibnamefont {Pereira}}\ and\ \bibinfo {author} {\bibfnamefont {D.}~\bibnamefont {Donadio}},\ }\bibfield  {title} {\bibinfo {title} {Divergence of the thermal conductivity in uniaxially strained graphene},\ }\href {https://doi.org/10.1103/PhysRevB.87.125424} {\bibfield  {journal} {\bibinfo  {journal} {Physical Review B}\ }\textbf {\bibinfo {volume} {87}},\ \bibinfo {pages} {125424} (\bibinfo {year} {2013})},\ \bibinfo {note} {publisher: American Physical Society}\BibitemShut {NoStop}%
\bibitem [{\citenamefont {Barbarino}\ \emph {et~al.}(2015)\citenamefont {Barbarino}, \citenamefont {Melis},\ and\ \citenamefont {Colombo}}]{barbarino_intrinsic_2015}%
  \BibitemOpen
  \bibfield  {author} {\bibinfo {author} {\bibfnamefont {G.}~\bibnamefont {Barbarino}}, \bibinfo {author} {\bibfnamefont {C.}~\bibnamefont {Melis}},\ and\ \bibinfo {author} {\bibfnamefont {L.}~\bibnamefont {Colombo}},\ }\bibfield  {title} {\bibinfo {title} {Intrinsic thermal conductivity in monolayer graphene is ultimately upper limited: {A} direct estimation by atomistic simulations},\ }\href {https://doi.org/10.1103/PhysRevB.91.035416} {\bibfield  {journal} {\bibinfo  {journal} {Physical Review B}\ }\textbf {\bibinfo {volume} {91}},\ \bibinfo {pages} {035416} (\bibinfo {year} {2015})},\ \bibinfo {note} {publisher: American Physical Society}\BibitemShut {NoStop}%
\bibitem [{\citenamefont {Majee}\ and\ \citenamefont {Aksamija}(2018)}]{majee_dynamical_2018}%
  \BibitemOpen
  \bibfield  {author} {\bibinfo {author} {\bibfnamefont {A.~K.}\ \bibnamefont {Majee}}\ and\ \bibinfo {author} {\bibfnamefont {Z.}~\bibnamefont {Aksamija}},\ }\bibfield  {title} {\bibinfo {title} {Dynamical thermal conductivity of suspended graphene ribbons in the hydrodynamic regime},\ }\href {https://link.aps.org/doi/10.1103/PhysRevB.98.024303} {\bibfield  {journal} {\bibinfo  {journal} {Phys. Rev. B}\ }\textbf {\bibinfo {volume} {98}},\ \bibinfo {pages} {024303} (\bibinfo {year} {2018})}\BibitemShut {NoStop}%
\bibitem [{\citenamefont {Braun}\ \emph {et~al.}(2022)\citenamefont {Braun}, \citenamefont {Furrer}, \citenamefont {Butti}, \citenamefont {Thodkar}, \citenamefont {Shorubalko}, \citenamefont {Zardo}, \citenamefont {Calame},\ and\ \citenamefont {Perrin}}]{braun_spatially_2022}%
  \BibitemOpen
  \bibfield  {author} {\bibinfo {author} {\bibfnamefont {O.}~\bibnamefont {Braun}}, \bibinfo {author} {\bibfnamefont {R.}~\bibnamefont {Furrer}}, \bibinfo {author} {\bibfnamefont {P.}~\bibnamefont {Butti}}, \bibinfo {author} {\bibfnamefont {K.}~\bibnamefont {Thodkar}}, \bibinfo {author} {\bibfnamefont {I.}~\bibnamefont {Shorubalko}}, \bibinfo {author} {\bibfnamefont {I.}~\bibnamefont {Zardo}}, \bibinfo {author} {\bibfnamefont {M.}~\bibnamefont {Calame}},\ and\ \bibinfo {author} {\bibfnamefont {M.~L.}\ \bibnamefont {Perrin}},\ }\bibfield  {title} {\bibinfo {title} {Spatially mapping thermal transport in graphene by an opto-thermal method},\ }\href {https://www.nature.com/articles/s41699-021-00277-2} {\bibfield  {journal} {\bibinfo  {journal} {NPJ 2D Mater. Appl.}\ }\textbf {\bibinfo {volume} {6}},\ \bibinfo {pages} {1} (\bibinfo {year} {2022})}\BibitemShut {NoStop}%
\bibitem [{\citenamefont {Han}\ and\ \citenamefont {Ruan}(2023)}]{han_thermal_2023}%
  \BibitemOpen
  \bibfield  {author} {\bibinfo {author} {\bibfnamefont {Z.}~\bibnamefont {Han}}\ and\ \bibinfo {author} {\bibfnamefont {X.}~\bibnamefont {Ruan}},\ }\bibfield  {title} {\bibinfo {title} {Thermal conductivity of monolayer graphene: {Convergent} and lower than diamond},\ }\href {https://doi.org/10.1103/PhysRevB.108.L121412} {\bibfield  {journal} {\bibinfo  {journal} {Physical Review B}\ }\textbf {\bibinfo {volume} {108}},\ \bibinfo {pages} {L121412} (\bibinfo {year} {2023})},\ \bibinfo {note} {publisher: American Physical Society}\BibitemShut {NoStop}%
\bibitem [{\citenamefont {Fugallo}\ \emph {et~al.}(2014)\citenamefont {Fugallo}, \citenamefont {Cepellotti}, \citenamefont {Paulatto}, \citenamefont {Lazzeri}, \citenamefont {Marzari},\ and\ \citenamefont {Mauri}}]{fugallo_thermal_2014}%
  \BibitemOpen
  \bibfield  {author} {\bibinfo {author} {\bibfnamefont {G.}~\bibnamefont {Fugallo}}, \bibinfo {author} {\bibfnamefont {A.}~\bibnamefont {Cepellotti}}, \bibinfo {author} {\bibfnamefont {L.}~\bibnamefont {Paulatto}}, \bibinfo {author} {\bibfnamefont {M.}~\bibnamefont {Lazzeri}}, \bibinfo {author} {\bibfnamefont {N.}~\bibnamefont {Marzari}},\ and\ \bibinfo {author} {\bibfnamefont {F.}~\bibnamefont {Mauri}},\ }\bibfield  {title} {\bibinfo {title} {Thermal {Conductivity} of {Graphene} and {Graphite}: {Collective} {Excitations} and {Mean} {Free} {Paths}},\ }\href {https://pubs.acs.org/doi/10.1021/nl502059f} {\bibfield  {journal} {\bibinfo  {journal} {Nano Lett.}\ }\textbf {\bibinfo {volume} {14}},\ \bibinfo {pages} {6109} (\bibinfo {year} {2014})}\BibitemShut {NoStop}%
\bibitem [{\citenamefont {Zhang}\ \emph {et~al.}(2016)\citenamefont {Zhang}, \citenamefont {Chen}, \citenamefont {Jho},\ and\ \citenamefont {Minnich}}]{zhang_temperature-dependent_2016}%
  \BibitemOpen
  \bibfield  {author} {\bibinfo {author} {\bibfnamefont {H.}~\bibnamefont {Zhang}}, \bibinfo {author} {\bibfnamefont {X.}~\bibnamefont {Chen}}, \bibinfo {author} {\bibfnamefont {Y.-D.}\ \bibnamefont {Jho}},\ and\ \bibinfo {author} {\bibfnamefont {A.~J.}\ \bibnamefont {Minnich}},\ }\bibfield  {title} {\bibinfo {title} {Temperature-{Dependent} {Mean} {Free} {Path} {Spectra} of {Thermal} {Phonons} {Along} the c-{Axis} of {Graphite}},\ }\href {https://doi.org/10.1021/acs.nanolett.5b04499} {\bibfield  {journal} {\bibinfo  {journal} {Nano Letters}\ }\textbf {\bibinfo {volume} {16}},\ \bibinfo {pages} {1643} (\bibinfo {year} {2016})},\ \bibinfo {note} {publisher: American Chemical Society}\BibitemShut {NoStop}%
\bibitem [{\citenamefont {Huberman}\ \emph {et~al.}(2019)\citenamefont {Huberman}, \citenamefont {Duncan}, \citenamefont {Chen}, \citenamefont {Song}, \citenamefont {Chiloyan}, \citenamefont {Ding}, \citenamefont {Maznev}, \citenamefont {Chen},\ and\ \citenamefont {Nelson}}]{huberman_observation_2019}%
  \BibitemOpen
  \bibfield  {author} {\bibinfo {author} {\bibfnamefont {S.}~\bibnamefont {Huberman}}, \bibinfo {author} {\bibfnamefont {R.~A.}\ \bibnamefont {Duncan}}, \bibinfo {author} {\bibfnamefont {K.}~\bibnamefont {Chen}}, \bibinfo {author} {\bibfnamefont {B.}~\bibnamefont {Song}}, \bibinfo {author} {\bibfnamefont {V.}~\bibnamefont {Chiloyan}}, \bibinfo {author} {\bibfnamefont {Z.}~\bibnamefont {Ding}}, \bibinfo {author} {\bibfnamefont {A.~A.}\ \bibnamefont {Maznev}}, \bibinfo {author} {\bibfnamefont {G.}~\bibnamefont {Chen}},\ and\ \bibinfo {author} {\bibfnamefont {K.~A.}\ \bibnamefont {Nelson}},\ }\bibfield  {title} {\bibinfo {title} {Observation of second sound in graphite at temperatures above 100 {K}},\ }\href {https://doi.org/10.1126/science.aav3548} {\bibfield  {journal} {\bibinfo  {journal} {Science}\ }\textbf {\bibinfo {volume} {364}},\ \bibinfo {pages} {375} (\bibinfo {year} {2019})}\BibitemShut {NoStop}%
\bibitem [{\citenamefont {Machida}\ \emph {et~al.}(2020)\citenamefont {Machida}, \citenamefont {Matsumoto}, \citenamefont {Isono},\ and\ \citenamefont {Behnia}}]{machida_phonon_2020}%
  \BibitemOpen
  \bibfield  {author} {\bibinfo {author} {\bibfnamefont {Y.}~\bibnamefont {Machida}}, \bibinfo {author} {\bibfnamefont {N.}~\bibnamefont {Matsumoto}}, \bibinfo {author} {\bibfnamefont {T.}~\bibnamefont {Isono}},\ and\ \bibinfo {author} {\bibfnamefont {K.}~\bibnamefont {Behnia}},\ }\bibfield  {title} {\bibinfo {title} {Phonon hydrodynamics and ultrahigh–room-temperature thermal conductivity in thin graphite},\ }\href {https://doi.org/10.1126/science.aaz8043} {\bibfield  {journal} {\bibinfo  {journal} {Science}\ }\textbf {\bibinfo {volume} {367}},\ \bibinfo {pages} {309} (\bibinfo {year} {2020})}\BibitemShut {NoStop}%
\bibitem [{\citenamefont {Jeong}\ \emph {et~al.}(2021)\citenamefont {Jeong}, \citenamefont {Li}, \citenamefont {Lee}, \citenamefont {Shi},\ and\ \citenamefont {Wang}}]{Jeong2021}%
  \BibitemOpen
  \bibfield  {author} {\bibinfo {author} {\bibfnamefont {J.}~\bibnamefont {Jeong}}, \bibinfo {author} {\bibfnamefont {X.}~\bibnamefont {Li}}, \bibinfo {author} {\bibfnamefont {S.}~\bibnamefont {Lee}}, \bibinfo {author} {\bibfnamefont {L.}~\bibnamefont {Shi}},\ and\ \bibinfo {author} {\bibfnamefont {Y.}~\bibnamefont {Wang}},\ }\bibfield  {title} {\bibinfo {title} {Transient hydrodynamic lattice cooling by picosecond laser irradiation of graphite},\ }\href {https://link.aps.org/doi/10.1103/PhysRevLett.127.085901} {\bibfield  {journal} {\bibinfo  {journal} {Phys. Rev. Lett.}\ }\textbf {\bibinfo {volume} {127}},\ \bibinfo {pages} {085901} (\bibinfo {year} {2021})}\BibitemShut {NoStop}%
\bibitem [{\citenamefont {Ding}\ \emph {et~al.}(2022)\citenamefont {Ding}, \citenamefont {Chen}, \citenamefont {Song}, \citenamefont {Shin}, \citenamefont {Maznev}, \citenamefont {Nelson},\ and\ \citenamefont {Chen}}]{ding_observation_2022}%
  \BibitemOpen
  \bibfield  {author} {\bibinfo {author} {\bibfnamefont {Z.}~\bibnamefont {Ding}}, \bibinfo {author} {\bibfnamefont {K.}~\bibnamefont {Chen}}, \bibinfo {author} {\bibfnamefont {B.}~\bibnamefont {Song}}, \bibinfo {author} {\bibfnamefont {J.}~\bibnamefont {Shin}}, \bibinfo {author} {\bibfnamefont {A.~A.}\ \bibnamefont {Maznev}}, \bibinfo {author} {\bibfnamefont {K.~A.}\ \bibnamefont {Nelson}},\ and\ \bibinfo {author} {\bibfnamefont {G.}~\bibnamefont {Chen}},\ }\bibfield  {title} {\bibinfo {title} {Observation of second sound in graphite over 200 {K}},\ }\href {https://doi.org/10.1038/s41467-021-27907-z} {\bibfield  {journal} {\bibinfo  {journal} {Nature Communications}\ }\textbf {\bibinfo {volume} {13}},\ \bibinfo {pages} {285} (\bibinfo {year} {2022})}\BibitemShut {NoStop}%
\bibitem [{\citenamefont {Huang}\ \emph {et~al.}(2023)\citenamefont {Huang}, \citenamefont {Guo}, \citenamefont {Wu}, \citenamefont {Masubuchi}, \citenamefont {Watanabe}, \citenamefont {Taniguchi}, \citenamefont {Zhang}, \citenamefont {Volz}, \citenamefont {Machida},\ and\ \citenamefont {Nomura}}]{huang_observation_2022}%
  \BibitemOpen
  \bibfield  {author} {\bibinfo {author} {\bibfnamefont {X.}~\bibnamefont {Huang}}, \bibinfo {author} {\bibfnamefont {Y.}~\bibnamefont {Guo}}, \bibinfo {author} {\bibfnamefont {Y.}~\bibnamefont {Wu}}, \bibinfo {author} {\bibfnamefont {S.}~\bibnamefont {Masubuchi}}, \bibinfo {author} {\bibfnamefont {K.}~\bibnamefont {Watanabe}}, \bibinfo {author} {\bibfnamefont {T.}~\bibnamefont {Taniguchi}}, \bibinfo {author} {\bibfnamefont {Z.}~\bibnamefont {Zhang}}, \bibinfo {author} {\bibfnamefont {S.}~\bibnamefont {Volz}}, \bibinfo {author} {\bibfnamefont {T.}~\bibnamefont {Machida}},\ and\ \bibinfo {author} {\bibfnamefont {M.}~\bibnamefont {Nomura}},\ }\bibfield  {title} {\bibinfo {title} {Observation of phonon {Poiseuille} flow in isotopically purified graphite ribbons},\ }\href {https://www.nature.com/articles/s41467-023-37380-5} {\bibfield  {journal} {\bibinfo  {journal} {Nat. Commun.}\ }\textbf {\bibinfo {volume} {14}},\ \bibinfo {pages} {2044} (\bibinfo {year} {2023})}\BibitemShut {NoStop}%
\bibitem [{\citenamefont {Huang}\ \emph {et~al.}(2024)\citenamefont {Huang}, \citenamefont {Anufriev}, \citenamefont {Jalabert}, \citenamefont {Watanabe}, \citenamefont {Taniguchi}, \citenamefont {Guo}, \citenamefont {Ni}, \citenamefont {Volz},\ and\ \citenamefont {Nomura}}]{huang_graphite_2024}%
  \BibitemOpen
  \bibfield  {author} {\bibinfo {author} {\bibfnamefont {X.}~\bibnamefont {Huang}}, \bibinfo {author} {\bibfnamefont {R.}~\bibnamefont {Anufriev}}, \bibinfo {author} {\bibfnamefont {L.}~\bibnamefont {Jalabert}}, \bibinfo {author} {\bibfnamefont {K.}~\bibnamefont {Watanabe}}, \bibinfo {author} {\bibfnamefont {T.}~\bibnamefont {Taniguchi}}, \bibinfo {author} {\bibfnamefont {Y.}~\bibnamefont {Guo}}, \bibinfo {author} {\bibfnamefont {Y.}~\bibnamefont {Ni}}, \bibinfo {author} {\bibfnamefont {S.}~\bibnamefont {Volz}},\ and\ \bibinfo {author} {\bibfnamefont {M.}~\bibnamefont {Nomura}},\ }\bibfield  {title} {\bibinfo {title} {A graphite thermal {Tesla} valve driven by hydrodynamic phonon transport},\ }\href {https://doi.org/10.1038/s41586-024-08052-1} {\bibfield  {journal} {\bibinfo  {journal} {Nature}\ ,\ \bibinfo {pages} {1}} (\bibinfo {year} {2024})},\ \bibinfo {note} {publisher: Nature Publishing Group}\BibitemShut {NoStop}%
\bibitem [{\citenamefont {Dragašević}\ and\ \citenamefont {Simoncelli}(2024)}]{dragasevic_viscous_2024}%
  \BibitemOpen
  \bibfield  {author} {\bibinfo {author} {\bibfnamefont {J.}~\bibnamefont {Dragašević}}\ and\ \bibinfo {author} {\bibfnamefont {M.}~\bibnamefont {Simoncelli}},\ }\href {http://arxiv.org/abs/2303.12777} {\bibinfo {title} {Viscous heat backflow and temperature resonances in extreme thermal conductors}} (\bibinfo {year} {2024}),\ \bibinfo {note} {arXiv:2303.12777 [cond-mat]}\BibitemShut {NoStop}%
\bibitem [{\citenamefont {Balandin}(2011)}]{balandin_thermal_2011}%
  \BibitemOpen
  \bibfield  {author} {\bibinfo {author} {\bibfnamefont {A.~A.}\ \bibnamefont {Balandin}},\ }\bibfield  {title} {\bibinfo {title} {Thermal properties of graphene and nanostructured carbon materials},\ }\href {https://doi.org/10.1038/nmat3064} {\bibfield  {journal} {\bibinfo  {journal} {Nature Materials}\ }\textbf {\bibinfo {volume} {10}},\ \bibinfo {pages} {569} (\bibinfo {year} {2011})},\ \bibinfo {note} {number: 8 Publisher: Nature Publishing Group}\BibitemShut {NoStop}%
\bibitem [{\citenamefont {Goblot}\ \emph {et~al.}(2024)\citenamefont {Goblot}, \citenamefont {Wu}, \citenamefont {Lucente}, \citenamefont {Zhu}, \citenamefont {Losero}, \citenamefont {Jobert}, \citenamefont {Concha}, \citenamefont {Marzari}, \citenamefont {Simoncelli},\ and\ \citenamefont {Galland}}]{goblot_imaging_2024}%
  \BibitemOpen
  \bibfield  {author} {\bibinfo {author} {\bibfnamefont {V.}~\bibnamefont {Goblot}}, \bibinfo {author} {\bibfnamefont {K.}~\bibnamefont {Wu}}, \bibinfo {author} {\bibfnamefont {E.~D.}\ \bibnamefont {Lucente}}, \bibinfo {author} {\bibfnamefont {Y.}~\bibnamefont {Zhu}}, \bibinfo {author} {\bibfnamefont {E.}~\bibnamefont {Losero}}, \bibinfo {author} {\bibfnamefont {Q.}~\bibnamefont {Jobert}}, \bibinfo {author} {\bibfnamefont {C.~J.}\ \bibnamefont {Concha}}, \bibinfo {author} {\bibfnamefont {N.}~\bibnamefont {Marzari}}, \bibinfo {author} {\bibfnamefont {M.}~\bibnamefont {Simoncelli}},\ and\ \bibinfo {author} {\bibfnamefont {C.}~\bibnamefont {Galland}},\ }\href {http://arxiv.org/abs/2411.04065} {\bibinfo {title} {Imaging heat transport in suspended diamond nanostructures with integrated spin defect thermometers}} (\bibinfo {year} {2024}),\ \bibinfo {note} {arXiv:2411.04065}\BibitemShut {NoStop}%
\bibitem [{\citenamefont {Morath}\ \emph {et~al.}(1994)\citenamefont {Morath}, \citenamefont {Maris}, \citenamefont {Cuomo}, \citenamefont {Pappas}, \citenamefont {Grill}, \citenamefont {Patel}, \citenamefont {Doyle},\ and\ \citenamefont {Saenger}}]{morath_picosecond_1994}%
  \BibitemOpen
  \bibfield  {author} {\bibinfo {author} {\bibfnamefont {C.~J.}\ \bibnamefont {Morath}}, \bibinfo {author} {\bibfnamefont {H.~J.}\ \bibnamefont {Maris}}, \bibinfo {author} {\bibfnamefont {J.~J.}\ \bibnamefont {Cuomo}}, \bibinfo {author} {\bibfnamefont {D.~L.}\ \bibnamefont {Pappas}}, \bibinfo {author} {\bibfnamefont {A.}~\bibnamefont {Grill}}, \bibinfo {author} {\bibfnamefont {V.~V.}\ \bibnamefont {Patel}}, \bibinfo {author} {\bibfnamefont {J.~P.}\ \bibnamefont {Doyle}},\ and\ \bibinfo {author} {\bibfnamefont {K.~L.}\ \bibnamefont {Saenger}},\ }\bibfield  {title} {\bibinfo {title} {Picosecond optical studies of amorphous diamond and diamondlike carbon: {Thermal} conductivity and longitudinal sound velocity},\ }\href {https://doi.org/10.1063/1.357560} {\bibfield  {journal} {\bibinfo  {journal} {Journal of Applied Physics}\ }\textbf {\bibinfo {volume} {76}},\ \bibinfo {pages} {2636} (\bibinfo {year} {1994})}\BibitemShut {NoStop}%
\bibitem [{\citenamefont {Hurler}\ \emph {et~al.}(1995)\citenamefont {Hurler}, \citenamefont {Pietralla},\ and\ \citenamefont {Hammerschmidt}}]{hurler_determination_1995}%
  \BibitemOpen
  \bibfield  {author} {\bibinfo {author} {\bibfnamefont {W.}~\bibnamefont {Hurler}}, \bibinfo {author} {\bibfnamefont {M.}~\bibnamefont {Pietralla}},\ and\ \bibinfo {author} {\bibfnamefont {A.}~\bibnamefont {Hammerschmidt}},\ }\bibfield  {title} {\bibinfo {title} {Determination of thermal properties of hydrogenated amorphous carbon films via mirage effect measurements},\ }\href {https://doi.org/10.1016/0925-9635(94)00259-2} {\bibfield  {journal} {\bibinfo  {journal} {Diamond and Related Materials}\ }\textbf {\bibinfo {volume} {4}},\ \bibinfo {pages} {954} (\bibinfo {year} {1995})}\BibitemShut {NoStop}%
\bibitem [{\citenamefont {Bullen}\ \emph {et~al.}(2000)\citenamefont {Bullen}, \citenamefont {O’Hara}, \citenamefont {Cahill}, \citenamefont {Monteiro},\ and\ \citenamefont {von Keudell}}]{bullen_thermal_2000}%
  \BibitemOpen
  \bibfield  {author} {\bibinfo {author} {\bibfnamefont {A.~J.}\ \bibnamefont {Bullen}}, \bibinfo {author} {\bibfnamefont {K.~E.}\ \bibnamefont {O’Hara}}, \bibinfo {author} {\bibfnamefont {D.~G.}\ \bibnamefont {Cahill}}, \bibinfo {author} {\bibfnamefont {O.}~\bibnamefont {Monteiro}},\ and\ \bibinfo {author} {\bibfnamefont {A.}~\bibnamefont {von Keudell}},\ }\bibfield  {title} {\bibinfo {title} {Thermal conductivity of amorphous carbon thin films},\ }\href {https://doi.org/10.1063/1.1314301} {\bibfield  {journal} {\bibinfo  {journal} {Journal of Applied Physics}\ }\textbf {\bibinfo {volume} {88}},\ \bibinfo {pages} {6317} (\bibinfo {year} {2000})}\BibitemShut {NoStop}%
\bibitem [{\citenamefont {Shamsa}\ \emph {et~al.}(2006)\citenamefont {Shamsa}, \citenamefont {Liu}, \citenamefont {Balandin}, \citenamefont {Casiraghi}, \citenamefont {Milne},\ and\ \citenamefont {Ferrari}}]{shamsa_thermal_2006}%
  \BibitemOpen
  \bibfield  {author} {\bibinfo {author} {\bibfnamefont {M.}~\bibnamefont {Shamsa}}, \bibinfo {author} {\bibfnamefont {W.~L.}\ \bibnamefont {Liu}}, \bibinfo {author} {\bibfnamefont {A.~A.}\ \bibnamefont {Balandin}}, \bibinfo {author} {\bibfnamefont {C.}~\bibnamefont {Casiraghi}}, \bibinfo {author} {\bibfnamefont {W.~I.}\ \bibnamefont {Milne}},\ and\ \bibinfo {author} {\bibfnamefont {A.~C.}\ \bibnamefont {Ferrari}},\ }\bibfield  {title} {\bibinfo {title} {Thermal conductivity of diamond-like carbon films},\ }\href {https://doi.org/10.1063/1.2362601} {\bibfield  {journal} {\bibinfo  {journal} {Applied Physics Letters}\ }\textbf {\bibinfo {volume} {89}},\ \bibinfo {pages} {161921} (\bibinfo {year} {2006})}\BibitemShut {NoStop}%
\bibitem [{\citenamefont {Scott}\ \emph {et~al.}(2021)\citenamefont {Scott}, \citenamefont {King}, \citenamefont {Jarenwattananon}, \citenamefont {Lanford}, \citenamefont {Li}, \citenamefont {Rhodes},\ and\ \citenamefont {Hopkins}}]{scott_thermal_2021}%
  \BibitemOpen
  \bibfield  {author} {\bibinfo {author} {\bibfnamefont {E.~A.}\ \bibnamefont {Scott}}, \bibinfo {author} {\bibfnamefont {S.~W.}\ \bibnamefont {King}}, \bibinfo {author} {\bibfnamefont {N.~N.}\ \bibnamefont {Jarenwattananon}}, \bibinfo {author} {\bibfnamefont {W.~A.}\ \bibnamefont {Lanford}}, \bibinfo {author} {\bibfnamefont {H.}~\bibnamefont {Li}}, \bibinfo {author} {\bibfnamefont {J.}~\bibnamefont {Rhodes}},\ and\ \bibinfo {author} {\bibfnamefont {P.~E.}\ \bibnamefont {Hopkins}},\ }\bibfield  {title} {\bibinfo {title} {Thermal {Conductivity} {Enhancement} in {Ion}-{Irradiated} {Hydrogenated} {Amorphous} {Carbon} {Films}},\ }\href {https://doi.org/10.1021/acs.nanolett.1c00616} {\bibfield  {journal} {\bibinfo  {journal} {Nano Letters}\ }\textbf {\bibinfo {volume} {21}},\ \bibinfo {pages} {3935} (\bibinfo {year} {2021})},\ \bibinfo {note} {publisher: American Chemical Society}\BibitemShut {NoStop}%
\bibitem [{\citenamefont {Arlein}\ \emph {et~al.}(2008)\citenamefont {Arlein}, \citenamefont {Palaich}, \citenamefont {Daly}, \citenamefont {Subramonium},\ and\ \citenamefont {Antonelli}}]{arlein_optical_2008}%
  \BibitemOpen
  \bibfield  {author} {\bibinfo {author} {\bibfnamefont {J.~L.}\ \bibnamefont {Arlein}}, \bibinfo {author} {\bibfnamefont {S.~E.~M.}\ \bibnamefont {Palaich}}, \bibinfo {author} {\bibfnamefont {B.~C.}\ \bibnamefont {Daly}}, \bibinfo {author} {\bibfnamefont {P.}~\bibnamefont {Subramonium}},\ and\ \bibinfo {author} {\bibfnamefont {G.~A.}\ \bibnamefont {Antonelli}},\ }\bibfield  {title} {\bibinfo {title} {Optical pump-probe measurements of sound velocity and thermal conductivity of hydrogenated amorphous carbon films},\ }\href {https://doi.org/10.1063/1.2963366} {\bibfield  {journal} {\bibinfo  {journal} {Journal of Applied Physics}\ }\textbf {\bibinfo {volume} {104}},\ \bibinfo {pages} {033508} (\bibinfo {year} {2008})}\BibitemShut {NoStop}%
\bibitem [{\citenamefont {Chen}\ \emph {et~al.}(2000)\citenamefont {Chen}, \citenamefont {Hui},\ and\ \citenamefont {Xu}}]{chen_thermal_2000}%
  \BibitemOpen
  \bibfield  {author} {\bibinfo {author} {\bibfnamefont {G.}~\bibnamefont {Chen}}, \bibinfo {author} {\bibfnamefont {P.}~\bibnamefont {Hui}},\ and\ \bibinfo {author} {\bibfnamefont {S.}~\bibnamefont {Xu}},\ }\bibfield  {title} {\bibinfo {title} {Thermal conduction in metalized tetrahedral amorphous carbon (ta–{C}) films on silicon},\ }\href {https://doi.org/10.1016/S0040-6090(99)01097-4} {\bibfield  {journal} {\bibinfo  {journal} {Thin Solid Films}\ }\textbf {\bibinfo {volume} {366}},\ \bibinfo {pages} {95} (\bibinfo {year} {2000})}\BibitemShut {NoStop}%
\bibitem [{\citenamefont {Maruyama}\ and\ \citenamefont {Harayama}(1992)}]{maruyama_neutron_1992}%
  \BibitemOpen
  \bibfield  {author} {\bibinfo {author} {\bibfnamefont {T.}~\bibnamefont {Maruyama}}\ and\ \bibinfo {author} {\bibfnamefont {M.}~\bibnamefont {Harayama}},\ }\bibfield  {title} {\bibinfo {title} {Neutron irradiation effect on the thermal conductivity and dimensional change of graphite materials},\ }\href {https://doi.org/10.1016/0022-3115(92)90362-O} {\bibfield  {journal} {\bibinfo  {journal} {Journal of Nuclear Materials}\ }\textbf {\bibinfo {volume} {195}},\ \bibinfo {pages} {44} (\bibinfo {year} {1992})}\BibitemShut {NoStop}%
\bibitem [{\citenamefont {Wu}\ \emph {et~al.}(1994)\citenamefont {Wu}, \citenamefont {Bonal}, \citenamefont {Thiele}, \citenamefont {Tsotridis}, \citenamefont {Kwast}, \citenamefont {Werle}, \citenamefont {Coad}, \citenamefont {Federici},\ and\ \citenamefont {Vieider}}]{wu_neutron_1994}%
  \BibitemOpen
  \bibfield  {author} {\bibinfo {author} {\bibfnamefont {C.~H.}\ \bibnamefont {Wu}}, \bibinfo {author} {\bibfnamefont {J.~P.}\ \bibnamefont {Bonal}}, \bibinfo {author} {\bibfnamefont {B.}~\bibnamefont {Thiele}}, \bibinfo {author} {\bibfnamefont {G.}~\bibnamefont {Tsotridis}}, \bibinfo {author} {\bibfnamefont {H.}~\bibnamefont {Kwast}}, \bibinfo {author} {\bibfnamefont {H.}~\bibnamefont {Werle}}, \bibinfo {author} {\bibfnamefont {J.~P.}\ \bibnamefont {Coad}}, \bibinfo {author} {\bibfnamefont {G.}~\bibnamefont {Federici}},\ and\ \bibinfo {author} {\bibfnamefont {G.}~\bibnamefont {Vieider}},\ }\bibfield  {title} {\bibinfo {title} {Neutron irradiation effects on the properties of carbon materials},\ }\href {https://doi.org/10.1016/0022-3115(94)90096-5} {\bibfield  {journal} {\bibinfo  {journal} {Journal of Nuclear Materials}\ }\textbf {\bibinfo {volume} {212-215}},\ \bibinfo {pages} {416} (\bibinfo {year} {1994})}\BibitemShut {NoStop}%
\bibitem [{\citenamefont {Snead}\ and\ \citenamefont {Burchell}(1995{\natexlab{a}})}]{snead_reduction_1995}%
  \BibitemOpen
  \bibfield  {author} {\bibinfo {author} {\bibfnamefont {L.}~\bibnamefont {Snead}}\ and\ \bibinfo {author} {\bibfnamefont {T.}~\bibnamefont {Burchell}},\ }\bibfield  {title} {\bibinfo {title} {Reduction in thermal conductivity due to neutron irradiation},\ }\href@noop {} {\bibfield  {journal} {\bibinfo  {journal} {22nd Biennial Conference on Carbon}\ } (\bibinfo {year} {1995}{\natexlab{a}})}\BibitemShut {NoStop}%
\bibitem [{\citenamefont {Snead}\ and\ \citenamefont {Burchell}(1995{\natexlab{b}})}]{snead_thermal_1995}%
  \BibitemOpen
  \bibfield  {author} {\bibinfo {author} {\bibfnamefont {L.~L.}\ \bibnamefont {Snead}}\ and\ \bibinfo {author} {\bibfnamefont {T.~D.}\ \bibnamefont {Burchell}},\ }\bibfield  {title} {\bibinfo {title} {Thermal conductivity degradation of graphites due to nuetron irradiation at low temperature},\ }\href {https://doi.org/10.1016/0022-3115(95)00071-2} {\bibfield  {journal} {\bibinfo  {journal} {Journal of Nuclear Materials}\ }\textbf {\bibinfo {volume} {224}},\ \bibinfo {pages} {222} (\bibinfo {year} {1995}{\natexlab{b}})}\BibitemShut {NoStop}%
\bibitem [{\citenamefont {Bonal}\ and\ \citenamefont {Wu}(1996)}]{bonal_neutron_1996}%
  \BibitemOpen
  \bibfield  {author} {\bibinfo {author} {\bibfnamefont {J.}~\bibnamefont {Bonal}}\ and\ \bibinfo {author} {\bibfnamefont {C.}~\bibnamefont {Wu}},\ }\bibfield  {title} {\bibinfo {title} {Neutron irradiation effects on the thermal conductivity and dimensional stability of carbon fiber composites at divertor conditions},\ }\href {https://doi.org/10.1016/S0022-3115(95)00247-2} {\bibfield  {journal} {\bibinfo  {journal} {Journal of Nuclear Materials}\ }\textbf {\bibinfo {volume} {228}},\ \bibinfo {pages} {155} (\bibinfo {year} {1996})}\BibitemShut {NoStop}%
\bibitem [{\citenamefont {Ishiyama}\ \emph {et~al.}(1996)\citenamefont {Ishiyama}, \citenamefont {Burchell}, \citenamefont {Strizak},\ and\ \citenamefont {Eto}}]{ishiyama_effect_1996}%
  \BibitemOpen
  \bibfield  {author} {\bibinfo {author} {\bibfnamefont {S.}~\bibnamefont {Ishiyama}}, \bibinfo {author} {\bibfnamefont {T.}~\bibnamefont {Burchell}}, \bibinfo {author} {\bibfnamefont {J.}~\bibnamefont {Strizak}},\ and\ \bibinfo {author} {\bibfnamefont {M.}~\bibnamefont {Eto}},\ }\bibfield  {title} {\bibinfo {title} {The effect of high fluence neutron irradiation on the properties of a fine-grained isotropic nuclear graphite},\ }\href {https://doi.org/10.1016/0022-3115(96)00005-0} {\bibfield  {journal} {\bibinfo  {journal} {Journal of Nuclear Materials}\ }\textbf {\bibinfo {volume} {230}},\ \bibinfo {pages} {1} (\bibinfo {year} {1996})}\BibitemShut {NoStop}%
\bibitem [{\citenamefont {Barabash}\ \emph {et~al.}(2002)\citenamefont {Barabash}, \citenamefont {Mazul}, \citenamefont {Latypov}, \citenamefont {Pokrovsky},\ and\ \citenamefont {Wu}}]{barabash_effect_2002}%
  \BibitemOpen
  \bibfield  {author} {\bibinfo {author} {\bibfnamefont {V.}~\bibnamefont {Barabash}}, \bibinfo {author} {\bibfnamefont {I.}~\bibnamefont {Mazul}}, \bibinfo {author} {\bibfnamefont {R.}~\bibnamefont {Latypov}}, \bibinfo {author} {\bibfnamefont {A.}~\bibnamefont {Pokrovsky}},\ and\ \bibinfo {author} {\bibfnamefont {C.}~\bibnamefont {Wu}},\ }\bibfield  {title} {\bibinfo {title} {The effect of low temperature neutron irradiation and annealing on the thermal conductivity of advanced carbon-based materials},\ }\href {https://doi.org/10.1016/S0022-3115(02)00961-3} {\bibfield  {journal} {\bibinfo  {journal} {Journal of Nuclear Materials}\ }\textbf {\bibinfo {volume} {307-311}},\ \bibinfo {pages} {1300} (\bibinfo {year} {2002})}\BibitemShut {NoStop}%
\bibitem [{\citenamefont {Snead}(2008)}]{snead_accumulation_2008}%
  \BibitemOpen
  \bibfield  {author} {\bibinfo {author} {\bibfnamefont {L.~L.}\ \bibnamefont {Snead}},\ }\bibfield  {title} {\bibinfo {title} {Accumulation of thermal resistance in neutron irradiated graphite materials},\ }\href {https://doi.org/10.1016/j.jnucmat.2008.07.017} {\bibfield  {journal} {\bibinfo  {journal} {Journal of Nuclear Materials}\ }\bibinfo {series} {Proceedings of the {Seventh} and {Eighth} {International} {Graphite} {Specialists} {Meetings} ({INGSM})},\ \textbf {\bibinfo {volume} {381}},\ \bibinfo {pages} {76} (\bibinfo {year} {2008})}\BibitemShut {NoStop}%
\bibitem [{\citenamefont {Campbell}\ \emph {et~al.}(2016)\citenamefont {Campbell}, \citenamefont {Katoh}, \citenamefont {Snead},\ and\ \citenamefont {Takizawa}}]{campbell_property_2016}%
  \BibitemOpen
  \bibfield  {author} {\bibinfo {author} {\bibfnamefont {A.~A.}\ \bibnamefont {Campbell}}, \bibinfo {author} {\bibfnamefont {Y.}~\bibnamefont {Katoh}}, \bibinfo {author} {\bibfnamefont {M.~A.}\ \bibnamefont {Snead}},\ and\ \bibinfo {author} {\bibfnamefont {K.}~\bibnamefont {Takizawa}},\ }\bibfield  {title} {\bibinfo {title} {Property changes of {G347A} graphite due to neutron irradiation},\ }\href {https://doi.org/10.1016/j.carbon.2016.08.042} {\bibfield  {journal} {\bibinfo  {journal} {Carbon}\ }\textbf {\bibinfo {volume} {109}},\ \bibinfo {pages} {860} (\bibinfo {year} {2016})}\BibitemShut {NoStop}%
\bibitem [{\citenamefont {Heijna}\ \emph {et~al.}(2017)\citenamefont {Heijna}, \citenamefont {De~Groot},\ and\ \citenamefont {Vreeling}}]{heijna_comparison_2017}%
  \BibitemOpen
  \bibfield  {author} {\bibinfo {author} {\bibfnamefont {M.}~\bibnamefont {Heijna}}, \bibinfo {author} {\bibfnamefont {S.}~\bibnamefont {De~Groot}},\ and\ \bibinfo {author} {\bibfnamefont {J.}~\bibnamefont {Vreeling}},\ }\bibfield  {title} {\bibinfo {title} {Comparison of irradiation behaviour of {HTR} graphite grades},\ }\href {https://doi.org/10.1016/j.jnucmat.2017.05.012} {\bibfield  {journal} {\bibinfo  {journal} {Journal of Nuclear Materials}\ }\textbf {\bibinfo {volume} {492}},\ \bibinfo {pages} {148} (\bibinfo {year} {2017})}\BibitemShut {NoStop}%
\bibitem [{\citenamefont {Maruyama}\ and\ \citenamefont {Li}(2019)}]{maruyama_dimensional_2019}%
  \BibitemOpen
  \bibfield  {author} {\bibinfo {author} {\bibfnamefont {T.}~\bibnamefont {Maruyama}}\ and\ \bibinfo {author} {\bibfnamefont {Z.}~\bibnamefont {Li}},\ }\bibfield  {title} {\bibinfo {title} {Dimensional changes and thermal conductivity by annealing and its relation to the defect concentration and stored energy release of neutron-irradiated graphite},\ }\href {https://doi.org/10.1080/00223131.2019.1633966} {\bibfield  {journal} {\bibinfo  {journal} {Journal of Nuclear Science and Technology}\ }\textbf {\bibinfo {volume} {56}},\ \bibinfo {pages} {1006} (\bibinfo {year} {2019})}\BibitemShut {NoStop}%
\bibitem [{\citenamefont {Galli}\ \emph {et~al.}(1989)\citenamefont {Galli}, \citenamefont {Martin}, \citenamefont {Car},\ and\ \citenamefont {Parrinello}}]{galli_structural_1989}%
  \BibitemOpen
  \bibfield  {author} {\bibinfo {author} {\bibfnamefont {G.}~\bibnamefont {Galli}}, \bibinfo {author} {\bibfnamefont {R.~M.}\ \bibnamefont {Martin}}, \bibinfo {author} {\bibfnamefont {R.}~\bibnamefont {Car}},\ and\ \bibinfo {author} {\bibfnamefont {M.}~\bibnamefont {Parrinello}},\ }\bibfield  {title} {\bibinfo {title} {Structural and {Electronic} {Properties} of {Amorphous} {Carbon}},\ }\href {https://doi.org/10.1103/PhysRevLett.62.555} {\bibfield  {journal} {\bibinfo  {journal} {Physical Review Letters}\ }\textbf {\bibinfo {volume} {62}},\ \bibinfo {pages} {555} (\bibinfo {year} {1989})},\ \bibinfo {note} {publisher: American Physical Society}\BibitemShut {NoStop}%
\bibitem [{\citenamefont {Thomas}\ \emph {et~al.}(2010)\citenamefont {Thomas}, \citenamefont {Turney}, \citenamefont {Iutzi}, \citenamefont {Amon},\ and\ \citenamefont {McGaughey}}]{thomas_predicting_2010}%
  \BibitemOpen
  \bibfield  {author} {\bibinfo {author} {\bibfnamefont {J.~A.}\ \bibnamefont {Thomas}}, \bibinfo {author} {\bibfnamefont {J.~E.}\ \bibnamefont {Turney}}, \bibinfo {author} {\bibfnamefont {R.~M.}\ \bibnamefont {Iutzi}}, \bibinfo {author} {\bibfnamefont {C.~H.}\ \bibnamefont {Amon}},\ and\ \bibinfo {author} {\bibfnamefont {A.~J.~H.}\ \bibnamefont {McGaughey}},\ }\bibfield  {title} {\bibinfo {title} {Predicting phonon dispersion relations and lifetimes from the spectral energy density},\ }\href {https://doi.org/10.1103/PhysRevB.81.081411} {\bibfield  {journal} {\bibinfo  {journal} {Physical Review B}\ }\textbf {\bibinfo {volume} {81}},\ \bibinfo {pages} {081411} (\bibinfo {year} {2010})},\ \bibinfo {note} {publisher: American Physical Society}\BibitemShut {NoStop}%
\bibitem [{\citenamefont {Zhou}\ \emph {et~al.}(2025)\citenamefont {Zhou}, \citenamefont {Liu}, \citenamefont {Tang}, \citenamefont {Wang}, \citenamefont {Dong}, \citenamefont {Xiu}, \citenamefont {Chen},\ and\ \citenamefont {Fan}}]{zhou_million-atom_2025}%
  \BibitemOpen
  \bibfield  {author} {\bibinfo {author} {\bibfnamefont {X.}~\bibnamefont {Zhou}}, \bibinfo {author} {\bibfnamefont {Y.}~\bibnamefont {Liu}}, \bibinfo {author} {\bibfnamefont {B.}~\bibnamefont {Tang}}, \bibinfo {author} {\bibfnamefont {J.}~\bibnamefont {Wang}}, \bibinfo {author} {\bibfnamefont {H.}~\bibnamefont {Dong}}, \bibinfo {author} {\bibfnamefont {X.}~\bibnamefont {Xiu}}, \bibinfo {author} {\bibfnamefont {S.}~\bibnamefont {Chen}},\ and\ \bibinfo {author} {\bibfnamefont {Z.}~\bibnamefont {Fan}},\ }\bibfield  {title} {\bibinfo {title} {Million-atom heat transport simulations of polycrystalline graphene approaching first-principles accuracy enabled by neuroevolution potential on desktop {GPUs}},\ }\href {https://doi.org/10.1063/5.0244987} {\bibfield  {journal} {\bibinfo  {journal} {Journal of Applied Physics}\ }\textbf {\bibinfo {volume} {137}},\ \bibinfo {pages} {014305} (\bibinfo {year} {2025})}\BibitemShut {NoStop}%
\bibitem [{\citenamefont {Giri}\ \emph {et~al.}(2022)\citenamefont {Giri}, \citenamefont {Dionne},\ and\ \citenamefont {Hopkins}}]{giri_atomic_2022}%
  \BibitemOpen
  \bibfield  {author} {\bibinfo {author} {\bibfnamefont {A.}~\bibnamefont {Giri}}, \bibinfo {author} {\bibfnamefont {C.~J.}\ \bibnamefont {Dionne}},\ and\ \bibinfo {author} {\bibfnamefont {P.~E.}\ \bibnamefont {Hopkins}},\ }\bibfield  {title} {\bibinfo {title} {Atomic coordination dictates vibrational characteristics and thermal conductivity in amorphous carbon},\ }\href {https://doi.org/10.1038/s41524-022-00741-7} {\bibfield  {journal} {\bibinfo  {journal} {npj Computational Materials}\ }\textbf {\bibinfo {volume} {8}},\ \bibinfo {pages} {1} (\bibinfo {year} {2022})},\ \bibinfo {note} {number: 1 Publisher: Nature Publishing Group}\BibitemShut {NoStop}%
\bibitem [{\citenamefont {Minamitani}\ \emph {et~al.}(2022)\citenamefont {Minamitani}, \citenamefont {Shiga}, \citenamefont {Kashiwagi},\ and\ \citenamefont {Obayashi}}]{minamitani_relationship_2022}%
  \BibitemOpen
  \bibfield  {author} {\bibinfo {author} {\bibfnamefont {E.}~\bibnamefont {Minamitani}}, \bibinfo {author} {\bibfnamefont {T.}~\bibnamefont {Shiga}}, \bibinfo {author} {\bibfnamefont {M.}~\bibnamefont {Kashiwagi}},\ and\ \bibinfo {author} {\bibfnamefont {I.}~\bibnamefont {Obayashi}},\ }\bibfield  {title} {\bibinfo {title} {Relationship between local coordinates and thermal conductivity in amorphous carbon},\ }\href {https://doi.org/10.1116/6.0001744} {\bibfield  {journal} {\bibinfo  {journal} {Journal of Vacuum Science \& Technology A}\ }\textbf {\bibinfo {volume} {40}},\ \bibinfo {pages} {033408} (\bibinfo {year} {2022})}\BibitemShut {NoStop}%
\bibitem [{\citenamefont {Moon}\ and\ \citenamefont {Tian}(2025)}]{moon_crystal-like_2025}%
  \BibitemOpen
  \bibfield  {author} {\bibinfo {author} {\bibfnamefont {J.}~\bibnamefont {Moon}}\ and\ \bibinfo {author} {\bibfnamefont {Z.}~\bibnamefont {Tian}},\ }\bibfield  {title} {\bibinfo {title} {Crystal-like thermal transport in amorphous carbon},\ }\href {https://doi.org/10.1038/s41524-025-01625-2} {\bibfield  {journal} {\bibinfo  {journal} {npj Computational Materials}\ }\textbf {\bibinfo {volume} {11}},\ \bibinfo {pages} {137} (\bibinfo {year} {2025})},\ \bibinfo {note} {publisher: Nature Publishing Group}\BibitemShut {NoStop}%
\bibitem [{\citenamefont {Lv}\ and\ \citenamefont {Henry}(2016{\natexlab{a}})}]{lv_phonon_2016}%
  \BibitemOpen
  \bibfield  {author} {\bibinfo {author} {\bibfnamefont {W.}~\bibnamefont {Lv}}\ and\ \bibinfo {author} {\bibfnamefont {A.}~\bibnamefont {Henry}},\ }\bibfield  {title} {\bibinfo {title} {Phonon transport in amorphous carbon using {Green}–{Kubo} modal analysis},\ }\href {https://doi.org/10.1063/1.4948605} {\bibfield  {journal} {\bibinfo  {journal} {Applied Physics Letters}\ }\textbf {\bibinfo {volume} {108}},\ \bibinfo {pages} {181905} (\bibinfo {year} {2016}{\natexlab{a}})}\BibitemShut {NoStop}%
\bibitem [{\citenamefont {Suarez-Martinez}\ and\ \citenamefont {Marks}(2011)}]{suarez-martinez_effect_2011}%
  \BibitemOpen
  \bibfield  {author} {\bibinfo {author} {\bibfnamefont {I.}~\bibnamefont {Suarez-Martinez}}\ and\ \bibinfo {author} {\bibfnamefont {N.~A.}\ \bibnamefont {Marks}},\ }\bibfield  {title} {\bibinfo {title} {Effect of microstructure on the thermal conductivity of disordered carbon},\ }\href {https://doi.org/10.1063/1.3607872} {\bibfield  {journal} {\bibinfo  {journal} {Applied Physics Letters}\ }\textbf {\bibinfo {volume} {99}},\ \bibinfo {pages} {033101} (\bibinfo {year} {2011})}\BibitemShut {NoStop}%
\bibitem [{\citenamefont {Zhang}\ \emph {et~al.}(2017)\citenamefont {Zhang}, \citenamefont {Ai}, \citenamefont {Chen},\ and\ \citenamefont {Xiong}}]{zhang_thermal_2017}%
  \BibitemOpen
  \bibfield  {author} {\bibinfo {author} {\bibfnamefont {X.-X.}\ \bibnamefont {Zhang}}, \bibinfo {author} {\bibfnamefont {L.-Q.}\ \bibnamefont {Ai}}, \bibinfo {author} {\bibfnamefont {M.}~\bibnamefont {Chen}},\ and\ \bibinfo {author} {\bibfnamefont {D.-X.}\ \bibnamefont {Xiong}},\ }\bibfield  {title} {\bibinfo {title} {Thermal conductive performance of deposited amorphous carbon materials by molecular dynamics simulation},\ }\href {https://doi.org/10.1080/00268976.2017.1288940} {\bibfield  {journal} {\bibinfo  {journal} {Molecular Physics}\ }\textbf {\bibinfo {volume} {115}},\ \bibinfo {pages} {831} (\bibinfo {year} {2017})}\BibitemShut {NoStop}%
\bibitem [{\citenamefont {Wang}\ \emph {et~al.}(2025)\citenamefont {Wang}, \citenamefont {Fan}, \citenamefont {Qian}, \citenamefont {Caro},\ and\ \citenamefont {Ala-Nissila}}]{wang_density_2025}%
  \BibitemOpen
  \bibfield  {author} {\bibinfo {author} {\bibfnamefont {Y.}~\bibnamefont {Wang}}, \bibinfo {author} {\bibfnamefont {Z.}~\bibnamefont {Fan}}, \bibinfo {author} {\bibfnamefont {P.}~\bibnamefont {Qian}}, \bibinfo {author} {\bibfnamefont {M.~A.}\ \bibnamefont {Caro}},\ and\ \bibinfo {author} {\bibfnamefont {T.}~\bibnamefont {Ala-Nissila}},\ }\bibfield  {title} {\bibinfo {title} {Density dependence of thermal conductivity in nanoporous and amorphous carbon with machine-learned molecular dynamics},\ }\href {https://doi.org/10.1103/PhysRevB.111.094205} {\bibfield  {journal} {\bibinfo  {journal} {Physical Review B}\ }\textbf {\bibinfo {volume} {111}},\ \bibinfo {pages} {094205} (\bibinfo {year} {2025})},\ \bibinfo {note} {publisher: American Physical Society}\BibitemShut {NoStop}%
\bibitem [{\citenamefont {Jung}\ \emph {et~al.}(2017)\citenamefont {Jung}, \citenamefont {Yeo}, \citenamefont {Tian}, \citenamefont {Qin},\ and\ \citenamefont {Buehler}}]{jung_unusually_2017}%
  \BibitemOpen
  \bibfield  {author} {\bibinfo {author} {\bibfnamefont {G.~S.}\ \bibnamefont {Jung}}, \bibinfo {author} {\bibfnamefont {J.}~\bibnamefont {Yeo}}, \bibinfo {author} {\bibfnamefont {Z.}~\bibnamefont {Tian}}, \bibinfo {author} {\bibfnamefont {Z.}~\bibnamefont {Qin}},\ and\ \bibinfo {author} {\bibfnamefont {M.~J.}\ \bibnamefont {Buehler}},\ }\bibfield  {title} {\bibinfo {title} {Unusually low and density-insensitive thermal conductivity of three-dimensional gyroid graphene},\ }\href {https://doi.org/10.1039/C7NR04455K} {\bibfield  {journal} {\bibinfo  {journal} {Nanoscale}\ }\textbf {\bibinfo {volume} {9}},\ \bibinfo {pages} {13477} (\bibinfo {year} {2017})},\ \bibinfo {note} {publisher: The Royal Society of Chemistry}\BibitemShut {NoStop}%
\bibitem [{\citenamefont {Simoncelli}\ \emph {et~al.}(2019)\citenamefont {Simoncelli}, \citenamefont {Marzari},\ and\ \citenamefont {Mauri}}]{simoncelli_unified_2019}%
  \BibitemOpen
  \bibfield  {author} {\bibinfo {author} {\bibfnamefont {M.}~\bibnamefont {Simoncelli}}, \bibinfo {author} {\bibfnamefont {N.}~\bibnamefont {Marzari}},\ and\ \bibinfo {author} {\bibfnamefont {F.}~\bibnamefont {Mauri}},\ }\bibfield  {title} {\bibinfo {title} {Unified theory of thermal transport in crystals and glasses},\ }\href {https://doi.org/10.1038/s41567-019-0520-x} {\bibfield  {journal} {\bibinfo  {journal} {Nature Physics}\ }\textbf {\bibinfo {volume} {15}},\ \bibinfo {pages} {809} (\bibinfo {year} {2019})}\BibitemShut {NoStop}%
\bibitem [{\citenamefont {Simoncelli}\ \emph {et~al.}(2022)\citenamefont {Simoncelli}, \citenamefont {Marzari},\ and\ \citenamefont {Mauri}}]{simoncelli_wigner_2022}%
  \BibitemOpen
  \bibfield  {author} {\bibinfo {author} {\bibfnamefont {M.}~\bibnamefont {Simoncelli}}, \bibinfo {author} {\bibfnamefont {N.}~\bibnamefont {Marzari}},\ and\ \bibinfo {author} {\bibfnamefont {F.}~\bibnamefont {Mauri}},\ }\bibfield  {title} {\bibinfo {title} {Wigner {Formulation} of {Thermal} {Transport} in {Solids}},\ }\href {https://doi.org/10.1103/PhysRevX.12.041011} {\bibfield  {journal} {\bibinfo  {journal} {Physical Review X}\ }\textbf {\bibinfo {volume} {12}},\ \bibinfo {pages} {041011} (\bibinfo {year} {2022})}\BibitemShut {NoStop}%
\bibitem [{\citenamefont {Simoncelli}\ \emph {et~al.}(2023)\citenamefont {Simoncelli}, \citenamefont {Mauri},\ and\ \citenamefont {Marzari}}]{simoncelli_thermal_2023}%
  \BibitemOpen
  \bibfield  {author} {\bibinfo {author} {\bibfnamefont {M.}~\bibnamefont {Simoncelli}}, \bibinfo {author} {\bibfnamefont {F.}~\bibnamefont {Mauri}},\ and\ \bibinfo {author} {\bibfnamefont {N.}~\bibnamefont {Marzari}},\ }\bibfield  {title} {\bibinfo {title} {Thermal conductivity of glasses: first-principles theory and applications},\ }\href {https://doi.org/10.1038/s41524-023-01033-4} {\bibfield  {journal} {\bibinfo  {journal} {npj Computational Materials}\ }\textbf {\bibinfo {volume} {9}},\ \bibinfo {pages} {1} (\bibinfo {year} {2023})},\ \bibinfo {note} {number: 1 Publisher: Nature Publishing Group}\BibitemShut {NoStop}%
\bibitem [{\citenamefont {Rowe}\ \emph {et~al.}(2020)\citenamefont {Rowe}, \citenamefont {Deringer}, \citenamefont {Gasparotto}, \citenamefont {Csányi},\ and\ \citenamefont {Michaelides}}]{rowe_accurate_2020}%
  \BibitemOpen
  \bibfield  {author} {\bibinfo {author} {\bibfnamefont {P.}~\bibnamefont {Rowe}}, \bibinfo {author} {\bibfnamefont {V.~L.}\ \bibnamefont {Deringer}}, \bibinfo {author} {\bibfnamefont {P.}~\bibnamefont {Gasparotto}}, \bibinfo {author} {\bibfnamefont {G.}~\bibnamefont {Csányi}},\ and\ \bibinfo {author} {\bibfnamefont {A.}~\bibnamefont {Michaelides}},\ }\bibfield  {title} {\bibinfo {title} {An accurate and transferable machine learning potential for carbon},\ }\href {https://doi.org/10.1063/5.0005084} {\bibfield  {journal} {\bibinfo  {journal} {The Journal of Chemical Physics}\ }\textbf {\bibinfo {volume} {153}},\ \bibinfo {pages} {034702} (\bibinfo {year} {2020})}\BibitemShut {NoStop}%
\bibitem [{\citenamefont {Schweinhart}\ \emph {et~al.}(2020)\citenamefont {Schweinhart}, \citenamefont {Rodney},\ and\ \citenamefont {Mason}}]{schweinhart_statistical_2020}%
  \BibitemOpen
  \bibfield  {author} {\bibinfo {author} {\bibfnamefont {B.}~\bibnamefont {Schweinhart}}, \bibinfo {author} {\bibfnamefont {D.}~\bibnamefont {Rodney}},\ and\ \bibinfo {author} {\bibfnamefont {J.~K.}\ \bibnamefont {Mason}},\ }\bibfield  {title} {\bibinfo {title} {Statistical topology of bond networks with applications to silica},\ }\href {https://doi.org/10.1103/PhysRevE.101.052312} {\bibfield  {journal} {\bibinfo  {journal} {Physical Review E}\ }\textbf {\bibinfo {volume} {101}},\ \bibinfo {pages} {052312} (\bibinfo {year} {2020})}\BibitemShut {NoStop}%
\bibitem [{\citenamefont {Chumakov}\ \emph {et~al.}(2014)\citenamefont {Chumakov}, \citenamefont {Monaco}, \citenamefont {Fontana}, \citenamefont {Bosak}, \citenamefont {Hermann}, \citenamefont {Bessas}, \citenamefont {Wehinger}, \citenamefont {Crichton}, \citenamefont {Krisch}, \citenamefont {R\"uffer}, \citenamefont {Baldi}, \citenamefont {Carini~Jr.}, \citenamefont {Carini}, \citenamefont {D'Angelo}, \citenamefont {Gilioli}, \citenamefont {Tripodo}, \citenamefont {Zanatta}, \citenamefont {Winkler}, \citenamefont {Milman}, \citenamefont {Refson}, \citenamefont {Dove}, \citenamefont {Dubrovinskaia}, \citenamefont {Dubrovinsky}, \citenamefont {Keding},\ and\ \citenamefont {Yue}}]{chumakov_role_2014}%
  \BibitemOpen
  \bibfield  {author} {\bibinfo {author} {\bibfnamefont {A.~I.}\ \bibnamefont {Chumakov}}, \bibinfo {author} {\bibfnamefont {G.}~\bibnamefont {Monaco}}, \bibinfo {author} {\bibfnamefont {A.}~\bibnamefont {Fontana}}, \bibinfo {author} {\bibfnamefont {A.}~\bibnamefont {Bosak}}, \bibinfo {author} {\bibfnamefont {R.~P.}\ \bibnamefont {Hermann}}, \bibinfo {author} {\bibfnamefont {D.}~\bibnamefont {Bessas}}, \bibinfo {author} {\bibfnamefont {B.}~\bibnamefont {Wehinger}}, \bibinfo {author} {\bibfnamefont {W.~A.}\ \bibnamefont {Crichton}}, \bibinfo {author} {\bibfnamefont {M.}~\bibnamefont {Krisch}}, \bibinfo {author} {\bibfnamefont {R.}~\bibnamefont {R\"uffer}}, \bibinfo {author} {\bibfnamefont {G.}~\bibnamefont {Baldi}}, \bibinfo {author} {\bibfnamefont {G.}~\bibnamefont {Carini~Jr.}}, \bibinfo {author} {\bibfnamefont {G.}~\bibnamefont {Carini}}, \bibinfo {author} {\bibfnamefont {G.}~\bibnamefont {D'Angelo}}, \bibinfo {author} {\bibfnamefont {E.}~\bibnamefont {Gilioli}}, \bibinfo {author} {\bibfnamefont {G.}~\bibnamefont {Tripodo}}, \bibinfo {author} {\bibfnamefont {M.}~\bibnamefont {Zanatta}}, \bibinfo {author} {\bibfnamefont {B.}~\bibnamefont {Winkler}}, \bibinfo {author} {\bibfnamefont {V.}~\bibnamefont {Milman}}, \bibinfo {author} {\bibfnamefont {K.}~\bibnamefont {Refson}}, \bibinfo {author} {\bibfnamefont {M.~T.}\ \bibnamefont {Dove}}, \bibinfo {author} {\bibfnamefont {N.}~\bibnamefont {Dubrovinskaia}}, \bibinfo {author} {\bibfnamefont {L.}~\bibnamefont {Dubrovinsky}}, \bibinfo {author} {\bibfnamefont {R.}~\bibnamefont {Keding}},\ and\ \bibinfo {author} {\bibfnamefont {Y.~Z.}\ \bibnamefont {Yue}},\ }\bibfield  {title} {\bibinfo {title} {Role of disorder in the thermodynamics and atomic dynamics of glasses},\ }\href {https://doi.org/10.1103/PhysRevLett.112.025502} {\bibfield  {journal} {\bibinfo  {journal} {Phys. Rev. Lett.}\ }\textbf {\bibinfo {volume} {112}},\ \bibinfo {pages} {025502} (\bibinfo {year} {2014})}\BibitemShut {NoStop}%
\bibitem [{\citenamefont {Fiorentino}\ \emph {et~al.}(2023)\citenamefont {Fiorentino}, \citenamefont {Pegolo},\ and\ \citenamefont {Baroni}}]{fiorentino_hydrodynamic_2023}%
  \BibitemOpen
  \bibfield  {author} {\bibinfo {author} {\bibfnamefont {A.}~\bibnamefont {Fiorentino}}, \bibinfo {author} {\bibfnamefont {P.}~\bibnamefont {Pegolo}},\ and\ \bibinfo {author} {\bibfnamefont {S.}~\bibnamefont {Baroni}},\ }\bibfield  {title} {\bibinfo {title} {Hydrodynamic finite-size scaling of the thermal conductivity in glasses},\ }\href {https://doi.org/10.1038/s41524-023-01116-2} {\bibfield  {journal} {\bibinfo  {journal} {npj Computational Materials}\ }\textbf {\bibinfo {volume} {9}},\ \bibinfo {pages} {1} (\bibinfo {year} {2023})},\ \bibinfo {note} {publisher: Nature Publishing Group}\BibitemShut {NoStop}%
\bibitem [{\citenamefont {Fiorentino}\ \emph {et~al.}(2025)\citenamefont {Fiorentino}, \citenamefont {Pegolo}, \citenamefont {Baroni},\ and\ \citenamefont {Donadio}}]{fiorentino_effects_2025}%
  \BibitemOpen
  \bibfield  {author} {\bibinfo {author} {\bibfnamefont {A.}~\bibnamefont {Fiorentino}}, \bibinfo {author} {\bibfnamefont {P.}~\bibnamefont {Pegolo}}, \bibinfo {author} {\bibfnamefont {S.}~\bibnamefont {Baroni}},\ and\ \bibinfo {author} {\bibfnamefont {D.}~\bibnamefont {Donadio}},\ }\bibfield  {title} {\bibinfo {title} {Effects of colored disorder on the heat conductivity of {SiGe} alloys from first principles},\ }\href {https://doi.org/10.1103/PhysRevB.111.134205} {\bibfield  {journal} {\bibinfo  {journal} {Physical Review B}\ }\textbf {\bibinfo {volume} {111}},\ \bibinfo {pages} {134205} (\bibinfo {year} {2025})},\ \bibinfo {note} {publisher: American Physical Society}\BibitemShut {NoStop}%
\bibitem [{\citenamefont {Zhang}\ \emph {et~al.}(2022)\citenamefont {Zhang}, \citenamefont {Guo}, \citenamefont {Bescond}, \citenamefont {Chen}, \citenamefont {Nomura},\ and\ \citenamefont {Volz}}]{zhang_how_2022}%
  \BibitemOpen
  \bibfield  {author} {\bibinfo {author} {\bibfnamefont {Z.}~\bibnamefont {Zhang}}, \bibinfo {author} {\bibfnamefont {Y.}~\bibnamefont {Guo}}, \bibinfo {author} {\bibfnamefont {M.}~\bibnamefont {Bescond}}, \bibinfo {author} {\bibfnamefont {J.}~\bibnamefont {Chen}}, \bibinfo {author} {\bibfnamefont {M.}~\bibnamefont {Nomura}},\ and\ \bibinfo {author} {\bibfnamefont {S.}~\bibnamefont {Volz}},\ }\bibfield  {title} {\bibinfo {title} {How coherence is governing diffuson heat transfer in amorphous solids},\ }\href {https://doi.org/10.1038/s41524-022-00776-w} {\bibfield  {journal} {\bibinfo  {journal} {npj Computational Materials}\ }\textbf {\bibinfo {volume} {8}},\ \bibinfo {pages} {96} (\bibinfo {year} {2022})}\BibitemShut {NoStop}%
\bibitem [{\citenamefont {Simoncelli}\ \emph {et~al.}(2025)\citenamefont {Simoncelli}, \citenamefont {Fournier}, \citenamefont {Marangolo}, \citenamefont {Balan}, \citenamefont {Béneut}, \citenamefont {Baptiste}, \citenamefont {Doisneau}, \citenamefont {Marzari},\ and\ \citenamefont {Mauri}}]{simoncelli_temperature-invariant_2025}%
  \BibitemOpen
  \bibfield  {author} {\bibinfo {author} {\bibfnamefont {M.}~\bibnamefont {Simoncelli}}, \bibinfo {author} {\bibfnamefont {D.}~\bibnamefont {Fournier}}, \bibinfo {author} {\bibfnamefont {M.}~\bibnamefont {Marangolo}}, \bibinfo {author} {\bibfnamefont {E.}~\bibnamefont {Balan}}, \bibinfo {author} {\bibfnamefont {K.}~\bibnamefont {Béneut}}, \bibinfo {author} {\bibfnamefont {B.}~\bibnamefont {Baptiste}}, \bibinfo {author} {\bibfnamefont {B.}~\bibnamefont {Doisneau}}, \bibinfo {author} {\bibfnamefont {N.}~\bibnamefont {Marzari}},\ and\ \bibinfo {author} {\bibfnamefont {F.}~\bibnamefont {Mauri}},\ }\bibfield  {title} {\bibinfo {title} {Temperature-invariant crystal–glass heat conduction: {From} meteorites to refractories},\ }\href {https://doi.org/10.1073/pnas.2422763122} {\bibfield  {journal} {\bibinfo  {journal} {Proceedings of the National Academy of Sciences}\ }\textbf {\bibinfo {volume} {122}},\ \bibinfo {pages} {e2422763122} (\bibinfo {year} {2025})},\ \bibinfo {note} {publisher: Proceedings of the National Academy of Sciences}\BibitemShut {NoStop}%
\bibitem [{\citenamefont {Harper}\ \emph {et~al.}(2024)\citenamefont {Harper}, \citenamefont {Iwanowski}, \citenamefont {Witt}, \citenamefont {Payne},\ and\ \citenamefont {Simoncelli}}]{harper_vibrational_2024}%
  \BibitemOpen
  \bibfield  {author} {\bibinfo {author} {\bibfnamefont {A.~F.}\ \bibnamefont {Harper}}, \bibinfo {author} {\bibfnamefont {K.}~\bibnamefont {Iwanowski}}, \bibinfo {author} {\bibfnamefont {W.~C.}\ \bibnamefont {Witt}}, \bibinfo {author} {\bibfnamefont {M.~C.}\ \bibnamefont {Payne}},\ and\ \bibinfo {author} {\bibfnamefont {M.}~\bibnamefont {Simoncelli}},\ }\bibfield  {title} {\bibinfo {title} {Vibrational and thermal properties of amorphous alumina from first principles},\ }\href {https://doi.org/10.1103/PhysRevMaterials.8.043601} {\bibfield  {journal} {\bibinfo  {journal} {Physical Review Materials}\ }\textbf {\bibinfo {volume} {8}},\ \bibinfo {pages} {043601} (\bibinfo {year} {2024})},\ \bibinfo {note} {publisher: American Physical Society}\BibitemShut {NoStop}%
\bibitem [{\citenamefont {Allen}\ and\ \citenamefont {Feldman}(1989)}]{allen_thermal_1989}%
  \BibitemOpen
  \bibfield  {author} {\bibinfo {author} {\bibfnamefont {P.~B.}\ \bibnamefont {Allen}}\ and\ \bibinfo {author} {\bibfnamefont {J.~L.}\ \bibnamefont {Feldman}},\ }\bibfield  {title} {\bibinfo {title} {Thermal {Conductivity} of {Glasses}: {Theory} and {Application} to {Amorphous} {Si}},\ }\href {https://doi.org/10.1103/PhysRevLett.62.645} {\bibfield  {journal} {\bibinfo  {journal} {Physical Review Letters}\ }\textbf {\bibinfo {volume} {62}},\ \bibinfo {pages} {645} (\bibinfo {year} {1989})}\BibitemShut {NoStop}%
\bibitem [{\citenamefont {Petkov}\ \emph {et~al.}(1999)\citenamefont {Petkov}, \citenamefont {Difrancesco}, \citenamefont {Billinge}, \citenamefont {Acharya},\ and\ \citenamefont {Foley}}]{petkov_local_1999}%
  \BibitemOpen
  \bibfield  {author} {\bibinfo {author} {\bibfnamefont {V.}~\bibnamefont {Petkov}}, \bibinfo {author} {\bibfnamefont {R.~G.}\ \bibnamefont {Difrancesco}}, \bibinfo {author} {\bibfnamefont {S.~J.~L.}\ \bibnamefont {Billinge}}, \bibinfo {author} {\bibfnamefont {M.}~\bibnamefont {Acharya}},\ and\ \bibinfo {author} {\bibfnamefont {H.~C.}\ \bibnamefont {Foley}},\ }\bibfield  {title} {\bibinfo {title} {Local structure of nanoporous carbons},\ }\href {https://doi.org/10.1080/13642819908218319} {\bibfield  {journal} {\bibinfo  {journal} {Philosophical Magazine B}\ }\textbf {\bibinfo {volume} {79}},\ \bibinfo {pages} {1519} (\bibinfo {year} {1999})}\BibitemShut {NoStop}%
\bibitem [{\citenamefont {Acharya}\ \emph {et~al.}(1999)\citenamefont {Acharya}, \citenamefont {Strano}, \citenamefont {Mathews}, \citenamefont {Billinge}, \citenamefont {Petkov}, \citenamefont {Subramoney},\ and\ \citenamefont {Foley}}]{acharya_simulation_1999}%
  \BibitemOpen
  \bibfield  {author} {\bibinfo {author} {\bibfnamefont {M.}~\bibnamefont {Acharya}}, \bibinfo {author} {\bibfnamefont {M.~S.}\ \bibnamefont {Strano}}, \bibinfo {author} {\bibfnamefont {J.~P.}\ \bibnamefont {Mathews}}, \bibinfo {author} {\bibfnamefont {S.~J.~L.}\ \bibnamefont {Billinge}}, \bibinfo {author} {\bibfnamefont {V.}~\bibnamefont {Petkov}}, \bibinfo {author} {\bibfnamefont {S.}~\bibnamefont {Subramoney}},\ and\ \bibinfo {author} {\bibfnamefont {H.~C.}\ \bibnamefont {Foley}},\ }\bibfield  {title} {\bibinfo {title} {Simulation of nanoporous carbons: {A} chemically constrained structure},\ }\href {https://doi.org/10.1080/13642819908218318} {\bibfield  {journal} {\bibinfo  {journal} {Philosophical Magazine B}\ }\textbf {\bibinfo {volume} {79}},\ \bibinfo {pages} {1499} (\bibinfo {year} {1999})}\BibitemShut {NoStop}%
\bibitem [{\citenamefont {Palmer}\ \emph {et~al.}(2010)\citenamefont {Palmer}, \citenamefont {Llobet}, \citenamefont {Yeon}, \citenamefont {Fischer}, \citenamefont {Shi}, \citenamefont {Gogotsi},\ and\ \citenamefont {Gubbins}}]{palmer_modeling_2010}%
  \BibitemOpen
  \bibfield  {author} {\bibinfo {author} {\bibfnamefont {J.~C.}\ \bibnamefont {Palmer}}, \bibinfo {author} {\bibfnamefont {A.}~\bibnamefont {Llobet}}, \bibinfo {author} {\bibfnamefont {S.~H.}\ \bibnamefont {Yeon}}, \bibinfo {author} {\bibfnamefont {J.~E.}\ \bibnamefont {Fischer}}, \bibinfo {author} {\bibfnamefont {Y.}~\bibnamefont {Shi}}, \bibinfo {author} {\bibfnamefont {Y.}~\bibnamefont {Gogotsi}},\ and\ \bibinfo {author} {\bibfnamefont {K.~E.}\ \bibnamefont {Gubbins}},\ }\bibfield  {title} {\bibinfo {title} {Modeling the structural evolution of carbide-derived carbons using quenched molecular dynamics},\ }\href {https://doi.org/10.1016/j.carbon.2009.11.033} {\bibfield  {journal} {\bibinfo  {journal} {Carbon}\ }\textbf {\bibinfo {volume} {48}},\ \bibinfo {pages} {1116} (\bibinfo {year} {2010})}\BibitemShut {NoStop}%
\bibitem [{\citenamefont {Deringer}\ \emph {et~al.}(2018)\citenamefont {Deringer}, \citenamefont {Merlet}, \citenamefont {Hu}, \citenamefont {Lee}, \citenamefont {Kattirtzi}, \citenamefont {Pecher}, \citenamefont {Csányi}, \citenamefont {Elliott},\ and\ \citenamefont {Grey}}]{deringer_towards_2018}%
  \BibitemOpen
  \bibfield  {author} {\bibinfo {author} {\bibfnamefont {V.~L.}\ \bibnamefont {Deringer}}, \bibinfo {author} {\bibfnamefont {C.}~\bibnamefont {Merlet}}, \bibinfo {author} {\bibfnamefont {Y.}~\bibnamefont {Hu}}, \bibinfo {author} {\bibfnamefont {T.~H.}\ \bibnamefont {Lee}}, \bibinfo {author} {\bibfnamefont {J.~A.}\ \bibnamefont {Kattirtzi}}, \bibinfo {author} {\bibfnamefont {O.}~\bibnamefont {Pecher}}, \bibinfo {author} {\bibfnamefont {G.}~\bibnamefont {Csányi}}, \bibinfo {author} {\bibfnamefont {S.~R.}\ \bibnamefont {Elliott}},\ and\ \bibinfo {author} {\bibfnamefont {C.~P.}\ \bibnamefont {Grey}},\ }\bibfield  {title} {\bibinfo {title} {Towards an atomistic understanding of disordered carbon electrode materials},\ }\href {https://doi.org/10.1039/C8CC01388H} {\bibfield  {journal} {\bibinfo  {journal} {Chemical Communications}\ }\textbf {\bibinfo {volume} {54}},\ \bibinfo {pages} {5988} (\bibinfo {year} {2018})},\ \bibinfo {note} {publisher: The Royal Society of Chemistry}\BibitemShut {NoStop}%
\bibitem [{\citenamefont {de~Tomas}\ \emph {et~al.}(2019)\citenamefont {de~Tomas}, \citenamefont {Aghajamali}, \citenamefont {Jones}, \citenamefont {Lim}, \citenamefont {López}, \citenamefont {Suarez-Martinez},\ and\ \citenamefont {Marks}}]{de_tomas_transferability_2019}%
  \BibitemOpen
  \bibfield  {author} {\bibinfo {author} {\bibfnamefont {C.}~\bibnamefont {de~Tomas}}, \bibinfo {author} {\bibfnamefont {A.}~\bibnamefont {Aghajamali}}, \bibinfo {author} {\bibfnamefont {J.~L.}\ \bibnamefont {Jones}}, \bibinfo {author} {\bibfnamefont {D.~J.}\ \bibnamefont {Lim}}, \bibinfo {author} {\bibfnamefont {M.~J.}\ \bibnamefont {López}}, \bibinfo {author} {\bibfnamefont {I.}~\bibnamefont {Suarez-Martinez}},\ and\ \bibinfo {author} {\bibfnamefont {N.~A.}\ \bibnamefont {Marks}},\ }\bibfield  {title} {\bibinfo {title} {Transferability in interatomic potentials for carbon},\ }\href {https://doi.org/10.1016/j.carbon.2019.07.074} {\bibfield  {journal} {\bibinfo  {journal} {Carbon}\ }\textbf {\bibinfo {volume} {155}},\ \bibinfo {pages} {624} (\bibinfo {year} {2019})}\BibitemShut {NoStop}%
\bibitem [{\citenamefont {de~Tomas}\ \emph {et~al.}(2016)\citenamefont {de~Tomas}, \citenamefont {Suarez-Martinez},\ and\ \citenamefont {Marks}}]{de_tomas_graphitization_2016}%
  \BibitemOpen
  \bibfield  {author} {\bibinfo {author} {\bibfnamefont {C.}~\bibnamefont {de~Tomas}}, \bibinfo {author} {\bibfnamefont {I.}~\bibnamefont {Suarez-Martinez}},\ and\ \bibinfo {author} {\bibfnamefont {N.~A.}\ \bibnamefont {Marks}},\ }\bibfield  {title} {\bibinfo {title} {Graphitization of amorphous carbons: {A} comparative study of interatomic potentials},\ }\href {https://doi.org/10.1016/j.carbon.2016.08.024} {\bibfield  {journal} {\bibinfo  {journal} {Carbon}\ }\textbf {\bibinfo {volume} {109}},\ \bibinfo {pages} {681} (\bibinfo {year} {2016})}\BibitemShut {NoStop}%
\bibitem [{\citenamefont {Allen}\ \emph {et~al.}(1999)\citenamefont {Allen}, \citenamefont {Feldman}, \citenamefont {Fabian},\ and\ \citenamefont {Wooten}}]{allen_diffusons_1999}%
  \BibitemOpen
  \bibfield  {author} {\bibinfo {author} {\bibfnamefont {P.~B.}\ \bibnamefont {Allen}}, \bibinfo {author} {\bibfnamefont {J.~L.}\ \bibnamefont {Feldman}}, \bibinfo {author} {\bibfnamefont {J.}~\bibnamefont {Fabian}},\ and\ \bibinfo {author} {\bibfnamefont {F.}~\bibnamefont {Wooten}},\ }\bibfield  {title} {\bibinfo {title} {Diffusons, {Locons}, {Propagons}: {Character} of {Atomic} {Vibrations} in {Amorphous} {Si}},\ }\href {https://doi.org/10.1080/014186399255836} {\bibfield  {journal} {\bibinfo  {journal} {Philosophical Magazine B}\ }\textbf {\bibinfo {volume} {79}},\ \bibinfo {pages} {1715} (\bibinfo {year} {1999})},\ \bibinfo {note} {arXiv:cond-mat/9907132}\BibitemShut {NoStop}%
\bibitem [{\citenamefont {Fiorentino}\ \emph {et~al.}(2024)\citenamefont {Fiorentino}, \citenamefont {Drigo}, \citenamefont {Baroni},\ and\ \citenamefont {Pegolo}}]{fiorentino_unearthing_2024}%
  \BibitemOpen
  \bibfield  {author} {\bibinfo {author} {\bibfnamefont {A.}~\bibnamefont {Fiorentino}}, \bibinfo {author} {\bibfnamefont {E.}~\bibnamefont {Drigo}}, \bibinfo {author} {\bibfnamefont {S.}~\bibnamefont {Baroni}},\ and\ \bibinfo {author} {\bibfnamefont {P.}~\bibnamefont {Pegolo}},\ }\bibfield  {title} {\bibinfo {title} {Unearthing the foundational role of anharmonicity in heat transport in glasses},\ }\href {https://doi.org/10.1103/PhysRevB.109.224202} {\bibfield  {journal} {\bibinfo  {journal} {Physical Review B}\ }\textbf {\bibinfo {volume} {109}},\ \bibinfo {pages} {224202} (\bibinfo {year} {2024})}\BibitemShut {NoStop}%
\bibitem [{\citenamefont {Puligheddu}\ \emph {et~al.}(2019)\citenamefont {Puligheddu}, \citenamefont {Xia}, \citenamefont {Chan},\ and\ \citenamefont {Galli}}]{puligheddu_computational_2019}%
  \BibitemOpen
  \bibfield  {author} {\bibinfo {author} {\bibfnamefont {M.}~\bibnamefont {Puligheddu}}, \bibinfo {author} {\bibfnamefont {Y.}~\bibnamefont {Xia}}, \bibinfo {author} {\bibfnamefont {M.}~\bibnamefont {Chan}},\ and\ \bibinfo {author} {\bibfnamefont {G.}~\bibnamefont {Galli}},\ }\bibfield  {title} {\bibinfo {title} {Computational prediction of lattice thermal conductivity: {A} comparison of molecular dynamics and {Boltzmann} transport approaches},\ }\href {https://doi.org/10.1103/PhysRevMaterials.3.085401} {\bibfield  {journal} {\bibinfo  {journal} {Physical Review Materials}\ }\textbf {\bibinfo {volume} {3}},\ \bibinfo {pages} {085401} (\bibinfo {year} {2019})}\BibitemShut {NoStop}%
\bibitem [{\citenamefont {Balandin}\ \emph {et~al.}(2008)\citenamefont {Balandin}, \citenamefont {Shamsa}, \citenamefont {Liu}, \citenamefont {Casiraghi},\ and\ \citenamefont {Ferrari}}]{balandin_thermal_2008}%
  \BibitemOpen
  \bibfield  {author} {\bibinfo {author} {\bibfnamefont {A.~A.}\ \bibnamefont {Balandin}}, \bibinfo {author} {\bibfnamefont {M.}~\bibnamefont {Shamsa}}, \bibinfo {author} {\bibfnamefont {W.~L.}\ \bibnamefont {Liu}}, \bibinfo {author} {\bibfnamefont {C.}~\bibnamefont {Casiraghi}},\ and\ \bibinfo {author} {\bibfnamefont {A.~C.}\ \bibnamefont {Ferrari}},\ }\bibfield  {title} {\bibinfo {title} {Thermal conductivity of ultrathin tetrahedral amorphous carbon films},\ }\href {https://doi.org/10.1063/1.2957041} {\bibfield  {journal} {\bibinfo  {journal} {Applied Physics Letters}\ }\textbf {\bibinfo {volume} {93}},\ \bibinfo {pages} {043115} (\bibinfo {year} {2008})}\BibitemShut {NoStop}%
\bibitem [{\citenamefont {Hanus}\ \emph {et~al.}(2021)\citenamefont {Hanus}, \citenamefont {Gurunathan}, \citenamefont {Lindsay}, \citenamefont {Agne}, \citenamefont {Shi}, \citenamefont {Graham},\ and\ \citenamefont {Jeffrey~Snyder}}]{hanus_thermal_2021}%
  \BibitemOpen
  \bibfield  {author} {\bibinfo {author} {\bibfnamefont {R.}~\bibnamefont {Hanus}}, \bibinfo {author} {\bibfnamefont {R.}~\bibnamefont {Gurunathan}}, \bibinfo {author} {\bibfnamefont {L.}~\bibnamefont {Lindsay}}, \bibinfo {author} {\bibfnamefont {M.~T.}\ \bibnamefont {Agne}}, \bibinfo {author} {\bibfnamefont {J.}~\bibnamefont {Shi}}, \bibinfo {author} {\bibfnamefont {S.}~\bibnamefont {Graham}},\ and\ \bibinfo {author} {\bibfnamefont {G.}~\bibnamefont {Jeffrey~Snyder}},\ }\bibfield  {title} {\bibinfo {title} {Thermal transport in defective and disordered materials},\ }\href {https://doi.org/10.1063/5.0055593} {\bibfield  {journal} {\bibinfo  {journal} {Applied Physics Reviews}\ }\textbf {\bibinfo {volume} {8}},\ \bibinfo {pages} {031311} (\bibinfo {year} {2021})}\BibitemShut {NoStop}%
\bibitem [{\citenamefont {Jurkiewicz}\ \emph {et~al.}(2017)\citenamefont {Jurkiewicz}, \citenamefont {Duber}, \citenamefont {Fischer},\ and\ \citenamefont {Burian}}]{jurkiewicz_modelling_2017}%
  \BibitemOpen
  \bibfield  {author} {\bibinfo {author} {\bibfnamefont {K.}~\bibnamefont {Jurkiewicz}}, \bibinfo {author} {\bibfnamefont {S.}~\bibnamefont {Duber}}, \bibinfo {author} {\bibfnamefont {H.}~\bibnamefont {Fischer}},\ and\ \bibinfo {author} {\bibfnamefont {A.}~\bibnamefont {Burian}},\ }\bibfield  {title} {\bibinfo {title} {Modelling of glass-like carbon structure and its experimental verification by neutron and {X}-ray diffraction},\ }\href {https://doi.org/10.1107/S1600576716017660} {\bibfield  {journal} {\bibinfo  {journal} {Journal of Applied Crystallography}\ }\textbf {\bibinfo {volume} {50}},\ \bibinfo {pages} {36} (\bibinfo {year} {2017})}\BibitemShut {NoStop}%
\bibitem [{\citenamefont {Elliott}(1991)}]{elliott_medium-range_1991}%
  \BibitemOpen
  \bibfield  {author} {\bibinfo {author} {\bibfnamefont {S.~R.}\ \bibnamefont {Elliott}},\ }\bibfield  {title} {\bibinfo {title} {Medium-range structural order in covalent amorphous solids},\ }\href {https://doi.org/10.1038/354445a0} {\bibfield  {journal} {\bibinfo  {journal} {Nature}\ }\textbf {\bibinfo {volume} {354}},\ \bibinfo {pages} {445} (\bibinfo {year} {1991})}\BibitemShut {NoStop}%
\bibitem [{\citenamefont {Wei}\ \emph {et~al.}(2019)\citenamefont {Wei}, \citenamefont {Yang}, \citenamefont {Jiang}, \citenamefont {Dai}, \citenamefont {Wang}, \citenamefont {Dyre}, \citenamefont {Douglass},\ and\ \citenamefont {Harrowell}}]{wei_assessing_2019}%
  \BibitemOpen
  \bibfield  {author} {\bibinfo {author} {\bibfnamefont {D.}~\bibnamefont {Wei}}, \bibinfo {author} {\bibfnamefont {J.}~\bibnamefont {Yang}}, \bibinfo {author} {\bibfnamefont {M.-Q.}\ \bibnamefont {Jiang}}, \bibinfo {author} {\bibfnamefont {L.-H.}\ \bibnamefont {Dai}}, \bibinfo {author} {\bibfnamefont {Y.-J.}\ \bibnamefont {Wang}}, \bibinfo {author} {\bibfnamefont {J.~C.}\ \bibnamefont {Dyre}}, \bibinfo {author} {\bibfnamefont {I.}~\bibnamefont {Douglass}},\ and\ \bibinfo {author} {\bibfnamefont {P.}~\bibnamefont {Harrowell}},\ }\bibfield  {title} {\bibinfo {title} {Assessing the utility of structure in amorphous materials},\ }\href {https://doi.org/10.1063/1.5064531} {\bibfield  {journal} {\bibinfo  {journal} {The Journal of Chemical Physics}\ }\textbf {\bibinfo {volume} {150}},\ \bibinfo {pages} {114502} (\bibinfo {year} {2019})}\BibitemShut {NoStop}%
\bibitem [{\citenamefont {Schwalbe-Koda}\ \emph {et~al.}(2025)\citenamefont {Schwalbe-Koda}, \citenamefont {Hamel}, \citenamefont {Sadigh}, \citenamefont {Zhou},\ and\ \citenamefont {Lordi}}]{schwalbe-koda_model-free_2025}%
  \BibitemOpen
  \bibfield  {author} {\bibinfo {author} {\bibfnamefont {D.}~\bibnamefont {Schwalbe-Koda}}, \bibinfo {author} {\bibfnamefont {S.}~\bibnamefont {Hamel}}, \bibinfo {author} {\bibfnamefont {B.}~\bibnamefont {Sadigh}}, \bibinfo {author} {\bibfnamefont {F.}~\bibnamefont {Zhou}},\ and\ \bibinfo {author} {\bibfnamefont {V.}~\bibnamefont {Lordi}},\ }\bibfield  {title} {\bibinfo {title} {Model-free estimation of completeness, uncertainties, and outliers in atomistic machine learning using information theory},\ }\href {https://doi.org/10.1038/s41467-025-59232-0} {\bibfield  {journal} {\bibinfo  {journal} {Nature Communications}\ }\textbf {\bibinfo {volume} {16}},\ \bibinfo {pages} {4014} (\bibinfo {year} {2025})},\ \bibinfo {note} {publisher: Nature Publishing Group}\BibitemShut {NoStop}%
\bibitem [{\citenamefont {Vink}\ and\ \citenamefont {Barkema}(2002)}]{vink_configurational_2002}%
  \BibitemOpen
  \bibfield  {author} {\bibinfo {author} {\bibfnamefont {R.~L.~C.}\ \bibnamefont {Vink}}\ and\ \bibinfo {author} {\bibfnamefont {G.~T.}\ \bibnamefont {Barkema}},\ }\bibfield  {title} {\bibinfo {title} {Configurational {Entropy} of {Network}-{Forming} {Materials}},\ }\href {https://doi.org/10.1103/PhysRevLett.89.076405} {\bibfield  {journal} {\bibinfo  {journal} {Physical Review Letters}\ }\textbf {\bibinfo {volume} {89}},\ \bibinfo {pages} {076405} (\bibinfo {year} {2002})},\ \bibinfo {note} {publisher: American Physical Society}\BibitemShut {NoStop}%
\bibitem [{\citenamefont {Mason}\ \emph {et~al.}(2012)\citenamefont {Mason}, \citenamefont {Lazar}, \citenamefont {MacPherson},\ and\ \citenamefont {Srolovitz}}]{masonStatisticalTopologyCellular2012}%
  \BibitemOpen
  \bibfield  {author} {\bibinfo {author} {\bibfnamefont {J.~K.}\ \bibnamefont {Mason}}, \bibinfo {author} {\bibfnamefont {E.~A.}\ \bibnamefont {Lazar}}, \bibinfo {author} {\bibfnamefont {R.~D.}\ \bibnamefont {MacPherson}},\ and\ \bibinfo {author} {\bibfnamefont {D.~J.}\ \bibnamefont {Srolovitz}},\ }\bibfield  {title} {\bibinfo {title} {Statistical topology of cellular networks in two and three dimensions},\ }\href {https://doi.org/10.1103/PhysRevE.86.051128} {\bibfield  {journal} {\bibinfo  {journal} {Physical Review E}\ }\textbf {\bibinfo {volume} {86}},\ \bibinfo {pages} {051128} (\bibinfo {year} {2012})}\BibitemShut {NoStop}%
\bibitem [{\citenamefont {Milkus}\ \emph {et~al.}(2018)\citenamefont {Milkus}, \citenamefont {Ness}, \citenamefont {Palyulin}, \citenamefont {Weber}, \citenamefont {Lapkin},\ and\ \citenamefont {Zaccone}}]{milkus_interpretation_2018}%
  \BibitemOpen
  \bibfield  {author} {\bibinfo {author} {\bibfnamefont {R.}~\bibnamefont {Milkus}}, \bibinfo {author} {\bibfnamefont {C.}~\bibnamefont {Ness}}, \bibinfo {author} {\bibfnamefont {V.~V.}\ \bibnamefont {Palyulin}}, \bibinfo {author} {\bibfnamefont {J.}~\bibnamefont {Weber}}, \bibinfo {author} {\bibfnamefont {A.}~\bibnamefont {Lapkin}},\ and\ \bibinfo {author} {\bibfnamefont {A.}~\bibnamefont {Zaccone}},\ }\bibfield  {title} {\bibinfo {title} {Interpretation of the {Vibrational} {Spectra} of {Glassy} {Polymers} {Using} {Coarse}-{Grained} {Simulations}},\ }\href {https://doi.org/10.1021/acs.macromol.7b02352} {\bibfield  {journal} {\bibinfo  {journal} {Macromolecules}\ }\textbf {\bibinfo {volume} {51}},\ \bibinfo {pages} {1559} (\bibinfo {year} {2018})},\ \bibinfo {note} {publisher: American Chemical Society}\BibitemShut {NoStop}%
\bibitem [{\citenamefont {Wallace}(1972)}]{wallace_thermodynamics_1972}%
  \BibitemOpen
  \bibfield  {author} {\bibinfo {author} {\bibfnamefont {D.~C.}\ \bibnamefont {Wallace}},\ }\href@noop {} {\emph {\bibinfo {title} {Thermodynamics of {Crystals}}}}\ (\bibinfo {year} {1972})\BibitemShut {NoStop}%
\bibitem [{\citenamefont {Kittel}(1949)}]{kittel_interpretation_1949}%
  \BibitemOpen
  \bibfield  {author} {\bibinfo {author} {\bibfnamefont {C.}~\bibnamefont {Kittel}},\ }\bibfield  {title} {\bibinfo {title} {Interpretation of the {Thermal} {Conductivity} of {Glasses}},\ }\href {https://doi.org/10.1103/PhysRev.75.972} {\bibfield  {journal} {\bibinfo  {journal} {Physical Review}\ }\textbf {\bibinfo {volume} {75}},\ \bibinfo {pages} {972} (\bibinfo {year} {1949})},\ \bibinfo {note} {publisher: American Physical Society}\BibitemShut {NoStop}%
\bibitem [{\citenamefont {Casimir}(1938)}]{casimir_note_1938}%
  \BibitemOpen
  \bibfield  {author} {\bibinfo {author} {\bibfnamefont {H.~B.~G.}\ \bibnamefont {Casimir}},\ }\bibfield  {title} {\bibinfo {title} {Note on the conduction of heat in crystals},\ }\href {https://doi.org/10.1016/S0031-8914(38)80162-2} {\bibfield  {journal} {\bibinfo  {journal} {Physica}\ }\textbf {\bibinfo {volume} {5}},\ \bibinfo {pages} {495} (\bibinfo {year} {1938})}\BibitemShut {NoStop}%
\bibitem [{\citenamefont {Peierls}(2001)}]{peierls1955quantum}%
  \BibitemOpen
  \bibfield  {author} {\bibinfo {author} {\bibfnamefont {R.~E.}\ \bibnamefont {Peierls}},\ }\href@noop {} {\emph {\bibinfo {title} {Quantum theory of solids}}}\ (\bibinfo  {publisher} {Oxford Classics Series},\ \bibinfo {year} {2001})\BibitemShut {NoStop}%
\bibitem [{\citenamefont {MARADUDIN}\ and\ \citenamefont {VOSKO}(1968)}]{maradudin_symmetry_1968}%
  \BibitemOpen
  \bibfield  {author} {\bibinfo {author} {\bibfnamefont {A.~A.}\ \bibnamefont {MARADUDIN}}\ and\ \bibinfo {author} {\bibfnamefont {S.~H.}\ \bibnamefont {VOSKO}},\ }\bibfield  {title} {\bibinfo {title} {Symmetry {Properties} of the {Normal} {Vibrations} of a {Crystal}},\ }\href {https://doi.org/10.1103/RevModPhys.40.1} {\bibfield  {journal} {\bibinfo  {journal} {Reviews of Modern Physics}\ }\textbf {\bibinfo {volume} {40}},\ \bibinfo {pages} {1} (\bibinfo {year} {1968})},\ \bibinfo {note} {publisher: American Physical Society}\BibitemShut {NoStop}%
\bibitem [{\citenamefont {Prat}\ \emph {et~al.}(2016)\citenamefont {Prat}, \citenamefont {Cherroret},\ and\ \citenamefont {Delande}}]{prat_semiclassical_2016}%
  \BibitemOpen
  \bibfield  {author} {\bibinfo {author} {\bibfnamefont {T.}~\bibnamefont {Prat}}, \bibinfo {author} {\bibfnamefont {N.}~\bibnamefont {Cherroret}},\ and\ \bibinfo {author} {\bibfnamefont {D.}~\bibnamefont {Delande}},\ }\bibfield  {title} {\bibinfo {title} {Semiclassical spectral function and density of states in speckle potentials},\ }\href {https://doi.org/10.1103/PhysRevA.94.022114} {\bibfield  {journal} {\bibinfo  {journal} {Physical Review A}\ }\textbf {\bibinfo {volume} {94}},\ \bibinfo {pages} {022114} (\bibinfo {year} {2016})},\ \bibinfo {note} {publisher: American Physical Society}\BibitemShut {NoStop}%
\bibitem [{\citenamefont {Chandrasekaran}\ and\ \citenamefont {Betouras}(2022)}]{chandrasekaran_effect_2022}%
  \BibitemOpen
  \bibfield  {author} {\bibinfo {author} {\bibfnamefont {A.}~\bibnamefont {Chandrasekaran}}\ and\ \bibinfo {author} {\bibfnamefont {J.~J.}\ \bibnamefont {Betouras}},\ }\bibfield  {title} {\bibinfo {title} {Effect of disorder on density of states and conductivity in higher-order {Van} {Hove} singularities in two-dimensional bands},\ }\href {https://doi.org/10.1103/PhysRevB.105.075144} {\bibfield  {journal} {\bibinfo  {journal} {Physical Review B}\ }\textbf {\bibinfo {volume} {105}},\ \bibinfo {pages} {075144} (\bibinfo {year} {2022})},\ \bibinfo {note} {publisher: American Physical Society}\BibitemShut {NoStop}%
\bibitem [{Note1()}]{Note1}%
  \BibitemOpen
  \bibinfo {note} {This statement is valid in the Lorentzian spectral function regime where the WTE can be applied \cite {caldarelli_many-body_2022,simoncelli_wigner_2022}.}\BibitemShut {Stop}%
\bibitem [{\citenamefont {Simkin}\ and\ \citenamefont {Mahan}(2000)}]{simkin_minimum_2000}%
  \BibitemOpen
  \bibfield  {author} {\bibinfo {author} {\bibfnamefont {M.~V.}\ \bibnamefont {Simkin}}\ and\ \bibinfo {author} {\bibfnamefont {G.~D.}\ \bibnamefont {Mahan}},\ }\bibfield  {title} {\bibinfo {title} {Minimum {Thermal} {Conductivity} of {Superlattices}},\ }\href {https://doi.org/10.1103/PhysRevLett.84.927} {\bibfield  {journal} {\bibinfo  {journal} {Physical Review Letters}\ }\textbf {\bibinfo {volume} {84}},\ \bibinfo {pages} {927} (\bibinfo {year} {2000})}\BibitemShut {NoStop}%
\bibitem [{\citenamefont {{J. M. Ziman}}(1960)}]{ziman_electrons_1960}%
  \BibitemOpen
  \bibfield  {author} {\bibinfo {author} {\bibnamefont {{J. M. Ziman}}},\ }\href@noop {} {\emph {\bibinfo {title} {Electrons and {Phonons}: {The} {Theory} of {Transport} Phenomena in Solids}}}\ (\bibinfo {year} {1960})\BibitemShut {NoStop}%
\bibitem [{\citenamefont {Osipov}\ and\ \citenamefont {Krasavin}(1998)}]{osipov_disclination_1998}%
  \BibitemOpen
  \bibfield  {author} {\bibinfo {author} {\bibfnamefont {V.~A.}\ \bibnamefont {Osipov}}\ and\ \bibinfo {author} {\bibfnamefont {S.~E.}\ \bibnamefont {Krasavin}},\ }\bibfield  {title} {\bibinfo {title} {Disclination dipoles as the basic structural elements of dielectric glasses},\ }\href {https://doi.org/10.1016/S0375-9601(98)00841-X} {\bibfield  {journal} {\bibinfo  {journal} {Physics Letters A}\ }\textbf {\bibinfo {volume} {250}},\ \bibinfo {pages} {369} (\bibinfo {year} {1998})}\BibitemShut {NoStop}%
\bibitem [{\citenamefont {Karthik}\ \emph {et~al.}(2011)\citenamefont {Karthik}, \citenamefont {Kane}, \citenamefont {Butt}, \citenamefont {Windes},\ and\ \citenamefont {Ubic}}]{karthik_situ_2011}%
  \BibitemOpen
  \bibfield  {author} {\bibinfo {author} {\bibfnamefont {C.}~\bibnamefont {Karthik}}, \bibinfo {author} {\bibfnamefont {J.}~\bibnamefont {Kane}}, \bibinfo {author} {\bibfnamefont {D.~P.}\ \bibnamefont {Butt}}, \bibinfo {author} {\bibfnamefont {W.~E.}\ \bibnamefont {Windes}},\ and\ \bibinfo {author} {\bibfnamefont {R.}~\bibnamefont {Ubic}},\ }\bibfield  {title} {\bibinfo {title} {In situ transmission electron microscopy of electron-beam induced damage process in nuclear grade graphite},\ }\href {https://doi.org/10.1016/j.jnucmat.2011.03.024} {\bibfield  {journal} {\bibinfo  {journal} {Journal of Nuclear Materials}\ }\textbf {\bibinfo {volume} {412}},\ \bibinfo {pages} {321} (\bibinfo {year} {2011})}\BibitemShut {NoStop}%
\bibitem [{\citenamefont {Tamura}(1983)}]{tamura_isotope_1983}%
  \BibitemOpen
  \bibfield  {author} {\bibinfo {author} {\bibfnamefont {S.-i.}\ \bibnamefont {Tamura}},\ }\bibfield  {title} {\bibinfo {title} {Isotope scattering of dispersive phonons in {Ge}},\ }\href {https://doi.org/10.1103/PhysRevB.27.858} {\bibfield  {journal} {\bibinfo  {journal} {Physical Review B}\ }\textbf {\bibinfo {volume} {27}},\ \bibinfo {pages} {858} (\bibinfo {year} {1983})},\ \bibinfo {note} {publisher: American Physical Society}\BibitemShut {NoStop}%
\bibitem [{Note2()}]{Note2}%
  \BibitemOpen
  \bibinfo {note} {Specifically, the form of the factor $a$ derives from: (i) considering an elementary Debye model having speed of sound $v_{\protect \rm sound}$ and Debye frequency $\omega _D^3 = 6 \pi ^2 \protect \frac {N_{at}}{\protect \mathcal {V}} v_{\protect \rm sound}^3$; (ii) assuming the main change in the frequencies when lowering the density of graphite is mainly due to changes in the volume and has negligible effect on $v_{\protect \rm sound}$. This implies that the frequencies scale with the cube root of the density: $\omega \propto \rho ^{1/3}$.}\BibitemShut {Stop}%
\bibitem [{\citenamefont {Conyuh}\ and\ \citenamefont {Beltukov}(2021)}]{conyuh_random_2021}%
  \BibitemOpen
  \bibfield  {author} {\bibinfo {author} {\bibfnamefont {D.~A.}\ \bibnamefont {Conyuh}}\ and\ \bibinfo {author} {\bibfnamefont {Y.~M.}\ \bibnamefont {Beltukov}},\ }\bibfield  {title} {\bibinfo {title} {Random matrix approach to the boson peak and {Ioffe}-{Regel} criterion in amorphous solids},\ }\href {https://doi.org/10.1103/PhysRevB.103.104204} {\bibfield  {journal} {\bibinfo  {journal} {Physical Review B}\ }\textbf {\bibinfo {volume} {103}},\ \bibinfo {pages} {104204} (\bibinfo {year} {2021})}\BibitemShut {NoStop}%
\bibitem [{\citenamefont {Ciarella}\ \emph {et~al.}(2023)\citenamefont {Ciarella}, \citenamefont {Khomenko}, \citenamefont {Berthier}, \citenamefont {Mocanu}, \citenamefont {Reichman}, \citenamefont {Scalliet},\ and\ \citenamefont {Zamponi}}]{ciarella_finding_2023}%
  \BibitemOpen
  \bibfield  {author} {\bibinfo {author} {\bibfnamefont {S.}~\bibnamefont {Ciarella}}, \bibinfo {author} {\bibfnamefont {D.}~\bibnamefont {Khomenko}}, \bibinfo {author} {\bibfnamefont {L.}~\bibnamefont {Berthier}}, \bibinfo {author} {\bibfnamefont {F.~C.}\ \bibnamefont {Mocanu}}, \bibinfo {author} {\bibfnamefont {D.~R.}\ \bibnamefont {Reichman}}, \bibinfo {author} {\bibfnamefont {C.}~\bibnamefont {Scalliet}},\ and\ \bibinfo {author} {\bibfnamefont {F.}~\bibnamefont {Zamponi}},\ }\bibfield  {title} {\bibinfo {title} {Finding defects in glasses through machine learning},\ }\href {https://doi.org/10.1038/s41467-023-39948-7} {\bibfield  {journal} {\bibinfo  {journal} {Nature Communications}\ }\textbf {\bibinfo {volume} {14}},\ \bibinfo {pages} {4229} (\bibinfo {year} {2023})},\ \bibinfo {note} {publisher: Nature Publishing Group}\BibitemShut {NoStop}%
\bibitem [{\citenamefont {Billinge}\ and\ \citenamefont {Levin}(2007)}]{billinge_problem_2007}%
  \BibitemOpen
  \bibfield  {author} {\bibinfo {author} {\bibfnamefont {S.~J.~L.}\ \bibnamefont {Billinge}}\ and\ \bibinfo {author} {\bibfnamefont {I.}~\bibnamefont {Levin}},\ }\bibfield  {title} {\bibinfo {title} {The {Problem} with {Determining} {Atomic} {Structure} at the {Nanoscale}},\ }\href {https://doi.org/10.1126/science.1135080} {\bibfield  {journal} {\bibinfo  {journal} {Science}\ }\textbf {\bibinfo {volume} {316}},\ \bibinfo {pages} {561} (\bibinfo {year} {2007})},\ \bibinfo {note} {publisher: American Association for the Advancement of Science}\BibitemShut {NoStop}%
\bibitem [{\citenamefont {Luo}\ \emph {et~al.}(2020)\citenamefont {Luo}, \citenamefont {Yang}, \citenamefont {Feng}, \citenamefont {Wang},\ and\ \citenamefont {Ruan}}]{luo_vibrational_2020}%
  \BibitemOpen
  \bibfield  {author} {\bibinfo {author} {\bibfnamefont {Y.}~\bibnamefont {Luo}}, \bibinfo {author} {\bibfnamefont {X.}~\bibnamefont {Yang}}, \bibinfo {author} {\bibfnamefont {T.}~\bibnamefont {Feng}}, \bibinfo {author} {\bibfnamefont {J.}~\bibnamefont {Wang}},\ and\ \bibinfo {author} {\bibfnamefont {X.}~\bibnamefont {Ruan}},\ }\bibfield  {title} {\bibinfo {title} {Vibrational hierarchy leads to dual-phonon transport in low thermal conductivity crystals},\ }\href {https://doi.org/10.1038/s41467-020-16371-w} {\bibfield  {journal} {\bibinfo  {journal} {Nature Communications}\ }\textbf {\bibinfo {volume} {11}},\ \bibinfo {pages} {2554} (\bibinfo {year} {2020})}\BibitemShut {NoStop}%
\bibitem [{\citenamefont {Legenstein}\ \emph {et~al.}(2025)\citenamefont {Legenstein}, \citenamefont {Reicht}, \citenamefont {Wieser}, \citenamefont {Simoncelli},\ and\ \citenamefont {Zojer}}]{legenstein_heat_2025}%
  \BibitemOpen
  \bibfield  {author} {\bibinfo {author} {\bibfnamefont {L.}~\bibnamefont {Legenstein}}, \bibinfo {author} {\bibfnamefont {L.}~\bibnamefont {Reicht}}, \bibinfo {author} {\bibfnamefont {S.}~\bibnamefont {Wieser}}, \bibinfo {author} {\bibfnamefont {M.}~\bibnamefont {Simoncelli}},\ and\ \bibinfo {author} {\bibfnamefont {E.}~\bibnamefont {Zojer}},\ }\bibfield  {title} {\bibinfo {title} {Heat transport in crystalline organic semiconductors: coexistence of phonon propagation and tunneling},\ }\href {https://doi.org/10.1038/s41524-025-01514-8} {\bibfield  {journal} {\bibinfo  {journal} {npj Computational Materials}\ }\textbf {\bibinfo {volume} {11}},\ \bibinfo {pages} {29} (\bibinfo {year} {2025})},\ \bibinfo {note} {publisher: Nature Publishing Group}\BibitemShut {NoStop}%
\bibitem [{\citenamefont {Hurley}\ \emph {et~al.}(2022)\citenamefont {Hurley}, \citenamefont {El-Azab}, \citenamefont {Bryan}, \citenamefont {Cooper}, \citenamefont {Dennett}, \citenamefont {Gofryk}, \citenamefont {He}, \citenamefont {Khafizov}, \citenamefont {Lander}, \citenamefont {Manley}, \citenamefont {Mann}, \citenamefont {Marianetti}, \citenamefont {Rickert}, \citenamefont {Selim}, \citenamefont {Tonks},\ and\ \citenamefont {Wharry}}]{hurley_thermal_2022}%
  \BibitemOpen
  \bibfield  {author} {\bibinfo {author} {\bibfnamefont {D.~H.}\ \bibnamefont {Hurley}}, \bibinfo {author} {\bibfnamefont {A.}~\bibnamefont {El-Azab}}, \bibinfo {author} {\bibfnamefont {M.~S.}\ \bibnamefont {Bryan}}, \bibinfo {author} {\bibfnamefont {M.~W.~D.}\ \bibnamefont {Cooper}}, \bibinfo {author} {\bibfnamefont {C.~A.}\ \bibnamefont {Dennett}}, \bibinfo {author} {\bibfnamefont {K.}~\bibnamefont {Gofryk}}, \bibinfo {author} {\bibfnamefont {L.}~\bibnamefont {He}}, \bibinfo {author} {\bibfnamefont {M.}~\bibnamefont {Khafizov}}, \bibinfo {author} {\bibfnamefont {G.~H.}\ \bibnamefont {Lander}}, \bibinfo {author} {\bibfnamefont {M.~E.}\ \bibnamefont {Manley}}, \bibinfo {author} {\bibfnamefont {J.~M.}\ \bibnamefont {Mann}}, \bibinfo {author} {\bibfnamefont {C.~A.}\ \bibnamefont {Marianetti}}, \bibinfo {author} {\bibfnamefont {K.}~\bibnamefont {Rickert}}, \bibinfo {author} {\bibfnamefont {F.~A.}\ \bibnamefont {Selim}}, \bibinfo {author} {\bibfnamefont {M.~R.}\ \bibnamefont {Tonks}},\ and\ \bibinfo {author} {\bibfnamefont {J.~P.}\ \bibnamefont {Wharry}},\ }\bibfield  {title} {\bibinfo {title} {Thermal {Energy} {Transport} in {Oxide} {Nuclear} {Fuel}},\ }\href {https://doi.org/10.1021/acs.chemrev.1c00262} {\bibfield  {journal} {\bibinfo  {journal} {Chemical Reviews}\ }\textbf {\bibinfo {volume} {122}},\ \bibinfo {pages} {3711} (\bibinfo {year} {2022})},\ \bibinfo {note} {publisher: American Chemical Society}\BibitemShut {NoStop}%
\bibitem [{\citenamefont {Dennett}\ \emph {et~al.}(2021)\citenamefont {Dennett}, \citenamefont {Deskins}, \citenamefont {Khafizov}, \citenamefont {Hua}, \citenamefont {Khanolkar}, \citenamefont {Bawane}, \citenamefont {Fu}, \citenamefont {Mann}, \citenamefont {Marianetti}, \citenamefont {He}, \citenamefont {Hurley},\ and\ \citenamefont {El-Azab}}]{dennett_integrated_2021}%
  \BibitemOpen
  \bibfield  {author} {\bibinfo {author} {\bibfnamefont {C.~A.}\ \bibnamefont {Dennett}}, \bibinfo {author} {\bibfnamefont {W.~R.}\ \bibnamefont {Deskins}}, \bibinfo {author} {\bibfnamefont {M.}~\bibnamefont {Khafizov}}, \bibinfo {author} {\bibfnamefont {Z.}~\bibnamefont {Hua}}, \bibinfo {author} {\bibfnamefont {A.}~\bibnamefont {Khanolkar}}, \bibinfo {author} {\bibfnamefont {K.}~\bibnamefont {Bawane}}, \bibinfo {author} {\bibfnamefont {L.}~\bibnamefont {Fu}}, \bibinfo {author} {\bibfnamefont {J.~M.}\ \bibnamefont {Mann}}, \bibinfo {author} {\bibfnamefont {C.~A.}\ \bibnamefont {Marianetti}}, \bibinfo {author} {\bibfnamefont {L.}~\bibnamefont {He}}, \bibinfo {author} {\bibfnamefont {D.~H.}\ \bibnamefont {Hurley}},\ and\ \bibinfo {author} {\bibfnamefont {A.}~\bibnamefont {El-Azab}},\ }\bibfield  {title} {\bibinfo {title} {An integrated experimental and computational investigation of defect and microstructural effects on thermal transport in thorium dioxide},\ }\href {https://doi.org/10.1016/j.actamat.2021.116934} {\bibfield  {journal} {\bibinfo  {journal} {Acta Materialia}\ }\textbf {\bibinfo {volume} {213}},\ \bibinfo {pages} {116934} (\bibinfo {year} {2021})}\BibitemShut {NoStop}%
\bibitem [{\citenamefont {Islamov}\ \emph {et~al.}(2023)\citenamefont {Islamov}, \citenamefont {Babaei}, \citenamefont {Anderson}, \citenamefont {Sezginel}, \citenamefont {Long}, \citenamefont {McGaughey}, \citenamefont {Gomez-Gualdron},\ and\ \citenamefont {Wilmer}}]{islamov_high-throughput_2023}%
  \BibitemOpen
  \bibfield  {author} {\bibinfo {author} {\bibfnamefont {M.}~\bibnamefont {Islamov}}, \bibinfo {author} {\bibfnamefont {H.}~\bibnamefont {Babaei}}, \bibinfo {author} {\bibfnamefont {R.}~\bibnamefont {Anderson}}, \bibinfo {author} {\bibfnamefont {K.~B.}\ \bibnamefont {Sezginel}}, \bibinfo {author} {\bibfnamefont {J.~R.}\ \bibnamefont {Long}}, \bibinfo {author} {\bibfnamefont {A.~J.~H.}\ \bibnamefont {McGaughey}}, \bibinfo {author} {\bibfnamefont {D.~A.}\ \bibnamefont {Gomez-Gualdron}},\ and\ \bibinfo {author} {\bibfnamefont {C.~E.}\ \bibnamefont {Wilmer}},\ }\bibfield  {title} {\bibinfo {title} {High-throughput screening of hypothetical metal-organic frameworks for thermal conductivity},\ }\href {https://doi.org/10.1038/s41524-022-00961-x} {\bibfield  {journal} {\bibinfo  {journal} {npj Computational Materials}\ }\textbf {\bibinfo {volume} {9}},\ \bibinfo {pages} {1} (\bibinfo {year} {2023})},\ \bibinfo {note} {publisher: Nature Publishing Group}\BibitemShut {NoStop}%
\bibitem [{\citenamefont {Haque}\ \emph {et~al.}(2020)\citenamefont {Haque}, \citenamefont {Kee}, \citenamefont {Villalva}, \citenamefont {Ong},\ and\ \citenamefont {Baran}}]{haque_halide_2020}%
  \BibitemOpen
  \bibfield  {author} {\bibinfo {author} {\bibfnamefont {M.~A.}\ \bibnamefont {Haque}}, \bibinfo {author} {\bibfnamefont {S.}~\bibnamefont {Kee}}, \bibinfo {author} {\bibfnamefont {D.~R.}\ \bibnamefont {Villalva}}, \bibinfo {author} {\bibfnamefont {W.-L.}\ \bibnamefont {Ong}},\ and\ \bibinfo {author} {\bibfnamefont {D.}~\bibnamefont {Baran}},\ }\bibfield  {title} {\bibinfo {title} {Halide {Perovskites}: {Thermal} {Transport} and {Prospects} for {Thermoelectricity}},\ }\href {https://doi.org/10.1002/advs.201903389} {\bibfield  {journal} {\bibinfo  {journal} {Advanced Science}\ }\textbf {\bibinfo {volume} {7}},\ \bibinfo {pages} {1903389} (\bibinfo {year} {2020})}\BibitemShut {NoStop}%
\bibitem [{\citenamefont {Bosoni}\ \emph {et~al.}(2020)\citenamefont {Bosoni}, \citenamefont {Campi}, \citenamefont {Donadio}, \citenamefont {Sosso}, \citenamefont {Behler},\ and\ \citenamefont {Bernasconi}}]{bosoni_atomistic_2020}%
  \BibitemOpen
  \bibfield  {author} {\bibinfo {author} {\bibfnamefont {E.}~\bibnamefont {Bosoni}}, \bibinfo {author} {\bibfnamefont {D.}~\bibnamefont {Campi}}, \bibinfo {author} {\bibfnamefont {D.}~\bibnamefont {Donadio}}, \bibinfo {author} {\bibfnamefont {G.~C.}\ \bibnamefont {Sosso}}, \bibinfo {author} {\bibfnamefont {J.}~\bibnamefont {Behler}},\ and\ \bibinfo {author} {\bibfnamefont {M.}~\bibnamefont {Bernasconi}},\ }\bibfield  {title} {\bibinfo {title} {Atomistic simulations of thermal conductivity in {GeTe} nanowires},\ }\href@noop {} {\bibfield  {journal} {\bibinfo  {journal} {J. Phys. D}\ }\textbf {\bibinfo {volume} {53}},\ \bibinfo {pages} {054001} (\bibinfo {year} {2020})}\BibitemShut {NoStop}%
\bibitem [{\citenamefont {Jain}(2020)}]{PhysRevB.102.201201}%
  \BibitemOpen
  \bibfield  {author} {\bibinfo {author} {\bibfnamefont {A.}~\bibnamefont {Jain}},\ }\bibfield  {title} {\bibinfo {title} {{Multichannel thermal transport in crystalline ${\mathrm{Tl}}_{3}{\mathrm{VSe}}_{4}$}},\ }\href {https://doi.org/10.1103/PhysRevB.102.201201} {\bibfield  {journal} {\bibinfo  {journal} {Phys. Rev. B}\ }\textbf {\bibinfo {volume} {102}},\ \bibinfo {pages} {201201} (\bibinfo {year} {2020})}\BibitemShut {NoStop}%
\bibitem [{\citenamefont {Verdi}\ \emph {et~al.}(2021)\citenamefont {Verdi}, \citenamefont {Karsai}, \citenamefont {Liu}, \citenamefont {Jinnouchi},\ and\ \citenamefont {Kresse}}]{verdi2021thermal}%
  \BibitemOpen
  \bibfield  {author} {\bibinfo {author} {\bibfnamefont {C.}~\bibnamefont {Verdi}}, \bibinfo {author} {\bibfnamefont {F.}~\bibnamefont {Karsai}}, \bibinfo {author} {\bibfnamefont {P.}~\bibnamefont {Liu}}, \bibinfo {author} {\bibfnamefont {R.}~\bibnamefont {Jinnouchi}},\ and\ \bibinfo {author} {\bibfnamefont {G.}~\bibnamefont {Kresse}},\ }\bibfield  {title} {\bibinfo {title} {Thermal transport and phase transitions of zirconia by on-the-fly machine-learned interatomic potentials},\ }\href@noop {} {\bibfield  {journal} {\bibinfo  {journal} {Npj Comput. Mater.}\ }\textbf {\bibinfo {volume} {7}},\ \bibinfo {pages} {1} (\bibinfo {year} {2021})}\BibitemShut {NoStop}%
\bibitem [{\citenamefont {Yang}\ \emph {et~al.}(2022)\citenamefont {Yang}, \citenamefont {Jain},\ and\ \citenamefont {Ong}}]{yang_inter-channel_2022}%
  \BibitemOpen
  \bibfield  {author} {\bibinfo {author} {\bibfnamefont {J.}~\bibnamefont {Yang}}, \bibinfo {author} {\bibfnamefont {A.}~\bibnamefont {Jain}},\ and\ \bibinfo {author} {\bibfnamefont {W.-L.}\ \bibnamefont {Ong}},\ }\bibfield  {title} {\bibinfo {title} {Inter-channel conversion between population-/coherence-channel dictates thermal transport in {MAPbI3} crystals},\ }\href {https://doi.org/10.1016/j.mtphys.2022.100892} {\bibfield  {journal} {\bibinfo  {journal} {Materials Today Physics}\ }\textbf {\bibinfo {volume} {28}},\ \bibinfo {pages} {100892} (\bibinfo {year} {2022})}\BibitemShut {NoStop}%
\bibitem [{\citenamefont {Xia}\ \emph {et~al.}(2023)\citenamefont {Xia}, \citenamefont {Gaines}, \citenamefont {He}, \citenamefont {Pal}, \citenamefont {Li}, \citenamefont {Kanatzidis}, \citenamefont {Ozoliņš},\ and\ \citenamefont {Wolverton}}]{xia_unified_2023}%
  \BibitemOpen
  \bibfield  {author} {\bibinfo {author} {\bibfnamefont {Y.}~\bibnamefont {Xia}}, \bibinfo {author} {\bibfnamefont {D.}~\bibnamefont {Gaines}}, \bibinfo {author} {\bibfnamefont {J.}~\bibnamefont {He}}, \bibinfo {author} {\bibfnamefont {K.}~\bibnamefont {Pal}}, \bibinfo {author} {\bibfnamefont {Z.}~\bibnamefont {Li}}, \bibinfo {author} {\bibfnamefont {M.~G.}\ \bibnamefont {Kanatzidis}}, \bibinfo {author} {\bibfnamefont {V.}~\bibnamefont {Ozoliņš}},\ and\ \bibinfo {author} {\bibfnamefont {C.}~\bibnamefont {Wolverton}},\ }\bibfield  {title} {\bibinfo {title} {A unified understanding of minimum lattice thermal conductivity},\ }\href {https://doi.org/10.1073/pnas.2302541120} {\bibfield  {journal} {\bibinfo  {journal} {Proceedings of the National Academy of Sciences}\ }\textbf {\bibinfo {volume} {120}},\ \bibinfo {pages} {e2302541120} (\bibinfo {year} {2023})}\BibitemShut {NoStop}%
\bibitem [{\citenamefont {Knoop}\ \emph {et~al.}(2023)\citenamefont {Knoop}, \citenamefont {Purcell}, \citenamefont {Scheffler},\ and\ \citenamefont {Carbogno}}]{knoop_anharmonicity_2023}%
  \BibitemOpen
  \bibfield  {author} {\bibinfo {author} {\bibfnamefont {F.}~\bibnamefont {Knoop}}, \bibinfo {author} {\bibfnamefont {T.~A.}\ \bibnamefont {Purcell}}, \bibinfo {author} {\bibfnamefont {M.}~\bibnamefont {Scheffler}},\ and\ \bibinfo {author} {\bibfnamefont {C.}~\bibnamefont {Carbogno}},\ }\bibfield  {title} {\bibinfo {title} {Anharmonicity in {Thermal} {Insulators}: {An} {Analysis} from {First} {Principles}},\ }\href {https://doi.org/10.1103/PhysRevLett.130.236301} {\bibfield  {journal} {\bibinfo  {journal} {Physical Review Letters}\ }\textbf {\bibinfo {volume} {130}},\ \bibinfo {pages} {236301} (\bibinfo {year} {2023})},\ \bibinfo {note} {publisher: American Physical Society}\BibitemShut {NoStop}%
\bibitem [{\citenamefont {Dangić}\ \emph {et~al.}(2025)\citenamefont {Dangić}, \citenamefont {Caldarelli}, \citenamefont {Bianco}, \citenamefont {Savić},\ and\ \citenamefont {Errea}}]{dangic_lattice_2025}%
  \BibitemOpen
  \bibfield  {author} {\bibinfo {author} {\bibfnamefont {D.}~\bibnamefont {Dangić}}, \bibinfo {author} {\bibfnamefont {G.}~\bibnamefont {Caldarelli}}, \bibinfo {author} {\bibfnamefont {R.}~\bibnamefont {Bianco}}, \bibinfo {author} {\bibfnamefont {I.}~\bibnamefont {Savić}},\ and\ \bibinfo {author} {\bibfnamefont {I.}~\bibnamefont {Errea}},\ }\bibfield  {title} {\bibinfo {title} {Lattice thermal conductivity in the anharmonic overdamped regime},\ }\href {https://doi.org/10.1103/PhysRevB.111.104314} {\bibfield  {journal} {\bibinfo  {journal} {Physical Review B}\ }\textbf {\bibinfo {volume} {111}},\ \bibinfo {pages} {104314} (\bibinfo {year} {2025})},\ \bibinfo {note} {publisher: American Physical Society}\BibitemShut {NoStop}%
\bibitem [{\citenamefont {Jasrasaria}\ and\ \citenamefont {Berkelbach}(2025)}]{jasrasaria_strong_2025}%
  \BibitemOpen
  \bibfield  {author} {\bibinfo {author} {\bibfnamefont {D.}~\bibnamefont {Jasrasaria}}\ and\ \bibinfo {author} {\bibfnamefont {T.~C.}\ \bibnamefont {Berkelbach}},\ }\bibfield  {title} {\bibinfo {title} {Strong anharmonicity dictates ultralow thermal conductivities of type-{I} clathrates},\ }\href {https://doi.org/10.1103/s9z1-htzl} {\bibfield  {journal} {\bibinfo  {journal} {Physical Review B}\ }\textbf {\bibinfo {volume} {112}},\ \bibinfo {pages} {014308} (\bibinfo {year} {2025})},\ \bibinfo {note} {publisher: American Physical Society}\BibitemShut {NoStop}%
\bibitem [{\citenamefont {Jin}\ \emph {et~al.}(2023)\citenamefont {Jin}, \citenamefont {Singer}, \citenamefont {Liang},\ and\ \citenamefont {Yang}}]{jin_structural_2023}%
  \BibitemOpen
  \bibfield  {author} {\bibinfo {author} {\bibfnamefont {T.}~\bibnamefont {Jin}}, \bibinfo {author} {\bibfnamefont {G.}~\bibnamefont {Singer}}, \bibinfo {author} {\bibfnamefont {K.}~\bibnamefont {Liang}},\ and\ \bibinfo {author} {\bibfnamefont {Y.}~\bibnamefont {Yang}},\ }\bibfield  {title} {\bibinfo {title} {Structural batteries: {Advances}, challenges and perspectives},\ }\href {https://doi.org/10.1016/j.mattod.2022.12.001} {\bibfield  {journal} {\bibinfo  {journal} {Materials Today}\ }\textbf {\bibinfo {volume} {62}},\ \bibinfo {pages} {151} (\bibinfo {year} {2023})}\BibitemShut {NoStop}%
\bibitem [{\citenamefont {Fedrigucci}\ \emph {et~al.}(2024)\citenamefont {Fedrigucci}, \citenamefont {Marzari},\ and\ \citenamefont {Ricci}}]{fedrigucciComprehensiveScreeningPlasmaFacing2024}%
  \BibitemOpen
  \bibfield  {author} {\bibinfo {author} {\bibfnamefont {A.}~\bibnamefont {Fedrigucci}}, \bibinfo {author} {\bibfnamefont {N.}~\bibnamefont {Marzari}},\ and\ \bibinfo {author} {\bibfnamefont {P.}~\bibnamefont {Ricci}},\ }\bibfield  {title} {\bibinfo {title} {Comprehensive {Screening} of {Plasma}-{Facing} {Materials} for {Nuclear} {Fusion}},\ }\href {https://doi.org/10.1103/PRXEnergy.3.043002} {\bibfield  {journal} {\bibinfo  {journal} {PRX Energy}\ }\textbf {\bibinfo {volume} {3}},\ \bibinfo {pages} {043002} (\bibinfo {year} {2024})},\ \bibinfo {note} {publisher: American Physical Society}\BibitemShut {NoStop}%
\bibitem [{\citenamefont {Lin}\ \emph {et~al.}(2025)\citenamefont {Lin}, \citenamefont {Koyanagi}, \citenamefont {Zinkle}, \citenamefont {Snead},\ and\ \citenamefont {Katoh}}]{lin_perspectives_2025}%
  \BibitemOpen
  \bibfield  {author} {\bibinfo {author} {\bibfnamefont {Y.-R.}\ \bibnamefont {Lin}}, \bibinfo {author} {\bibfnamefont {T.}~\bibnamefont {Koyanagi}}, \bibinfo {author} {\bibfnamefont {S.~J.}\ \bibnamefont {Zinkle}}, \bibinfo {author} {\bibfnamefont {L.~L.}\ \bibnamefont {Snead}},\ and\ \bibinfo {author} {\bibfnamefont {Y.}~\bibnamefont {Katoh}},\ }\bibfield  {title} {\bibinfo {title} {Perspectives and challenges of ultra-high temperature ceramics for fusion plasma-facing applications},\ }\href {https://doi.org/10.1016/j.cossms.2025.101223} {\bibfield  {journal} {\bibinfo  {journal} {Current Opinion in Solid State and Materials Science}\ }\textbf {\bibinfo {volume} {36}},\ \bibinfo {pages} {101223} (\bibinfo {year} {2025})}\BibitemShut {NoStop}%
\bibitem [{\citenamefont {Nygren}\ \emph {et~al.}(2016)\citenamefont {Nygren}, \citenamefont {Youchison}, \citenamefont {Wirth},\ and\ \citenamefont {Snead}}]{nygren_new_2016}%
  \BibitemOpen
  \bibfield  {author} {\bibinfo {author} {\bibfnamefont {R.~E.}\ \bibnamefont {Nygren}}, \bibinfo {author} {\bibfnamefont {D.~L.}\ \bibnamefont {Youchison}}, \bibinfo {author} {\bibfnamefont {B.~D.}\ \bibnamefont {Wirth}},\ and\ \bibinfo {author} {\bibfnamefont {L.~L.}\ \bibnamefont {Snead}},\ }\bibfield  {title} {\bibinfo {title} {A new vision of plasma facing components},\ }\href {https://doi.org/10.1016/j.fusengdes.2016.03.031} {\bibfield  {journal} {\bibinfo  {journal} {Fusion Engineering and Design}\ }\bibinfo {series} {Proceedings of the 12th {International} {Symposium} on {Fusion} {Nuclear} {Technology}-12 ({ISFNT}-12)},\ \textbf {\bibinfo {volume} {109-111}},\ \bibinfo {pages} {192} (\bibinfo {year} {2016})}\BibitemShut {NoStop}%
\bibitem [{\citenamefont {Abrams}\ \emph {et~al.}(2021)\citenamefont {Abrams}, \citenamefont {Bringuier}, \citenamefont {Thomas}, \citenamefont {Sinclair}, \citenamefont {Gonderman}, \citenamefont {Holland}, \citenamefont {Rudakov}, \citenamefont {Wilcox}, \citenamefont {Unterberg},\ and\ \citenamefont {Scotti}}]{abrams_evaluation_2021}%
  \BibitemOpen
  \bibfield  {author} {\bibinfo {author} {\bibfnamefont {T.}~\bibnamefont {Abrams}}, \bibinfo {author} {\bibfnamefont {S.}~\bibnamefont {Bringuier}}, \bibinfo {author} {\bibfnamefont {D.}~\bibnamefont {Thomas}}, \bibinfo {author} {\bibfnamefont {G.}~\bibnamefont {Sinclair}}, \bibinfo {author} {\bibfnamefont {S.}~\bibnamefont {Gonderman}}, \bibinfo {author} {\bibfnamefont {L.}~\bibnamefont {Holland}}, \bibinfo {author} {\bibfnamefont {D.}~\bibnamefont {Rudakov}}, \bibinfo {author} {\bibfnamefont {R.}~\bibnamefont {Wilcox}}, \bibinfo {author} {\bibfnamefont {E.}~\bibnamefont {Unterberg}},\ and\ \bibinfo {author} {\bibfnamefont {F.}~\bibnamefont {Scotti}},\ }\bibfield  {title} {\bibinfo {title} {Evaluation of silicon carbide as a divertor armor material in {DIII}-{D} {H}-mode discharges},\ }\href {https://doi.org/10.1088/1741-4326/abecee} {\bibfield  {journal} {\bibinfo  {journal} {Nuclear Fusion}\ }\textbf {\bibinfo {volume} {61}},\ \bibinfo {pages} {066005} (\bibinfo {year} {2021})},\ \bibinfo {note} {publisher: IOP Publishing}\BibitemShut {NoStop}%
\bibitem [{\citenamefont {Linke}\ \emph {et~al.}(2019)\citenamefont {Linke}, \citenamefont {Du}, \citenamefont {Loewenhoff}, \citenamefont {Pintsuk}, \citenamefont {Spilker}, \citenamefont {Steudel},\ and\ \citenamefont {Wirtz}}]{linke_challenges_2019}%
  \BibitemOpen
  \bibfield  {author} {\bibinfo {author} {\bibfnamefont {J.}~\bibnamefont {Linke}}, \bibinfo {author} {\bibfnamefont {J.}~\bibnamefont {Du}}, \bibinfo {author} {\bibfnamefont {T.}~\bibnamefont {Loewenhoff}}, \bibinfo {author} {\bibfnamefont {G.}~\bibnamefont {Pintsuk}}, \bibinfo {author} {\bibfnamefont {B.}~\bibnamefont {Spilker}}, \bibinfo {author} {\bibfnamefont {I.}~\bibnamefont {Steudel}},\ and\ \bibinfo {author} {\bibfnamefont {M.}~\bibnamefont {Wirtz}},\ }\bibfield  {title} {\bibinfo {title} {Challenges for plasma-facing components in nuclear fusion},\ }\href {https://doi.org/10.1063/1.5090100} {\bibfield  {journal} {\bibinfo  {journal} {Matter and Radiation at Extremes}\ }\textbf {\bibinfo {volume} {4}},\ \bibinfo {pages} {056201} (\bibinfo {year} {2019})}\BibitemShut {NoStop}%
\bibitem [{\citenamefont {Lv}\ and\ \citenamefont {Henry}(2016{\natexlab{b}})}]{lv_non-negligible_2016}%
  \BibitemOpen
  \bibfield  {author} {\bibinfo {author} {\bibfnamefont {W.}~\bibnamefont {Lv}}\ and\ \bibinfo {author} {\bibfnamefont {A.}~\bibnamefont {Henry}},\ }\bibfield  {title} {\bibinfo {title} {Non-negligible {Contributions} to {Thermal} {Conductivity} {From} {Localized} {Modes} in {Amorphous} {Silicon} {Dioxide}},\ }\href {https://doi.org/10.1038/srep35720} {\bibfield  {journal} {\bibinfo  {journal} {Scientific Reports}\ }\textbf {\bibinfo {volume} {6}},\ \bibinfo {pages} {35720} (\bibinfo {year} {2016}{\natexlab{b}})}\BibitemShut {NoStop}%
\bibitem [{\citenamefont {Liang}\ \emph {et~al.}(2023)\citenamefont {Liang}, \citenamefont {Ying}, \citenamefont {Xu}, \citenamefont {Ye}, \citenamefont {Ling}, \citenamefont {Fan},\ and\ \citenamefont {Xu}}]{liang_mechanisms_2023}%
  \BibitemOpen
  \bibfield  {author} {\bibinfo {author} {\bibfnamefont {T.}~\bibnamefont {Liang}}, \bibinfo {author} {\bibfnamefont {P.}~\bibnamefont {Ying}}, \bibinfo {author} {\bibfnamefont {K.}~\bibnamefont {Xu}}, \bibinfo {author} {\bibfnamefont {Z.}~\bibnamefont {Ye}}, \bibinfo {author} {\bibfnamefont {C.}~\bibnamefont {Ling}}, \bibinfo {author} {\bibfnamefont {Z.}~\bibnamefont {Fan}},\ and\ \bibinfo {author} {\bibfnamefont {J.}~\bibnamefont {Xu}},\ }\bibfield  {title} {\bibinfo {title} {Mechanisms of temperature-dependent thermal transport in amorphous silica from machine-learning molecular dynamics},\ }\href {https://doi.org/10.1103/PhysRevB.108.184203} {\bibfield  {journal} {\bibinfo  {journal} {Physical Review B}\ }\textbf {\bibinfo {volume} {108}},\ \bibinfo {pages} {184203} (\bibinfo {year} {2023})},\ \bibinfo {note} {publisher: American Physical Society}\BibitemShut {NoStop}%
\bibitem [{\citenamefont {Le~Roux}\ and\ \citenamefont {Jund}(2010)}]{le_roux_ring_2010}%
  \BibitemOpen
  \bibfield  {author} {\bibinfo {author} {\bibfnamefont {S.}~\bibnamefont {Le~Roux}}\ and\ \bibinfo {author} {\bibfnamefont {P.}~\bibnamefont {Jund}},\ }\bibfield  {title} {\bibinfo {title} {Ring statistics analysis of topological networks: {New} approach and application to amorphous {GeS2} and {SiO2} systems},\ }\href {https://doi.org/10.1016/j.commatsci.2010.04.023} {\bibfield  {journal} {\bibinfo  {journal} {Computational Materials Science}\ }\textbf {\bibinfo {volume} {49}},\ \bibinfo {pages} {70} (\bibinfo {year} {2010})}\BibitemShut {NoStop}%
\bibitem [{Note3()}]{Note3}%
  \BibitemOpen
  \bibinfo {note} {\protect \url {https://github.com/bschweinhart/Swatches}}\BibitemShut {NoStop}%
\bibitem [{\citenamefont {Togo}(2023)}]{togo_first-principles_2023}%
  \BibitemOpen
  \bibfield  {author} {\bibinfo {author} {\bibfnamefont {A.}~\bibnamefont {Togo}},\ }\bibfield  {title} {\bibinfo {title} {First-principles {Phonon} {Calculations} with {Phonopy} and {Phono3py}},\ }\href {https://doi.org/10.7566/JPSJ.92.012001} {\bibfield  {journal} {\bibinfo  {journal} {Journal of the Physical Society of Japan}\ }\textbf {\bibinfo {volume} {92}},\ \bibinfo {pages} {012001} (\bibinfo {year} {2023})}\BibitemShut {NoStop}%
\bibitem [{\citenamefont {Paszke}\ \emph {et~al.}(2019)\citenamefont {Paszke}, \citenamefont {Gross}, \citenamefont {Massa}, \citenamefont {Lerer}, \citenamefont {Bradbury}, \citenamefont {Chanan}, \citenamefont {Killeen}, \citenamefont {Lin}, \citenamefont {Gimelshein}, \citenamefont {Antiga}, \citenamefont {Desmaison}, \citenamefont {Kopf}, \citenamefont {Yang}, \citenamefont {DeVito}, \citenamefont {Raison}, \citenamefont {Tejani}, \citenamefont {Chilamkurthy}, \citenamefont {Steiner}, \citenamefont {Fang}, \citenamefont {Bai},\ and\ \citenamefont {Chintala}}]{NEURIPS2019_9015}%
  \BibitemOpen
  \bibfield  {author} {\bibinfo {author} {\bibfnamefont {A.}~\bibnamefont {Paszke}}, \bibinfo {author} {\bibfnamefont {S.}~\bibnamefont {Gross}}, \bibinfo {author} {\bibfnamefont {F.}~\bibnamefont {Massa}}, \bibinfo {author} {\bibfnamefont {A.}~\bibnamefont {Lerer}}, \bibinfo {author} {\bibfnamefont {J.}~\bibnamefont {Bradbury}}, \bibinfo {author} {\bibfnamefont {G.}~\bibnamefont {Chanan}}, \bibinfo {author} {\bibfnamefont {T.}~\bibnamefont {Killeen}}, \bibinfo {author} {\bibfnamefont {Z.}~\bibnamefont {Lin}}, \bibinfo {author} {\bibfnamefont {N.}~\bibnamefont {Gimelshein}}, \bibinfo {author} {\bibfnamefont {L.}~\bibnamefont {Antiga}}, \bibinfo {author} {\bibfnamefont {A.}~\bibnamefont {Desmaison}}, \bibinfo {author} {\bibfnamefont {A.}~\bibnamefont {Kopf}}, \bibinfo {author} {\bibfnamefont {E.}~\bibnamefont {Yang}}, \bibinfo {author} {\bibfnamefont {Z.}~\bibnamefont {DeVito}}, \bibinfo {author} {\bibfnamefont {M.}~\bibnamefont {Raison}}, \bibinfo {author} {\bibfnamefont {A.}~\bibnamefont {Tejani}}, \bibinfo {author} {\bibfnamefont {S.}~\bibnamefont {Chilamkurthy}}, \bibinfo {author} {\bibfnamefont {B.}~\bibnamefont {Steiner}}, \bibinfo {author} {\bibfnamefont {L.}~\bibnamefont {Fang}}, \bibinfo {author} {\bibfnamefont {J.}~\bibnamefont {Bai}},\ and\ \bibinfo {author} {\bibfnamefont {S.}~\bibnamefont {Chintala}},\ }\bibfield  {title} {\bibinfo {title} {Pytorch: An imperative style, high-performance deep learning library},\ }in\ \href {http://papers.neurips.cc/paper/9015-pytorch-an-imperative-style-high-performance-deep-learning-library.pdf} {\emph {\bibinfo {booktitle} {Advances in Neural Information Processing Systems 32}}}\ (\bibinfo  {publisher} {Curran Associates, Inc.},\ \bibinfo {year} {2019})\ pp.\ \bibinfo {pages} {8024--8035}\BibitemShut {NoStop}%
\bibitem [{Note4()}]{Note4}%
  \BibitemOpen
  \bibinfo {note} {We recall that stronger disorder-induced repulsion between energy levels promotes a smoother VDOS.}\BibitemShut {Stop}%
\bibitem [{\citenamefont {Thébaud}\ \emph {et~al.}(2023)\citenamefont {Thébaud}, \citenamefont {Lindsay},\ and\ \citenamefont {Berlijn}}]{thebaud_breaking_2023}%
  \BibitemOpen
  \bibfield  {author} {\bibinfo {author} {\bibfnamefont {S.}~\bibnamefont {Thébaud}}, \bibinfo {author} {\bibfnamefont {L.}~\bibnamefont {Lindsay}},\ and\ \bibinfo {author} {\bibfnamefont {T.}~\bibnamefont {Berlijn}},\ }\bibfield  {title} {\bibinfo {title} {Breaking {Rayleigh}'s {Law} with {Spatially} {Correlated} {Disorder} to {Control} {Phonon} {Transport}},\ }\href {https://doi.org/10.1103/PhysRevLett.131.026301} {\bibfield  {journal} {\bibinfo  {journal} {Physical Review Letters}\ }\textbf {\bibinfo {volume} {131}},\ \bibinfo {pages} {026301} (\bibinfo {year} {2023})},\ \bibinfo {note} {publisher: American Physical Society}\BibitemShut {NoStop}%
\bibitem [{\citenamefont {Caldarelli}\ \emph {et~al.}(2022)\citenamefont {Caldarelli}, \citenamefont {Simoncelli}, \citenamefont {Marzari}, \citenamefont {Mauri},\ and\ \citenamefont {Benfatto}}]{caldarelli_many-body_2022}%
  \BibitemOpen
  \bibfield  {author} {\bibinfo {author} {\bibfnamefont {G.}~\bibnamefont {Caldarelli}}, \bibinfo {author} {\bibfnamefont {M.}~\bibnamefont {Simoncelli}}, \bibinfo {author} {\bibfnamefont {N.}~\bibnamefont {Marzari}}, \bibinfo {author} {\bibfnamefont {F.}~\bibnamefont {Mauri}},\ and\ \bibinfo {author} {\bibfnamefont {L.}~\bibnamefont {Benfatto}},\ }\bibfield  {title} {\bibinfo {title} {Many-body {Green}'s function approach to lattice thermal transport},\ }\href {https://doi.org/10.1103/PhysRevB.106.024312} {\bibfield  {journal} {\bibinfo  {journal} {Physical Review B}\ }\textbf {\bibinfo {volume} {106}},\ \bibinfo {pages} {024312} (\bibinfo {year} {2022})},\ \bibinfo {note} {publisher: American Physical Society}\BibitemShut {NoStop}%
\end{thebibliography}
\providecommand{\noopsort}[1]{}\providecommand{\singleletter}[1]{#1}%

\end{document}


\title{Supplementary Materials for ``Bond-Network Entropy Governs Heat Transport in Coordination-Disordered Solids''
}

\author{Kamil Iwanowski}
\affiliation{Department of Applied Physics and Applied Mathematics, Columbia University, New York (USA)}
\affiliation{Theory of Condensed Matter Group, Cavendish Laboratory, University of Cambridge (UK)}

\author{Gábor Csányi}
\affiliation{Applied Mechanics Group, Mechanics, Materials and Design, Department of Engineering, University of Cambridge (UK)}

\author{Michele Simoncelli}
\email{michele.simoncelli@columbia.edu}
\affiliation{Department of Applied Physics and Applied Mathematics, Columbia University, New York (USA)}
\affiliation{Theory of Condensed Matter Group, Cavendish Laboratory, University of Cambridge (UK)}

\maketitle

\section{Temperature dependence of the conductivity}
\label{sec:k_vs_T}
\begin{figure}[H]
  \centering
\includegraphics[width=\textwidth]{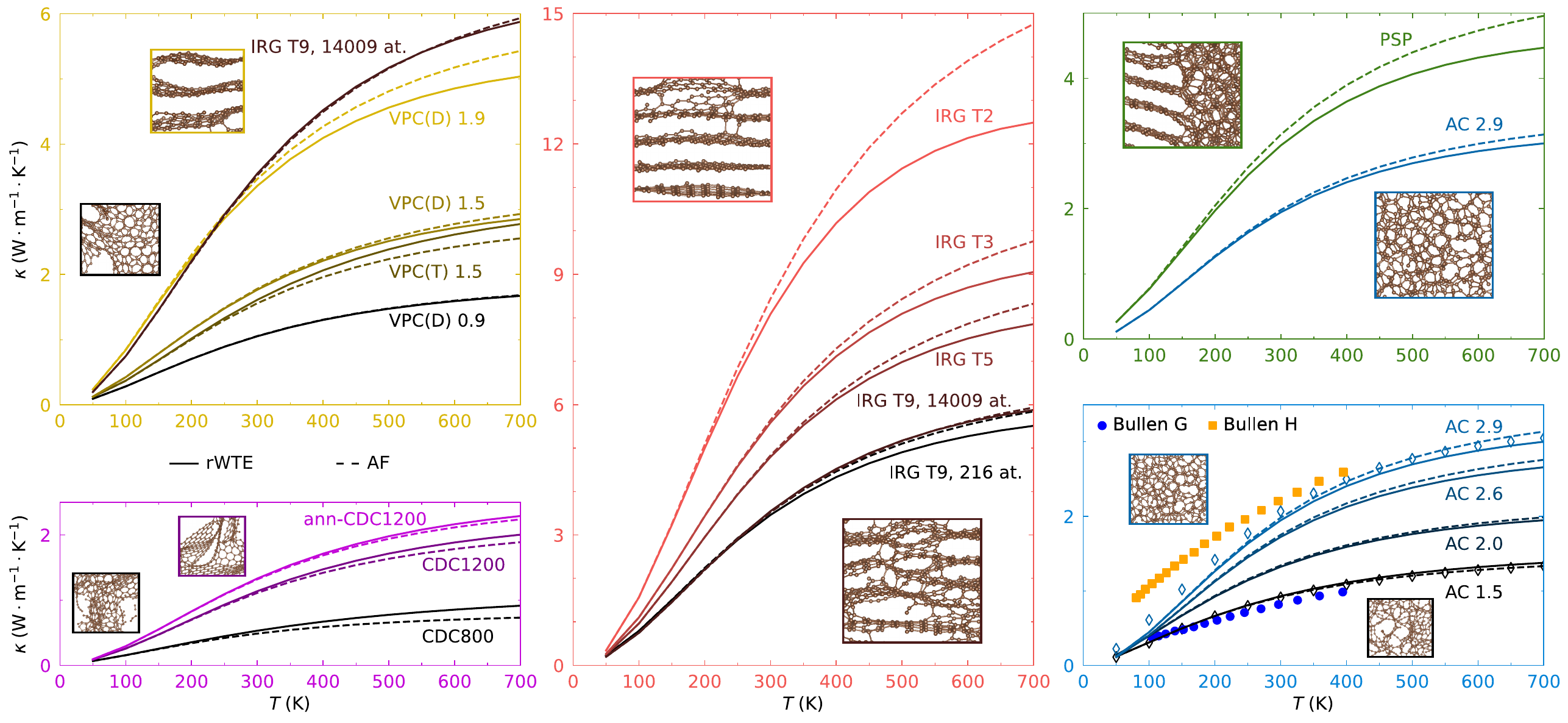}\\[-3mm]
  \caption{\textbf{Dependence of thermal conductivity of carbon polymorphs on temperature.} Variable-porosity carbon (VPC) is yellow, carbide-derived carbon (CDC) is purple, irradiated graphite (IRG) is red, phase-separated phase (PSP) is green, and amorphous carbon (AC) is blue. The solid and dashed lines are conductivities calculated using rWTE and AF formulations, respectively. 
  The empty diamonds in the blue AC panel are rWTE conductivity calculations for 8000-atom sized models of amorphous carbon, the solid and dashed lines are rWTE and AF conductivity calculations for models with 216 atoms. The dark blue circles and orange squares are  conductivity experiments taken from Bullen \textit{et al.} \cite{bullen_thermal_2000} of samples G (density 1.7 g/cm$^3$) and H (density 2.8 g/cm$^3$), respectively.
  }
  \label{fig:conductivity_temperature}
\end{figure}

In Fig. \ref{fig:conductivity_temperature} we analyze the impact of temperature on the thermal conductivity of disordered carbon. 
We note that conductivity for all structures increases in the temperature range 50$-$700 K, and such increase is faster in the temperature range between ${\sim} 100$ and  ${\sim }300$ K. We can understand this behavior by resolving the conductivity in terms of the frequency integral involving VDOS $g(\omega)$, specific heat $C(\omega, T)$ and diffusivity $D(\omega)$ (Eq.~(9)). Then, approximating the diffusivity to be in temperature-independent AF limit (Eq.~(13)), we obtain
\begin{equation}
    \kappa(T) \approx \kappa_{AF}(T)=\int_{0}^{\omega_{\rm max}} g(\omega) C(\omega,T) D_{AF}(\omega) d\omega\;,
\label{eq:conductivity_AF}
\end{equation}
Using Eq.~(\ref{eq:conductivity_AF}), the increase of conductivity with temperature is caused by increase in specific heat $C = \int_{0}^{\omega_{\rm max}} g(\omega) C(\omega,T) d\omega$ with temperature.
This increase saturates due to saturation of specific heat in the limit of large temperature.
\begin{equation}
    C(T=\infty) = \int_{0}^{\omega_{\rm max}} g(\omega) C(\omega,\infty) d\omega = \frac{3 N_{at}}{\mathcal{V}} k_{\rm B}
\end{equation}
The difference between AF and rWTE conductivities, caused by intrinsic linewidths due to anharmonicity and mass isotope disorder, depends both on the structure and temperature range. Specifically, at and below room temperature two formulations give equivalent results for all structures considered.
For VPC(D) 0.9, AC with density $\le 2.0$ g/cm$^3$ and IRG T9 14009 anharmonicity has a negligible effect over the entire temperature range 50-700 K. We observe the strongest effect of anharmonicity appears in the least disordered IRG T2 structure at 700 K, where rWTE conductivity is $\approx 15$\% smaller compared to AF. In the other structures anharmonicity has a weaker effect on the conductivity.

In AC we compare our theoretical predictions with available experimental data from \citet{bullen_thermal_2000} (sample G a-C:H with density 1.7 g/cm$^3$ and 2:1 C:H ratio; and sample H a-C with density 2.8 g/cm$^3$). We obtain a reasonable agreement with experiments obtaining the same positive trend of conductivity with temperature and positive trend of conductivity with density.
We note, in passing, that the amorphous-carbon samples used in experiments can contain hydrogen impurities \cite{bullen_thermal_2000}, especially at low density.  
Previous studies estimated that the main impact of hydrogen is to promote disorder in the bond network \cite{balandin_thermal_2008}, as well as to marginally increase the effective mass of carbon atoms \cite{morath_picosecond_1994}. The second effect can be explained as follows. Due to the low mass of hydrogen atoms, the typical vibrational frequencies of C-H and H-H bonds are 90 and above 100 THz \cite{morath_picosecond_1994} and so at room temperature they are very weakly activated given that $k_B T \approx 6$ THz for $T = 300$K. Hence we can treat C-H pair as a single molecule with increased mass compared to carbon, and at room temperature ignore its internal degrees of freedom. Since this mass increase effect is weak, we can directly compare conductivity of pure and hydrogenated carbon structures.

We focus in Fig. \ref{fig:conductivity_temperature} on the temperature region 50-700 K, as it explores a significant change in temperature and at the same time can be accurately described using models and techniques we employed. 
In particular, we employ the Wigner formulation of thermal transport, which accounts for anharmonic and mass isotope disorder \cite{tamura_isotope_1983} sources of intrinsic linewidth. Anharmonicity is calculated using the standard 3rd-order perturbative treatment, which is increasingly more accurate as temperature is lowered --- for this reason we limited the range of temperatures to below 700 K, and focused the analysis in the main text to room temperature.
Additionally, we note that we limited our analysis to the above-the-plateau regime $T > 50$K, since the finite-size models employed in this work do not allow to accurately sample the low-frequency ($\hbar\omega\lesssim 50$ cm$^-1$) vibrational modes that are relevant for transport at cryogenic temperatures. 
The influence of these poorly sampled low-frequency modes is negligible at room temperature, as at this temperature we do not observe significant changes in the thermal conductivity of models having different size (Fig.~1).

\section{Notes on the experimental measurements of thermal conductivity}
\subsection{Negligible contribution of conducting electrons to thermal conductivity in carbon}
The literature on the thermal transport in the considered carbon polymorphs shows that the thermal conductivity is dominated by the lattice contribution \cite{balandin_thermal_2011}. Even in ideal graphite, simulations predict negligible electronic contribution at both low and high doping levels \cite{coulter_coupled_2025}.
In experiments, where the electronic contribution was measured explicitly, it was found to be one order of magnitude smaller than phonon contribution at 100 K for glassy carbon \cite{katerberg_low-temperature_1978}. Additionally, assuming Wiedemann-Franz law $\kappa = LT\sigma$, $L = 2.44 \cdot 10^{-8} \,\, {\rm V}^2 \cdot {\rm K}^{-2}$, the electronic contribution of carbide-derived carbons \cite{kern_thermal_2016} contributes around 1-9\% to the total conductivity at room temperature.

\subsection{Density of solids in Hurler et al., Kern et al., Barabash et al. and Maruyama and Harayama}
Hurler et al. \cite{hurler_determination_1995} data is included as reported in Fig. 3 of Shamsa et al. \cite{shamsa_thermal_2006}.
Experimental density data from Ref.~\cite{kern_thermal_2016} was obtained by converting porosity data to density using equation: $\rho = \rho_{\rm skeleton} (1-{\rm porosity)}$ g/cm$^3$, where $\rho_{\rm skeleton} = 2.2$ g/cm$^3$ is the assumed skeleton density in Ref.~\cite{kern_thermal_2016}. 
For \cite{barabash_effect_2002} the reported densities are approximate as reported in \cite{barabash_effect_2002}.
The reported densities of the experimental structures from \cite{maruyama_neutron_1992} were the initial densities before irradiation, since the dimensional changes observed under irradiation were below 1 \%.

\subsection{Note about how the type of irradiation affects structural defects in graphite}
Past work \cite{snead_thermal_1995} discussed how the structural properties of irradiated graphite depend on: (i) the total irradiation dose (measured in terms of number of displacements per atom, DPA); (ii) the irradiation rate (\textit{i.e.}, dose per time); (iii) the type of irradiation (e.g., neutrons or electrons); (iv) the irradiation temperature (higher temperature allows partial annealing of the changes caused by irradiation). 
Even if the value of DPA alone is not sufficient to fully characterize irradiation, there are important similarities between impact of neutron and electron irradiation on the structure of graphite samples. 
Both neutron and electron irradiation cause swelling in the out-of-plane direction of graphite, and both can yield order-of-magnitude reduction of the thermal conductivity \cite{farbos_time-dependent_2017}. 
However the mechanism behind those changes is slightly different. As pointed out by previous literature \cite{koike_dimensional_1994, karthik_situ_2011}, ions and neutrons cause displacement cascades upon hitting the sample; 
in contrast, electron irradiation causes mainly isolated point defects, dislocation dipoles and interstitial atoms. 
Nevertheless, the irradiation rate
in neutron-based experiments is much lower compared to electron-based experiments, and in practice few minutes of very strong electron irradiation can yield similar effects to much longer times of neutron irradiation.\\

\FloatBarrier

\section{Extended information about bond-network entropy}

\subsection{Comparison with alternative descriptors of local environments}

\begin{figure*}
  \centering
  \includegraphics[width=\textwidth]{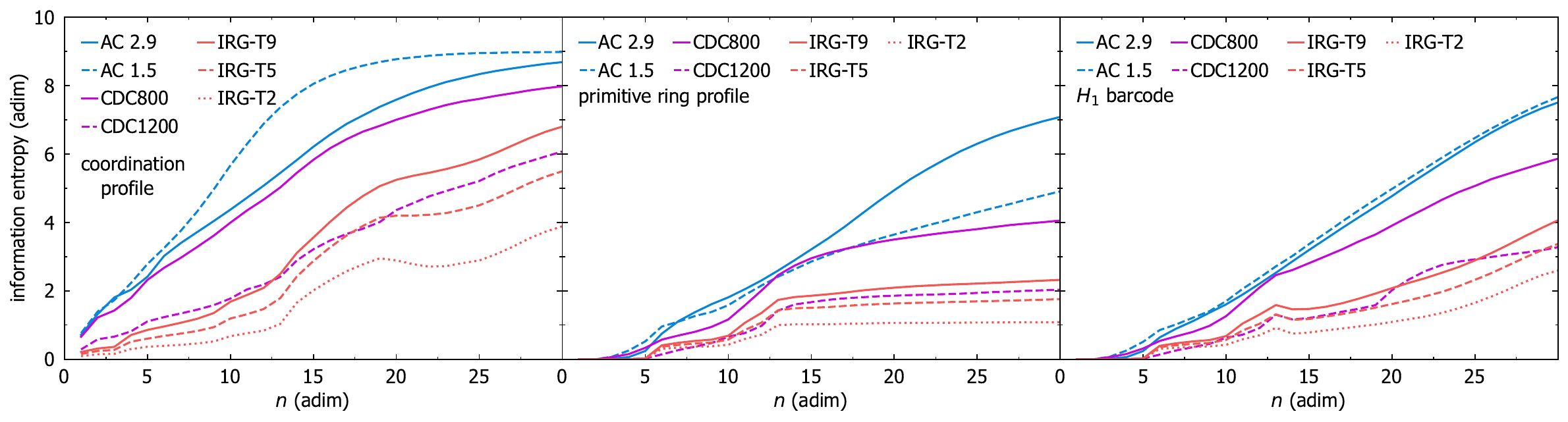}
  \caption{\textbf{Dependence of information entropy on the size of LAEs $n$ for different descriptors.} The three panels indicate entropy calculated with either coordination profile (left), primitive ring profile (center) or $H_1$ barcode (right) for seven exemplary structures of carbon. The structures span the range of disorder shown in Fig.~1 of the main text, from the highest (AC 2.9, AC 1.5: blue, solid and dashed lines respectively; CDC800: purple solid line) through intermediate (CDC1200: purple dashed line; IRG-T9: red solid line) to lowest (IRG-T5, IRG-T2: red, dashed and dotted lines respectively).}
  \label{fig:LAE_description}
\end{figure*}

Within the Wigner formulation of thermal transport \cite{simoncelli_unified_2019, simoncelli_wigner_2022, simoncelli_thermal_2023}, the thermal conductivity is obtained through the following steps: (i)  Calculation of second (FC2) and third (FC3) order force constants, and calculation of the dynamical matrix from the FC2. (ii) Diagonalization of the dynamical matrix to obtain vibrational frequencies and eigenvectors. (iii) Computation of the velocity operator that couples vibrational modes at different frequencies. (iv) Calculation of the linewidths due to anharmonicity and isotopic-mass disorder. (v) Evaluation of the thermal conductivity expression using frequencies, velocity operators, and linewidths. Hence, within rWTE, $\kappa$ is determined by the FC2 and FC3 matrices, which depend on the LAEs and how they are organized in space. In this work, we found ${\rm BNE}(n)/ n$ to be a descriptor of the material structure that is correlated with the thermal conductivity evaluated using full rWTE treatment. The simplicity of the descriptor necessarily entails discarding some of the information present in the structure. By testing different descriptors, we can check which information is the most useful for conductivity predictions.

For this purpose we evaluated information entropy of two additional descriptors of LAEs from \citet{schweinhart_statistical_2020}: coordination profile (CP) and primitive ring profile (PRP) and we plot the results in Fig.~\ref{fig:LAE_description}. Both CP and PRP contain different information about LAEs compared with $H_1$ barcode. CP characterizes coordination number of atoms within each edge distance from the central atom. Hence it has only direct information on the coordination number and is not directly informative about the rings. PRP on the other hand characterizes the number of atoms in primitive rings which pass through the central atom in the LAE. Because of this, it is short-ranged and does not inform about any rings in the LAE that do not include the central atom, in contrast to the $H_1$ barcode which contains this information. 

We find that both CP and PRP are not suitable to accurately quantify disorder in carbon systems. CP classifies AC 1.5 as more disordered than AC 2.9 due to broader distribution of coordination number. This prevents a straightforward relationship between average growth rate of CP entropy and conductivity divided by the density ($\kappa / \rho$), since AC structures have similar $\kappa / \rho$, which is explained by ${\rm BNE}(n)/ n$ based on the $H_1$ barcode. CP is also visibly saturating for highly disordered structures of AC before $n=30$. This limits the extent to which CP can quantify differences in disorder between all structures of carbon to short ranges ($n \le 13$).
PRP has also problems with short-ranged information, as for weakly disordered solids (IRG, CDC1200) it saturates after $n \approx 15$, and hence it is only sensitive to features within $n \le 15$. Therefore PRP entropy is not extensive and the growth rate is not approximately constant as for the case of $H_1$ barcode. Finally, it predicts AC 2.9 to be more disordered than AC 1.5 and hence leading to the same problems as in the case of CP.

Overall, we find the predictive power of $H_1$ barcode to be greater than other descriptors. This is in line with initial reports of Schweinhart et al. \cite{schweinhart_statistical_2020}, where $H_1$ barcode was found to be able to distinguish between local environments in crystalline structures of silica and silica glasses produced at different quench rates.
Additionally, $H_1$ barcode is faster to evaluate than PRP and unlike CP contains easily accessible information about rings, which are relevant for vibrational properties.

\subsection{Relations between BNE's growth rate, density and volume specific heat}

\begin{figure*}
  \centering
  \includegraphics[width=\textwidth]{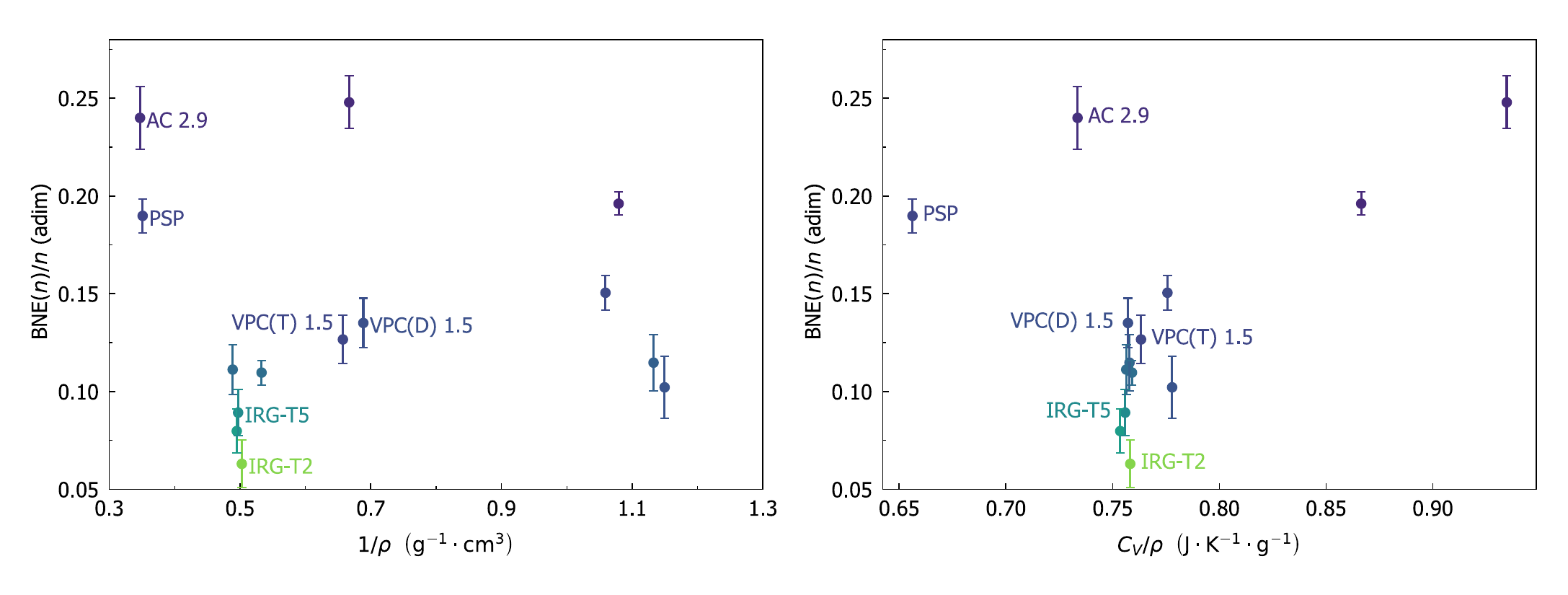}
  \caption{\textbf{Relation between ${\rm BNE}(n)/ n$, inverse of density, $1/\rho$ and mass specific heat.} Left panel: BNE$(n)/n$ vs inverse of density $1/ \rho$. Right panel: BNE$(n)/n$ vs mass specific heat at 300 K. Mass specific heat $C_m$ is equal to volume specific heat divided by the density $C_m = C_V / \rho$.}
  \label{fig:entropy_vs_other}
\end{figure*}

We calculate the volume specific heat $C_V$ at 300 K using Eq.~\ref{eq:heat_capacity}, with $C_{\bm{q} s}$ as defined in Eq.~2 in the main manuscript.
\begin{equation}\label{eq:heat_capacity}
    C_V = \frac{1}{\mathcal{V} N_c} \sum_{\bm{q} s} C_{\bm{q} s}
\end{equation}
We find no correlation between ${\rm BNE}(n)/n$ and density as we explored structures with both variable degree of disorder and variable density. For specific heat we find that structures with lowest amount of disorder have mass specific heat equal to around 0.76 ${\rm J} \cdot {\rm K}^{-1} \cdot {\rm g}^{-1}$. This is because majority of the atoms in those structures have coordination number equal to three (see Fig. \ref{fig:coordination_all_structures}) as well as coordination number in carbon is heavily correlated with the VDOS shape and hence its specific heat (see Fig. 3 in the main manuscript).

\subsection{Low and high temperature limits of BNE's growth rate-conductivity/density correlation}

\begin{figure*}
  \centering
  \includegraphics[width=\textwidth]{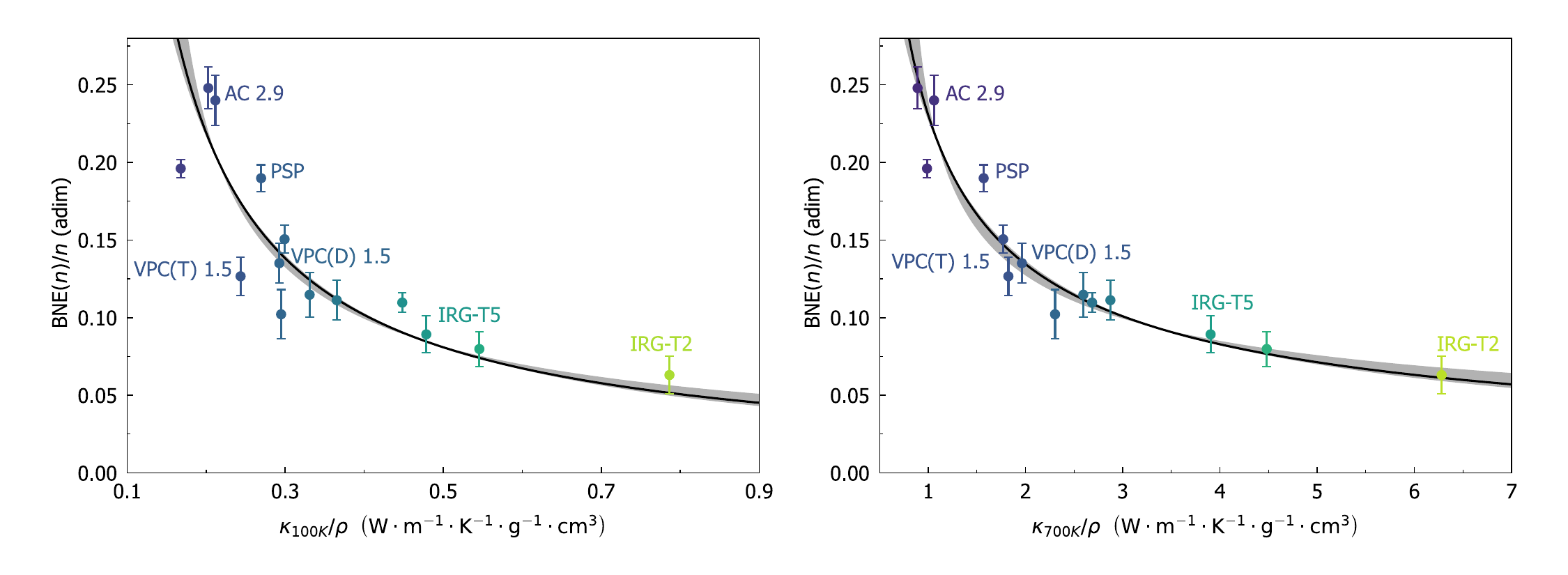}
  \caption{\textbf{Correlations between conductivities at 100 and 700 K divided by density, and average value of bond-network entropy divided by the size of LAEs, ${\rm{BNE}}(n)/n$.} Left and right panels show the correlations at 100 and 700 K respectively. The black lines are best-fit power laws with a constant ($\kappa/\rho = ({\rm BNE}(n)/n)^a \times e^b + c$) and the best-fit parameters were $a = -1.082$, $b = -3.508$, $c = 0.045$ for 100 K and $a = -1.574$, $b = -2.599$, $c = 0.250$ for 700 K. Units of $c$ are ${\rm W \cdot m^{-1} \cdot K^{-1} \cdot g^{-1} \cdot cm^3}$. The shaded areas show changes in the best-fit power laws when the constant $c$ is changed from 0 to 0.12 and from 0 to 0.7 for 100 and 700 K respectively.}
  \label{fig:conductivity_entropy_100_700K}
\end{figure*}

At 100 and 700 K, we also observe the inverse dependence between ${\rm BNE}(n)/n$ and thermal conductivity in addition to the correlation found at room temperature, see Fig.~\ref{fig:conductivity_entropy_100_700K}. The structures and temperatures analyzed in this manuscript are in the regime where short-range and medium-range order, and hence vibrations' damping caused by them, is the dominant limiting factor for thermal conduction; this is the underlying reason why a quantifier of disorder, ${\rm BNE}(n)/n$, correlates with thermal conductivity.

\subsection{Dependence of BNE's growth rate on grain size in polycrystalline graphene}

\begin{figure}
  \centering
\includegraphics[width=0.5\textwidth]{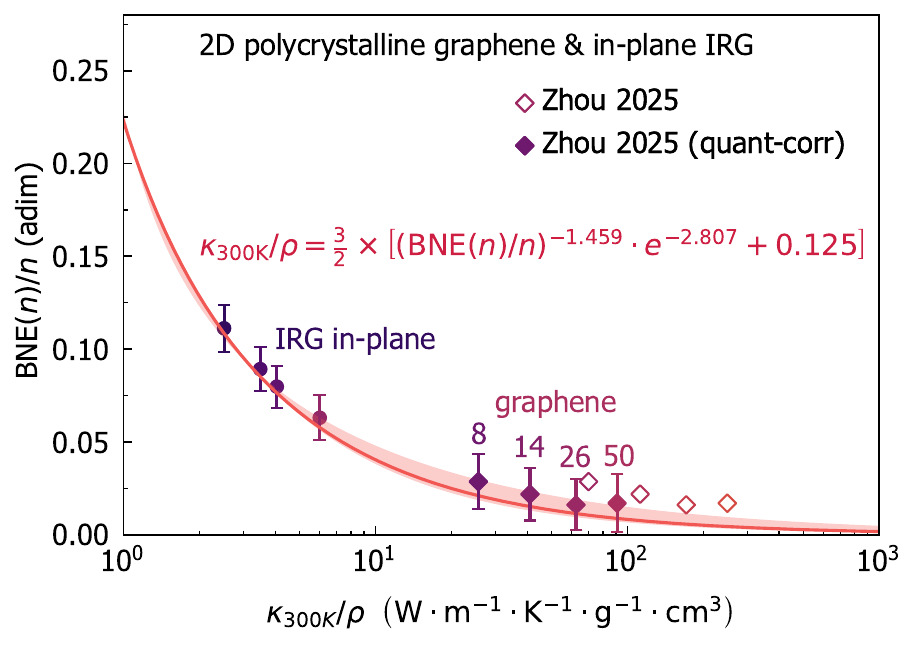}\\[-3mm]
  \caption{\textbf{Correlation between room-temperature in-plane conductivity divided by density, $\kappa_{\rm 300K}/\rho$, and average value of bond-network entropy divided by the size of LAEs, ${\rm{BNE}}(n)/n$.} We showcase the ${\rm BNE}(n)/ n$-$\kappa/\rho$ correlation for polycrystalline graphene and in-plane irradiated graphite.
  The red line is the best-fit power law in Fig.~2c multiplied by a factor of 3/2 (see main text). The shaded area shows changes in the best-fit power laws when the constant $c$ is changed from 0 to 0.4 $\rm W \cdot m^{-1} \cdot K^{-1} \cdot g^{-1} \cdot cm^3$. The empty diamond markers were calculated using $\kappa_{\rm 300K}$ computed using classical MD simulations by \citet{zhou_million-atom_2025}, and computing BNE$(n)/n$ from the atomic structures released with this reference; filled diamond markers are the same data after applying quantum corrections to account for quantum Bose-Einstein statistics of vibrations (using specific heat of IRG T2, see Appendix~A for details). Numbers on top of the diamond markers "Zhou 2025 (quant-corr)" show grain sizes in nanometres. For circles marked "IRG in-plane", the conductivity was calculated as the average of two largest eigenvalues of the conductivity tensor $\kappa^{\alpha \beta}$, which correspond to in-plane directions of the IRG structures.
  } 
  \label{fig:BNE-tc-polygraphene}
\end{figure}

In this section we discuss ${\rm BNE}(n)/ n$ dependence on the grain size by analyzing structures of polycrystalline graphene (PCG) \cite{zhou_million-atom_2025} with grain sizes ranging from 8 to 50nm. We plot the BNE$(n)/n$ together with $\kappa / \rho$ in Fig.~\ref{fig:BNE-tc-polygraphene}.
We compare results for PCG with conductivity of IRG structures predicted in the in-plane direction. We find in IRG that the in-plane conductivity is approximately $3/2$ times larger than the average trace of $\kappa$ tensor stemming from a negligible conduction in the out-of-plane direction. For this reason, we also report the ${\rm BNE}(n)/ n$-$\kappa/\rho$ correlation from Fig.~2c multiplied by a factor of $3/2$.
We find that the quantum-corrected predictions of \cite{zhou_million-atom_2025} are in quantitative agreement with the ${\rm BNE}(n)/ n$-$\kappa/\rho$ correlation. 
We also find that ${\rm BNE}(n)/ n$, as a descriptor of medium-range order, is most useful for predicting thermal conductivity in structures for which thermal transport is limited by disorder (e.g., room-temperature thermal conductivity for structures analyzed in the main manuscript). In the weak-disorder regime which polycrystalline graphene enters, the influence of intrinsic anharmonicity becomes more important. The correlation also becomes steeper, where small changes in the ${\rm BNE}(n)/ n$ lead to bigger changes in the thermal conductivity, hence prediction of thermal conductivity from BNE$(n)/n$ becomes less precise. This is consistent with how small irradiation dose increases lead to significant thermal conductivity reductions in weakly-irradiated graphite \cite{barabash_effect_2002}.

\FloatBarrier

\section{Similarity of the normalized partial VDOS across different structures}
\begin{figure}[H]
  \centering
\includegraphics[width=\textwidth]{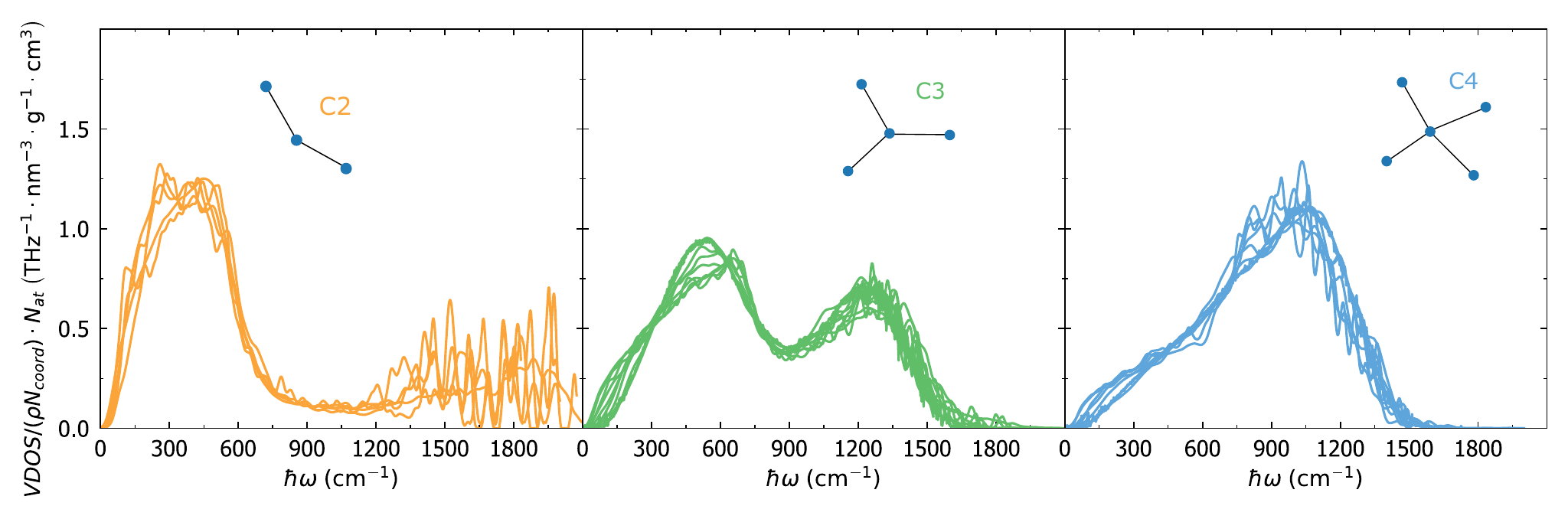}\\[-3mm]
  \caption{\textbf{Comparison of shapes of partial VDOS in AC.} Left, middle and right panels show the normalized PDOS contributions from atoms having coordination number equal to two, three and four respectively. The normalized coordination-resolved PDOS is shown only if at least 5\% of atoms in the structure have the corresponding coordination number. 
  }
  \label{fig:AC_VDOS_shapes}
\end{figure}

In Sec.~V of the main text we anticipated that when the VDOS of AC is resolved in terms of contributions from atoms belonging to a certain coordination environment, these contributions are remarkably similar in models with different density (Fig. \ref{fig:AC_VDOS_shapes}). This suggests that the vibrational properties of AC are mainly determined by short-range order, and in AC the shape of the total VDOS is determined by linear superimposition of coordination-resolved PDOS. This statement is quantitatively confirmed in Fig.~\ref{fig:AC_VDOS_shapes}, where we show that if the coordination-resolved PDOS are normalized to have the same area, their shapes overlap significantly.

\begin{figure}[H]
  \centering
\includegraphics[width=0.4\textwidth]{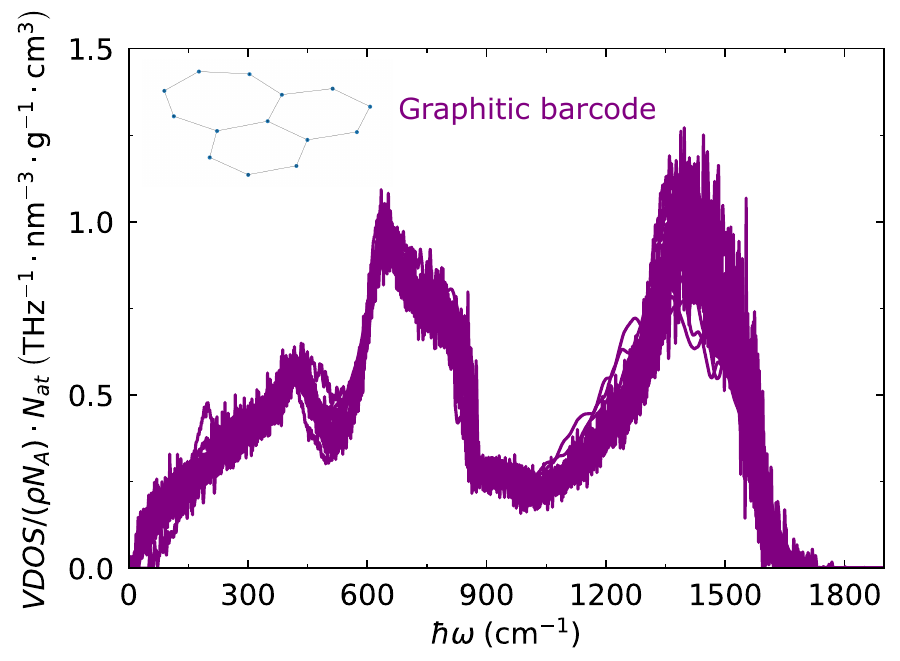}\\[-3mm]
  \caption{\textbf{Comparison of shapes of graphitic-barcode PDOS in CDC, VPC, IRG and PSP.} 
  The normalized graphitic-barcode PDOS is shown only if at least 5\% of atoms in the structure have this barcode.
  }
  \label{fig:Barcode_G_VDOS_shapes}
\end{figure}

Additionally, in Fig.~3 in the main text we showed that, in structures less disordered than AC, the coordination-resolved PDOS have different shapes across different structures. Therefore, we mentioned the necessity to resolve the PDOS in terms of structural motifs having lengthscale longer than the SRO, and showed in Fig.~3 that the $C_3$-PDOS of CDC, VPC and IRG receives significant contributions from a common MRO graphitic barcode.
Fig. \ref{fig:Barcode_G_VDOS_shapes} shows that the normalized graphitic-barcode PDOS has a very similar shape across the CDC, VPC, IRG and PSP structures.
This analysis confirms that the shape of the VDOS contains information on the structural features of disordered solids.

\section{Computational details}
\label{sub:computational_details}

\subsection{Structures}
We analyzed the following five different classes of disorder-carbon phases:
\begin{enumerate}
    \item Amorphous Carbon (AC). We employed the structures discussed in Ref. \cite{deringer_machine_2017}, including: a) two 8000-atom models having densities 1.5 and 3.0 g/cm$^3$, respectively; b) one 1000-atom model having density 3.0 g/cm$^3$; c) 7 216-atom models having densities from 1.5 to 3.0 g/cm$^3$. 
    \item Carbide-Derived Carbon (CDC). We employed the CDC800 and CDC1200 generated by \citet{palmer_modeling_2010} and used in Ref.~\cite{simoncelli_blue_2018}. The ann-CDC1200 structure was obtained by annealing CDC1200 using the GAP potential \cite{rowe_accurate_2020} (more on this later) with the following protocol \cite{deringer_machine_2017}: (i) heating the structure from 0 to 3000 K over 10 ps, and keeping the structure at constant volume; (ii) equilibrating the structure at constant temperature and volume for 10 ps; (iii) annealing the structure from 3000 to 300 K over 20 ps at constant volume.
    \item Variable-Porosity Carbon (VPC). We employed the structures generated in Ref.~\cite{deringer_towards_2018}, in particular: a) three $\sim$ 1000-atom structures (VPC(D)); b) four $\sim$ 200-atom models (these models were necessary for the calculation of the anharmonic linewidths). The 200-atom models were: a) samples 4 and 9 at density 1.0 g/cm$^3$; b) sample 8 at density 1.5 g/cm$^3$; c) and sample 9 at density 2.0 g/cm$^3$.
    The VPC(T) structure was obtained from Ref.\cite{de_tomas_transferability_2019}, \footnote{\url{http://www.carbonpotentials.org/}}.
    \item IRradiated Graphite (IRG), which is the product of neutron irradiation of graphite moderator in nuclear reactors. We employed the structure generated by \citet{farbos_time-dependent_2017}, in particular: a) one 216-atom structure of IRG T9; b) four 14009-atom structures of IRG T2,T3,T5 and T9.
    \item Phase-Separated Phase (PSP). We employed the structure generated by \citet{de_tomas_transferability_2019}.
\end{enumerate}
All densities mentioned above were initial densities of carbon structures before relaxation.

\subsection{Vibrational properties}
\begin{figure*}
  \centering
\includegraphics[width=\textwidth]{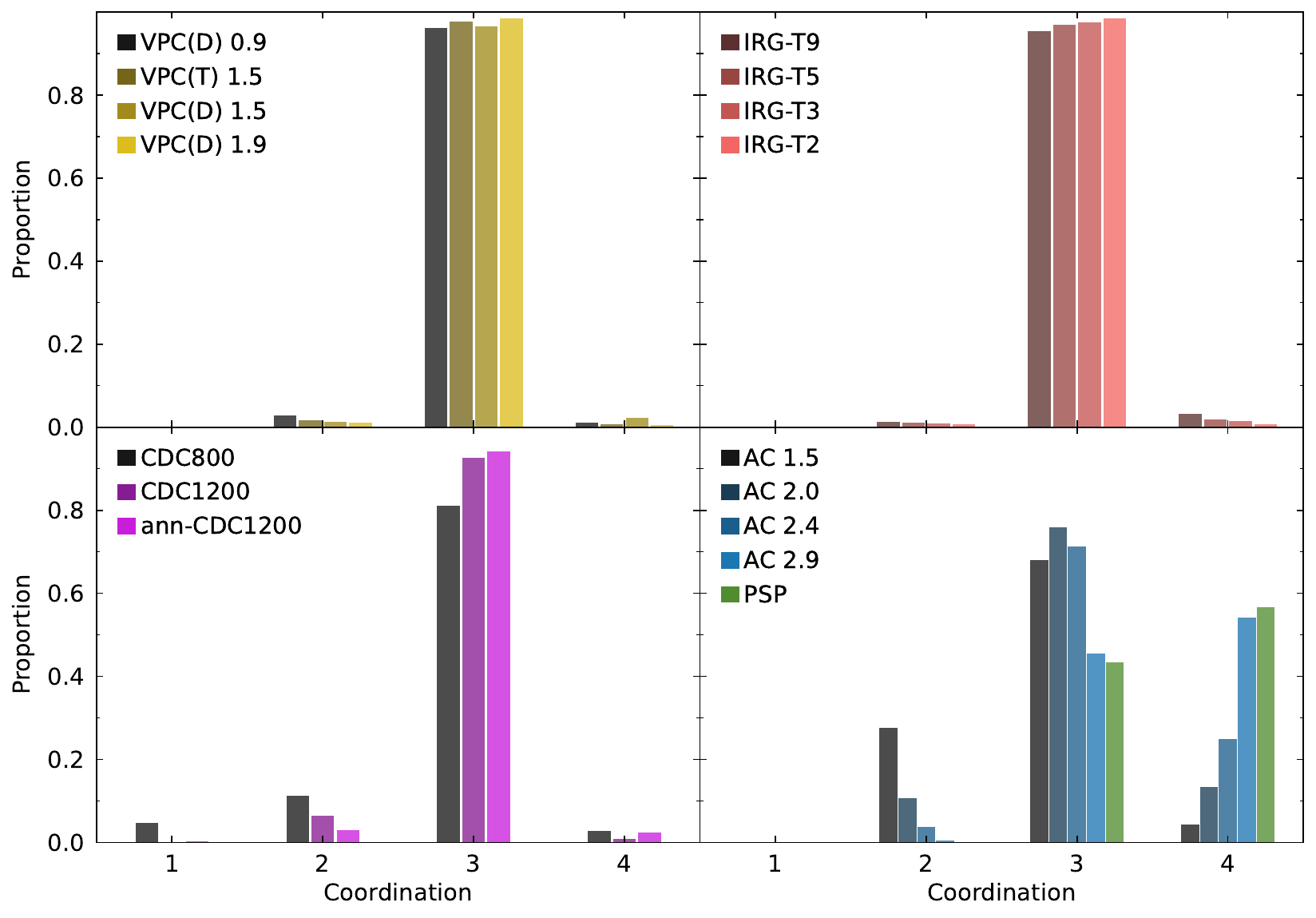}\\[-3mm]
  \caption{\textbf{Coordination number distributions for carbon polymorphs.} For structures, where we have models with different sizes, we report coordination number distribution for the largest model.
  }
  \label{fig:coordination_all_structures}
\end{figure*}

To simulate the vibrational properties of the models above, we employed the Gaussian Approximation Potential (GAP) discussed in Ref. \cite{rowe_accurate_2020}. We obtained interatomic forces and stress tensor using \texttt{LAMMPS} \cite{brown_implementing_2011}, and the \texttt{QUIP} \cite{csanyi_expressive_2007, bartok_gaussian_2010, kermode_f90wrap_2020} interface to call the GAP potential routines. We relaxed the cell parameters and interatomic forces with threshold 0.02 eV/\AA\space for all structures. The details on the residual forces and stresses of the relaxed structures are reported in Table~\ref{tab:relaxation}.
For completeness, we report the coordination number distributions for representative relaxed models from five classes of carbon polymorphs in Fig.~\ref{fig:coordination_all_structures}. To determine the coordination topology we calculated the number of atoms in a sphere of radius equal to the first minimum of the interatomic distance distribution, which is about 1.8 \AA \space for all the structures considered.

We obtained interatomic force constants using 2x2x2 supercells in all the models smaller than 1500 atoms, and a 1x1x1 for all the larger models.
To obtain vibrational frequencies and velocity operator elements of AC, CDC, VPC, PSP and IRG T9 with 216 atoms we used direct diagonalisation and implementation available in \texttt{Phonopy} and \texttt{Phono3py}. For structures of IRG with 14009 atoms we used a direct diagonalization implemented in an in-house code.\\

\begin{table*}[h]
    \centering
    \begin{tabular}{ |p{4cm}||p{4cm}|p{4cm}|p{4cm}| }
     \hline
     \multicolumn{4}{|c|}{Relaxation parameters} \\
     \hline
     Model & Pressure [bar] & Force two-norm [eV/Angstrom] & Force max component [eV/Angstrom]\\
     \hline
     AC 1.5 216 & 0.388 & 0.00214 & 0.00126\\
     AC 1.8 216 & -0.732 & 0.00318 & 0.00148\\
     AC 2.0 216 & -0.551 & 0.00210 & 0.00103\\
     AC 2.1 216 & -0.597 & 0.00208 & 0.000942\\
     AC 2.4 216 & 1.638 & 0.00466 & 0.00282\\
     AC 2.6 216 & 0.403 & 0.00250 & 0.00185\\
     AC 2.9 216 (estimate) & 2.899 & 0.00966 & 0.00717\\
     AC 1.5 8000 & -161.3 & 0.00443 & 0.000523\\
     AC 2.9 1000 & -2.695 & 0.0251 & 0.0184\\
     AC 2.9 8000 & 0.406 & 0.00250 & 0.000655\\
     CDC800 & -70.661 & 0.00335 & 0.000657\\
     CDC1200 & 25.108 & 0.00295 & 0.000743\\
     ann-CDC1200 & -157.956 & 0.00331 & 0.000867 \\
     VPC(D) 0.9 & -0.209 & 0.00310 & 0.000482 \\
     VPC(D) 1.5 & -11.206 & 0.00102 & 0.000272 \\
     VPC(D) 1.9 & 1.438 & 0.00169 & 0.000287 \\
     VPC(T) 1.5 & -12.918 & 0.00417 & 0.000903 \\
     PSP & -26.247 & 0.00406 & 0.000613 \\
     IRG T2 14009 & 0.0538 & 0.00405 & 0.000815 \\
     IRG T3 14009 & 0.693 & 0.00260 & 0.000396\\
     IRG T5 14009 & 1.911 & 0.00489 & 0.000754\\
     IRG T9 14009 & 1391.9 & 0.00709 & 0.00181\\
     IRG T9 216 & 0.721 & 0.00104 & 0.000287\\
     VPC(D) 0.9 (model 4) & -15.672 & 0.000911 & 0.000135\\
     VPC(D) 0.9 (model 9) & -1.270 & 0.000917 & 0.000131\\
     VPC(D) 1.4 (model 8) & 2.893 & 0.000866 & 0.000156\\
     VPC(D) 2.0 (model 9) & 3.095 & 0.00137 & 0.000308\\
     \hline
    \end{tabular}
    \caption{\textbf{Residual stresses and forces in the carbon polymorphs analyzed.} We report the pressure (average trace of the stress tensor) as well as force two-norm, and max component of forces after relaxation.}
    \label{tab:relaxation}
\end{table*}

\subsection{Thermal conductivity}
The thermal conductivity was computed using the approach discussed in Ref.~\cite{simoncelli_thermal_2023}. 
For the 216-atom models of AC, the 1000-atom model of AC, and the 216-atom model of IRG T9 we employed the convergence-acceleration protocol based on the Fourier interpolation \cite{simoncelli_thermal_2023}, sampling vibrational properties on a 5x5x5 $\bm{q}$-mesh. 
For CDC, VPC(D), VPC(T) and PSP, which have a reference cell containing from $\sim$ 1000 to 4000 atoms, we employed the convergence-acceleration protocol using a 3x3x3 $\bm{q}$-mesh.
Finally, for two 8000-atom models of AC and four 14009-atom models of IRG we evaluated the conductivity at $\bm{q}=\bm{0}$ only. The applicability of such protocol to amorphous carbon is confirmed by the good agreement between the following two conductivity calculations (see Fig.~1): (i)  the rWTE conductivity computed using the convergence-acceleration protocol in the small 216-atom model of AC; (ii) the rWTE and bare WTE conductivities in an 8000-atom model of AC calculated at $\bm{q}=\bm{0}$ only (for equivalence between WTE and rWTE in large models of IRG, see Fig.~11).

\begin{figure*}
  \centering
\includegraphics[width=0.5\textwidth]{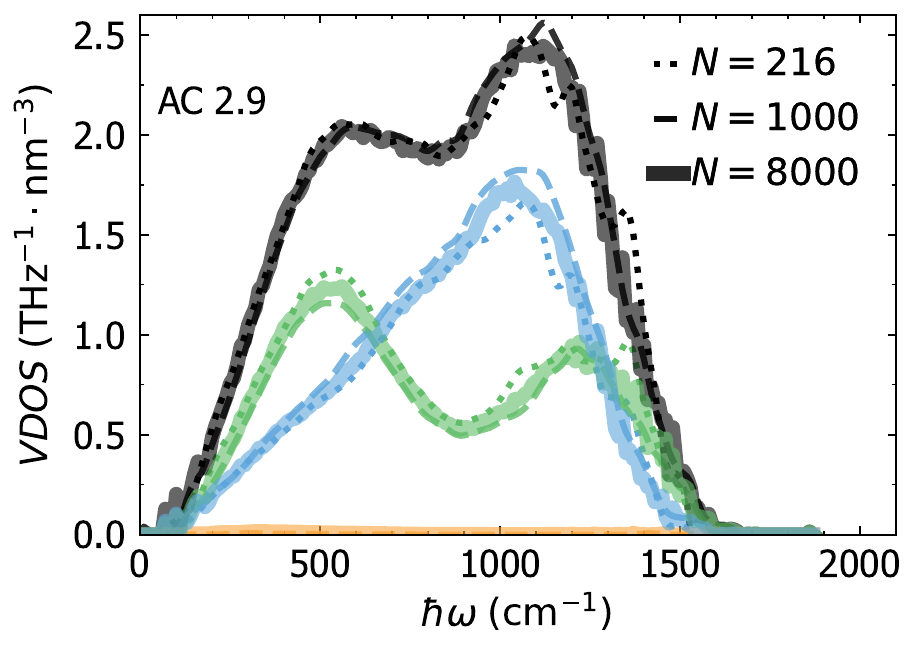}\\[-3mm]
  \caption{\textbf{Dependence of VDOS of AC on the model size.} We show total VDOS (black) and its coordination decompositions (yellow: 2 neighbors, green: 3 neighbors, blue: 4 neighbors) for models of AC 2.9 with 216 (dotted), 1000 (dashed) and 8000 (solid) atoms. The parameter $\eta$ is the broadening of the Gaussian distribution used for the numerical implementation for the Dirac delta appearing in the VDOS expression (Eq.~(6) in the main text), and the values used are 16.0, 16.0 and 1.0 cm$^{-1}$ for atomistic models having size 216, 1000, and 8000 atoms, respectively.
  }
  \label{fig:VDOS_size_analysis}
\end{figure*}

\begin{figure*}
  \centering
\includegraphics[width=\textwidth]{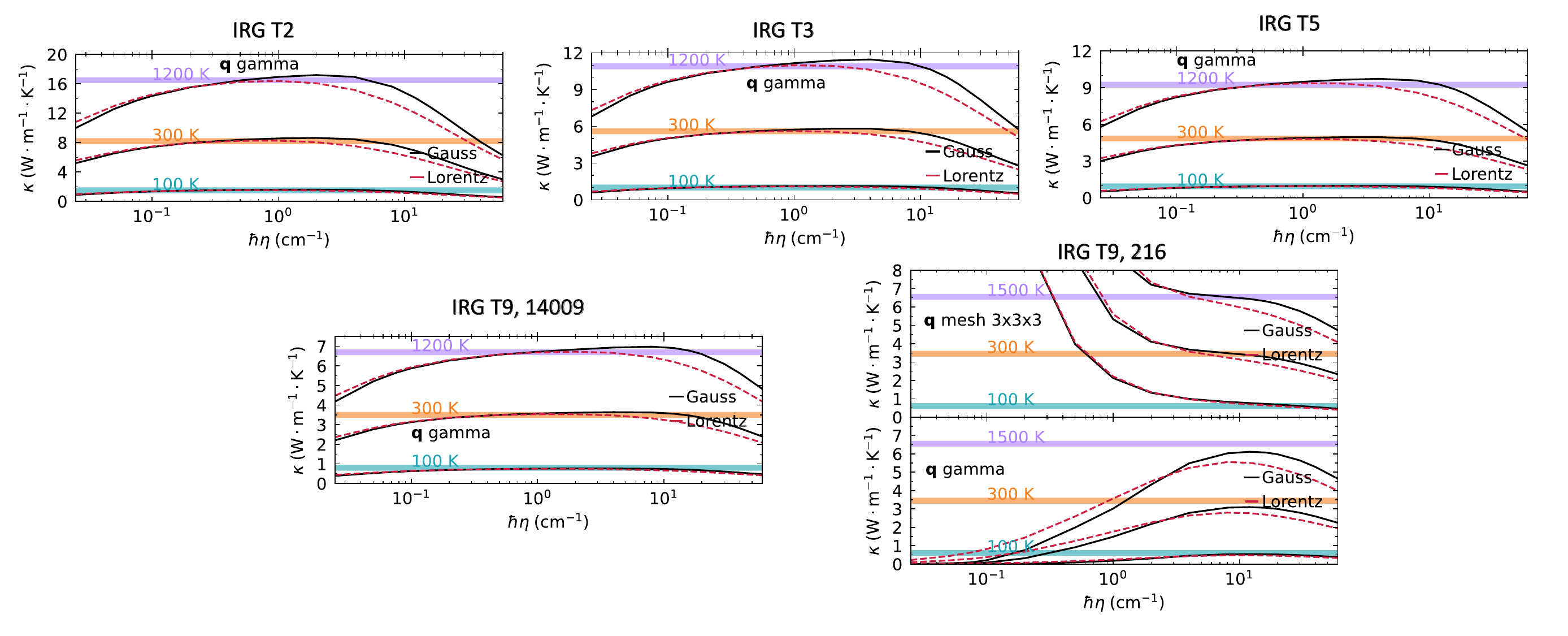}\\[-3mm]
  \caption{\textbf{Convergence plateau tests for IRG}. `Gauss' and `Lorentz' stand for Gaussian and Lorentzian representations of the Dirac delta, respectively.
  }
  \label{fig:conv_plateau_IRG}
\end{figure*}

\begin{figure*}
  \centering
\includegraphics[width=\textwidth]{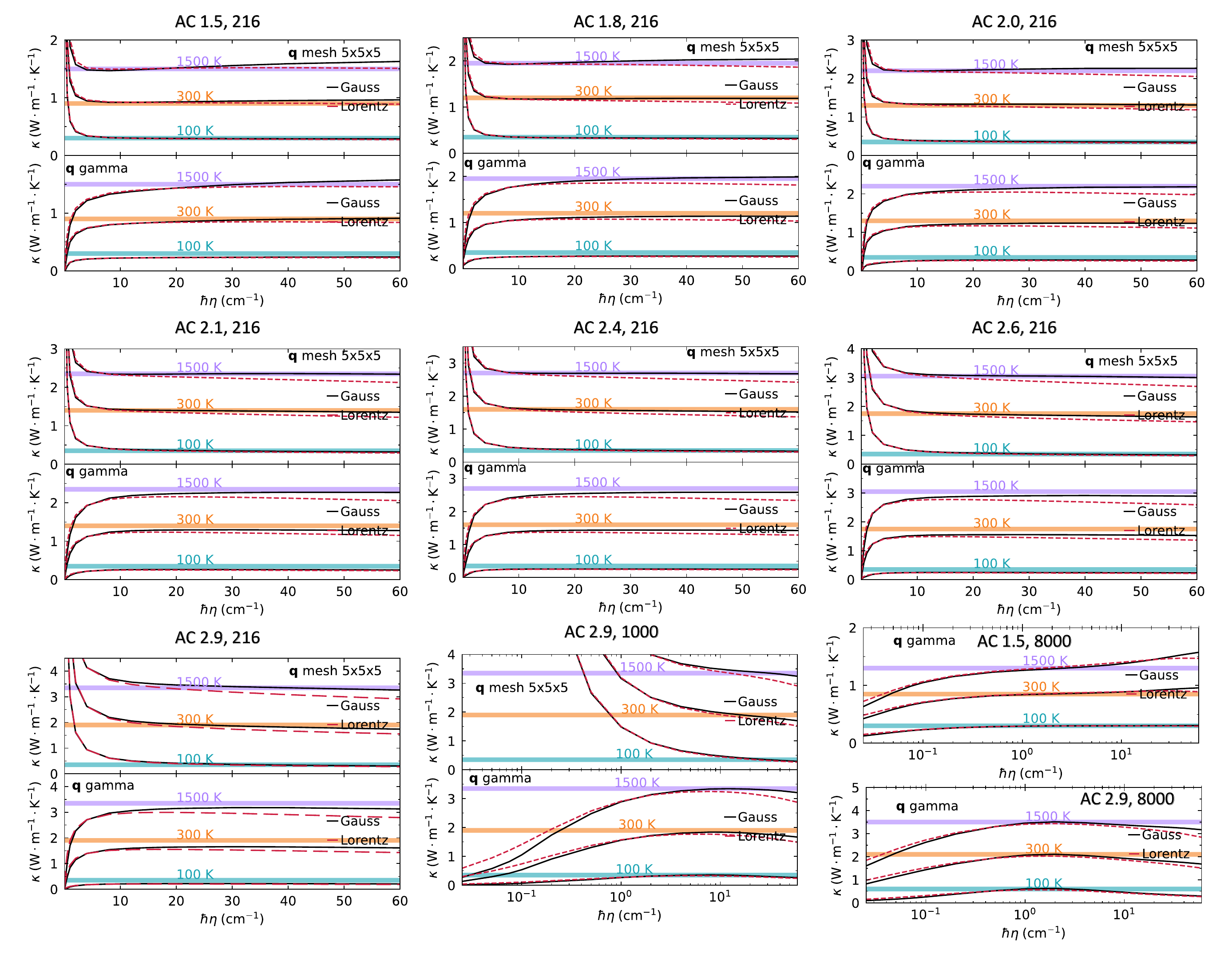}\\[-3mm]
  \caption{\textbf{Convergence plateau tests for AC}. `Gauss' and `Lorentz' stand for Gaussian and Lorentzian representations of the Dirac delta, respectively.
  }
  \label{fig:conv_plateau_AC}
\end{figure*}

\begin{figure*}
  \centering
\includegraphics[width=0.9\textwidth]{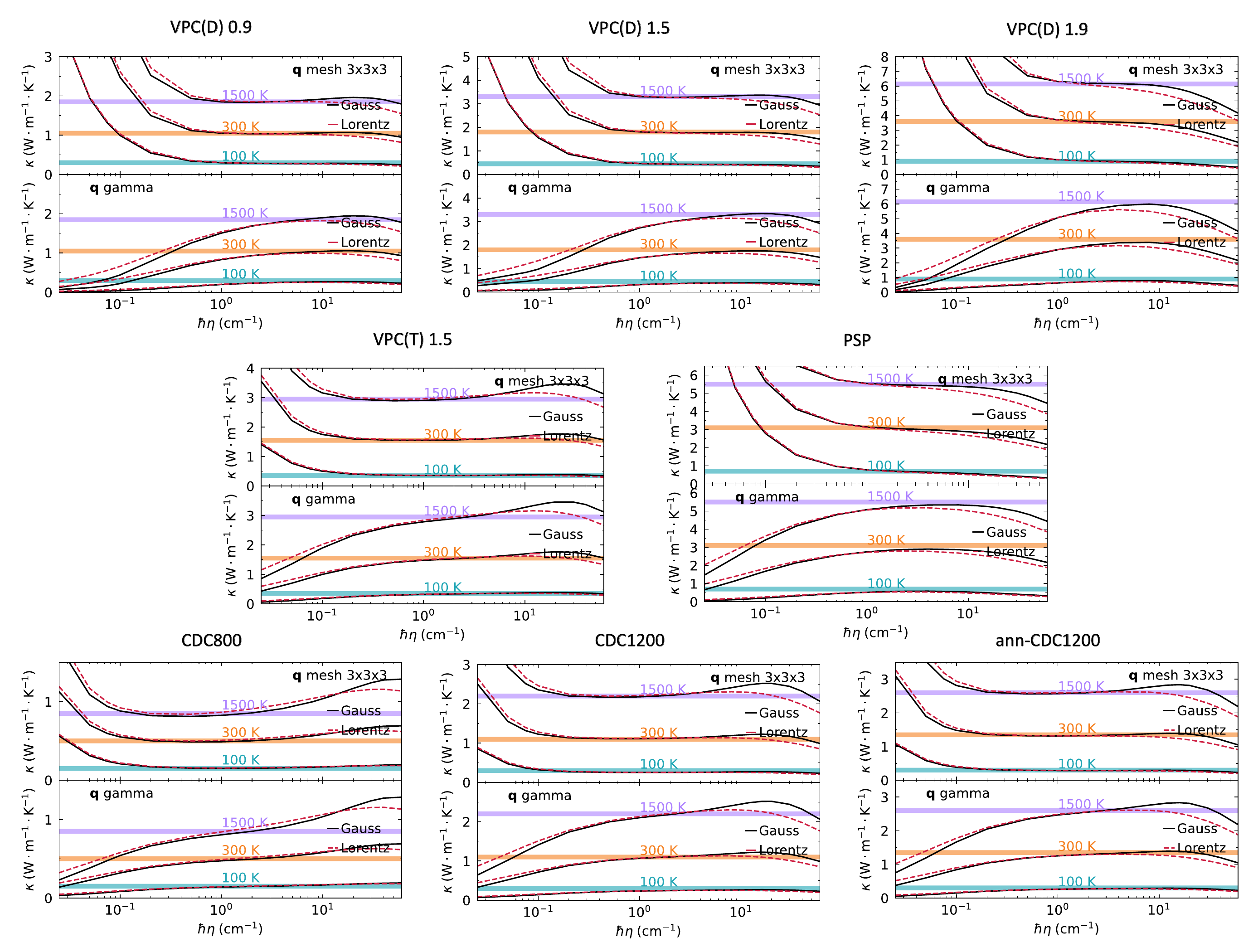}\\[-3mm]
  \caption{\textbf{Convergence plateau tests for VPC, PSP and CDC}. `Gauss' and `Lorentz' stand for Gaussian and Lorentzian representations of Dirac delta respectively.
  }
  \label{fig:conv_plateau_NC}
\end{figure*}

The application of the convergence-acceleration protocol requires finding the value of the numerical broadening parameter $\eta$ appearing in Eq.~(1).  
The parameter $\eta$ is the smallest value for which the AF conductivity does not depend on the value of $\eta$ \cite{allen_thermal_1993}. In Table~\ref{tab:convergence_plateau} we report the values of $\eta$ used in this study, and in Figs.~\ref{fig:conv_plateau_IRG}, \ref{fig:conv_plateau_AC} and \ref{fig:conv_plateau_NC} we show the convergence plateau plots used to determine them. In Fig.~\ref{fig:VDOS_size_analysis} we also show that model size has a small influence on the distribution of predicted frequencies, i.e., small models have VDOS which is a good approximation of the VDOS of larger models if appropriate values of the parameter $\eta$ are chosen. Specifically, for heavily disordered amorphous carbon we find in line with previous work \cite{harper_vibrational_2024} that larger models have VDOS of similar shape and lower roughness compared to VDOS of smaller models.

\begin{figure}[H]
  \centering
\includegraphics[width=\textwidth]{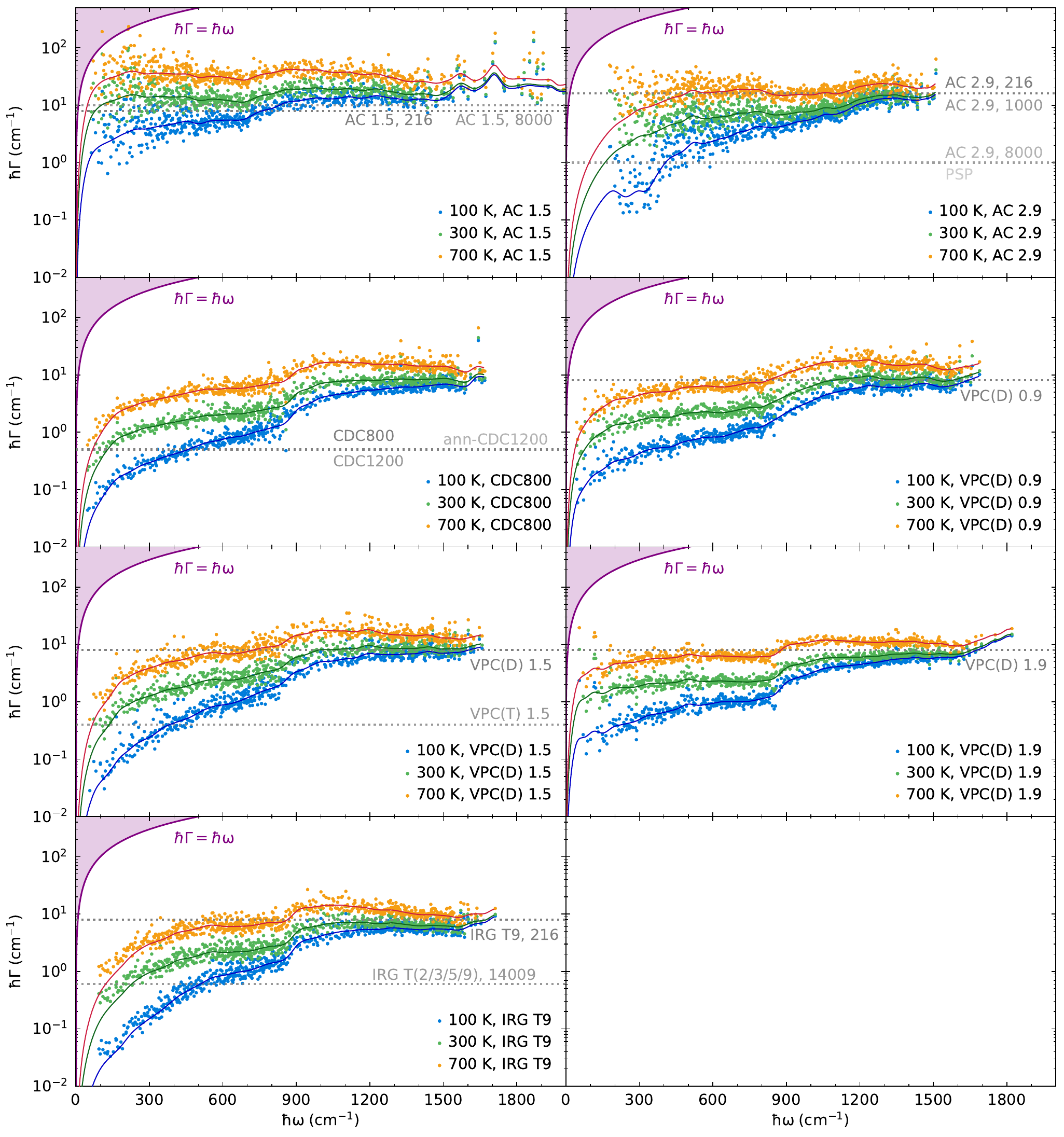}\\[-3mm]
  \caption{\textbf{Comparison of intrinsic linewidths.} The blue, green and orange colors are linewidths calculated at 100, 300 and 700 K, respectively. The dotted horizontal lines show the values of $\eta$ chosen from the convergence plateaus, see Table \ref{tab:convergence_plateau}.
  }
  \label{fig:linewidth}
\end{figure}

We computed the intrinsic linewidths accounting for third-order anharmonicity and isotopic-mass disorder at natural abundance.
Third-order force constants were computed in the reference cell using \texttt{ShengBTE} \cite{li_shengbte_2014} with 8$^{th}$ nearest-neighbour cutoff, and then converted to \texttt{Phono3py} \cite{togo_distributions_2015} format using the \texttt{hiphive} \cite{eriksson_hiphive_2019} package. Using \texttt{Phono3py} we calculated intrinsic linewidths only at $\bm{q} = \bm{0}$, and using a Gaussian smearing of $\hbar \sigma = 0.24$ cm$^{-1}$.
This approach was employed in two amorphous carbon 216-atom models at densities 1.5 and 3.0 g/cm$^3$, four 200-atom models of VPC, and for one 216-atom model of IRG T9.
The obtained linewidths were then extrapolated as a function of frequency using the approach discussed in Refs.~\cite{simoncelli_thermal_2023,pazhedath_first-principles_2024} and used for calculation of conductivity in larger models. The energy-linewidth distributions and the frequency-extrapolated coarse-grained functions for all the structures studied are shown in Fig. \ref{fig:linewidth}.

For amorphous carbon with 216 atoms we linearly extrapolated the value of the linewidth from two AC structures at initial density 1.5 and 3.0 g/cm$^3$ using the density. 
For AC structures with 8000 atoms at initial densities 1.5 and 3.0 g/cm$^3$ and 1000-atom structure at 3.0 g/cm$^3$ we used the linewidth of AC models with 216 atoms at similar initial densities.
For CDC structures we used the linewidth from 200-atom VPC model number 4 from Ref.~\cite{deringer_towards_2018} at initial density 1.0 g/cm$^3$. For VPC structures with $\sim$ 1000 atoms\cite{deringer_towards_2018} we employed the linewidths computed from their $\sim$ 200-atom equivalents: the atomistic model number 9 with density 1.0 g/cm$^3$, the model number 8 with density 1.5 g/cm$^3$, and the model number 9 with density 2.0 g/cm$^3$. The linewidths of atomistic model number 8 with density 1.5 g/cm$^3$ were also used for VPC(T) with density 1.5 g/cm$^3$.
For PSP we used linewidth from 216-atom model of AC with initial density of 3.0 g/cm$^3$. For IRG we used the linewidth from 216-atom structure of IRG T9. 

We conclude by noting the approximated treatment of anharmonicity discussed above is justified in Fig.~11, where we showed that, as long as the linewidth are finite and enable overlap between neighboring eigenstates, the values of the intrinsic linewidth do not significantly affect the value of the thermal conductivity in the temperature range 50-300 K.

\begin{table*}[h]
    \centering
    \begin{tabular}{ |p{4cm}||p{4cm}|p{4cm}|p{4cm}|  }
     \hline
     \multicolumn{4}{|c|}{Convergence plateau parameters} \\
     \hline
     Model & $\eta$ (cm$^{-1}$) & Model & $\eta$ (cm$^{-1}$)\\
     \hline
     AC 1.5 216 & 8.0 & ann-CDC1200 & 0.5 \\
     AC 1.8 216 & 8.0 & VPC(D) 0.9 & 8.0 \\
     AC 2.0 216 & 8.0 & VPC(D) 1.5 & 8.0 \\
     AC 2.1 216 & 12.0 & VPC(D) 1.9 & 8.0 \\
     AC 2.4 216 & 12.0 & VPC(T) 1.5 & 0.4 \\
     AC 2.6 216 & 16.0 & PSP & 1.0 \\
     AC 2.9 216 & 16.0 & IRG T2 14009 & 0.6 \\
     AC 1.5 8000 & 10.0 & IRG T3 14009 & 0.6 \\
     AC 2.9 1000 & 16.0 & IRG T5 14009 & 0.6 \\
     AC 2.9 8000 & 1.0 & IRG T9 14009 & 0.6 \\
     CDC800 & 0.5 & IRG T9 216 & 8.0 \\
     CDC1200 & 0.5 & & \\
     \hline
    \end{tabular}
    \caption{\textbf{Convergence plateau values of the broadening $\eta$ appearing in the AF and rWTE conductivities.} The parameter $\eta$ is determined as the beginning of the convergence plateau shown in Figs.~\ref{fig:conv_plateau_IRG}, \ref{fig:conv_plateau_AC}, and \ref{fig:conv_plateau_NC}. }
    \label{tab:convergence_plateau}
\end{table*}

\subsection{Green-Kubo molecular dynamics simulation}

\begin{figure*}
  \centering
\includegraphics[width=\textwidth]{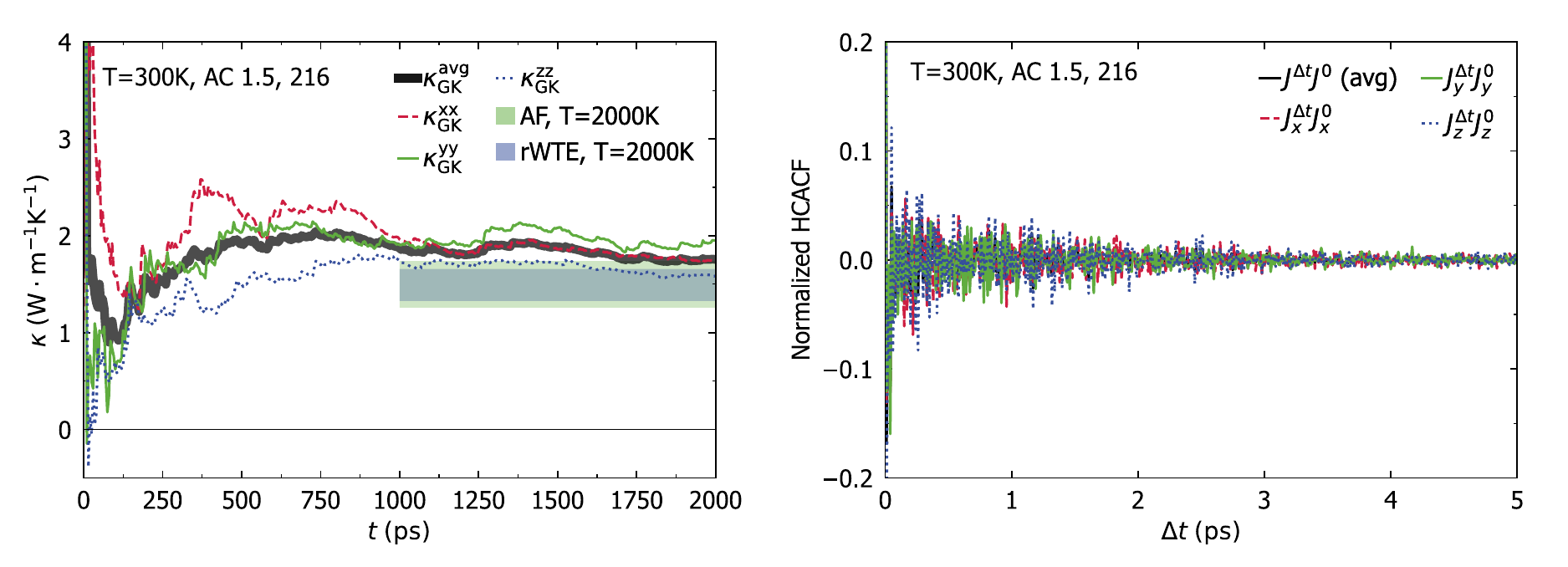}\\[-3mm]
  \caption{\textbf{Convergence of GK thermal conductivity and normalized heat current autocorrelation function (HCACF) with the correlation time.} On the left panel we show the convergence of GK conductivity prediction at room temperature for model of AC 1.5 with 216 atoms. The solid black, dashed red, solid green and dotted blue lines are respectively the trace of GK conductivity tensor and GK conductivity in the x, y and z directions as a function of simulation time. The shaded green and blue areas show the variations of the eigenvalues of AF and rWTE conductivity tensors in the classical limit ($T=2000$ K). The right panel shows the room temperature HCACF $\langle J_{\alpha}(t) J_{\alpha}(0) \rangle$ averaged over samples from the complete simulation, which is normalized by the $t=0$ product: $\langle J_{\alpha}(0) J_{\alpha}(0) \rangle$. The solid black, dashed red, solid green and dotted blue lines are respectively the average normalized HCACF and normalized HCACF in x, y and z directions.
  }
  \label{fig:GK_vs_ALD_conv}
\end{figure*}

To calculate GK thermal conductivity, we used \texttt{LAMMPS} \cite{brown_implementing_2011}, and the \texttt{QUIP} \cite{csanyi_expressive_2007, bartok_gaussian_2010, kermode_f90wrap_2020} interface to call the Gaussian Approximation Potential (GAP) \cite{rowe_accurate_2020}. At each temperature $T$, we first set the distribution of initial velocities to a gaussian distribution with temperature $T$ and then equilibrated the structure for 100ps with a timestep of 1fs in NVT ensemble. The heat flux simulation was done in NVT ensemble (as was similarly done for amorphous carbon e.g., in Ref.~\cite{moon_crystal-like_2025}) with total time equal to 2ns, heat flux sampled every 10fs and the sampling length of the heat current (flux) autocorrelation function (HCACF) $\langle J(t) J(0) \rangle$ set to 500, such that the total correlation time is $\approx 5$ps.
In Fig.~\ref{fig:GK_vs_ALD_conv} we report the convergence of the GK thermal conductivity prediction with the simulation time. The GK conductivities were extracted from the average predictions during the last 250 ps of the simulation. We found that the anisotropy of the GK $\kappa$ vector is similar to the anisotropy found in the rWTE prediction and is of the order of 20\% for the AC 1.5 with 216 atoms. We additionally checked that the HCACF \cite{giri_atomic_2022} $\langle J(t) J(0)\rangle$ decays asymptotically to zero within the correlation time of $\approx 5$ps set in the simulation.\\\\
Under GK formalism \cite{giri_atomic_2022, grasselli_invariance_2021}, the diagonal elements of thermal conductivity tensor are given by time integral:
\begin{equation}
    \kappa^{\alpha \alpha} = \frac{\mathcal{V}}{k_{\rm B} T^2} \int_0^{\infty} \langle J_{\alpha}(t) J_{\alpha}(0) \rangle {\rm d}t.
\end{equation}
For the heat flux we used the standard heat-flux expression implemented in \texttt{LAMMPS} \footnote{\url{https://docs.lammps.org/compute_heat_flux.html}. We note that other definitions of the heat flux are possible and discussed in the literature \cite{ercole_gauge_2016, grasselli_invariance_2021, caldarelli_many-body_2022, tai_revisiting_2025}. The investigation of the differences between these different choices goes beyond the scope of the present work.}:
\begin{equation}
    \bm{J} = \frac{1}{\mathcal{V}} \left[ \sum_b e_b \bm{v}_b - \sum_b \bm{S}_b \bm{v}_b\right],
\end{equation}
where $e_b$ is the per-atom energy, $\bm{v}_b$ is the velocity of atom $b$, and $\bm{S}_b$ is the per-atom stress tensor, which was computed using the following command in LAMMPS:
\verb|compute   myStress all stress/atom NULL virial|, 
where stress/atom method was used as it is available for the many-body potentials.

\subsection{Other computational details}
The structures shown in Figs.~1, 2, and \ref{fig:conductivity_temperature} were visualized using VESTA \cite{momma_vesta_2011}.
The power-law fits in Fig. 2 were calculated using \texttt{statsmodels} \cite{seabold2010statsmodels}. Specifically, only the structures for which thermal conductivity was computed using rWTE in this paper were used to fit the power-laws (they are visible as circles in the Fig.~2). The best-fit power law $\kappa/\rho = ({\rm BNE}(n)/n)^a \times e^b + c$ was determined by transforming it a linear problem $ \ln{[\kappa/\rho - c}] = a \ln{[{\rm BNE}(n)/n]} + b$ and then solving for values of $a$ and $b$ using linear regression for a given value of $c$. The final values of $a$, $b$ and $c$ are extracted from the power-law fit with the highest $R^2$ taken from the fits with $c$ in range from 0 to 0.4 $\rm W \cdot m^{-1} \cdot K^{-1} \cdot g^{-1} \cdot cm^3$. The best power-law fits for ${\rm BNE}(n)/ n$-$\kappa/\rho$ correlations at 100 and 700 K (Fig.~\ref{fig:conductivity_entropy_100_700K}) were done in a similar manner.
The $H_1$ barcode visualisations in Figs. 2, 3, and \ref{fig:Barcode_G_VDOS_shapes} were generated using \texttt{NetworkX} \cite{SciPyProceedings_11}. The Spearman rank correlation coefficient for the data in Fig.~2 was calculated using \texttt{Scipy}\cite{2020SciPy-NMeth}.
The VDOS in Fig. 3 were calculated by replacing the delta functions with Gaussian distributions having variance $\sigma^2$ related to the value of $\eta$ from convergence plateau ($\sigma^2 = \frac{\pi}{2} \eta^2$, Eq.~(3)), with exception for AC 2.0, where we chose higher $\eta = 16$ cm$^{-1}$. 
The VDOS and its decompositions in Fig. 3 were calculated on a 5x5x5 $\bm{q}$-mesh for AC structures with 216 atoms, and at $\bm{q}=\bm{0}$ point only for the rest of the structures. The calculation of coordination profile in Fig.~\ref{fig:LAE_description} was done using in-house python implementation, for primitive ring profile we used the implementation in \texttt{Swatches} software together with in-house python pre- and post-processing scripts.

\section{Data availability} 
The atomistic models of the carbon polymorphs studied in this work are available on the Materials Cloud Archive 
\footnote{K. Iwanowski, G. Csányi, and M. Simoncelli, Supporting data for `Bond-Network Entropy Governs Heat Transport in Coordination-Disordered Solids'. \textit{Materials Cloud Archive} \url{https://doi.org/10.24435/materialscloud:nh-ep} (2025)}
\cite{talirz_materials_2020}

\section{Code availability} 
\texttt{LAMMPS}  \cite{brown_implementing_2011} is available at \url{www.lammps.org} and the \texttt{QUIP}  \cite{csanyi_expressive_2007, bartok_gaussian_2010, kermode_f90wrap_2020} interface between \texttt{LAMMPS} and GAP potential is available at \url{github.com/libAtoms/QUIP}. \texttt{Phonopy} and \texttt{Phono3py} are available at \url{github.com/phonopy} and the scripts to calculate third order force constants using finite difference method are available at \url{bitbucket.org/sousaw/thirdorder}.  \texttt{statsmodels}, \texttt{NetworkX} and  \texttt{Scipy} are available at \url{http://statsmodels.sourceforge.net/}, \url{https://networkx.org/} and \url{https://scipy.org/} respectively. \texttt{Swatches} is available at \url{https://github.com/bschweinhart/Swatches/tree/master}.


\providecommand{\noopsort}[1]{}\providecommand{\singleletter}[1]{#1}%